%% file: main.tex
\documentclass[a4paper, amsfonts, amssymb, amsmath, reprint, showkeys, nofootinbib,twoside]{revtex4-1}
\usepackage[english]{babel}
\usepackage{subfig}
\usepackage{relsize}
\usepackage{chngcntr}
\usepackage{ragged2e}
\DeclareCaptionJustification{straight}{\justify}
\usepackage[format=plain,justification=straight]{caption}
\usepackage[utf8]{inputenc}
\usepackage[colorinlistoftodos, color=green!40, prependcaption]{todonotes}
\input{preamble}
\usepackage[pdftex, pdftitle={Article}, pdfauthor={Author}]{hyperref}

\newcommand*{\qrq}{\quad \Rightarrow \quad}

\newcommand*{\pd}{\partial}
\newcommand{\msun}{\mathrm{M}_\odot}

\newcommand{\un}[1]{\,\mathrm{#1}}
\newcommand{\E}[1]{\times 10^{#1}}
\usepackage{soul}

\def\apj{{\rm ApJ}}                 
\def\apjl{{\rm ApJ}}                
\def\mnras{{\rm MNRAS}}             
\def\prd{{\rm Phys.~Rev.~D}}        

\def\mnras{Mon.\ Not.\ R.\ Astron.\ Soc.\ }

\begin{document}
\title{Quantifying the Line-of-Sight Halo Contribution to the Dark Matter Convergence Power Spectrum from Strong Gravitational Lenses}

\author{Atınç Çağan Şengül}
    \email{sengul@g.harvard.edu}
    \affiliation{Harvard University, Department of Physics, Cambridge, MA}
\author{Arthur Tsang}
    \email{atsang@g.harvard.edu}
    \affiliation{Harvard University, Department of Physics, Cambridge, MA}
\author{Ana Diaz Rivero}
    \affiliation{Harvard University, Department of Physics, Cambridge, MA}   
\author{Cora Dvorkin}
    \affiliation{Harvard University, Department of Physics, Cambridge, MA}
\author{Hong-Ming Zhu}
    \affiliation{University of California, Berkeley, Department of Physics, Berkeley, CA}
\author{Uroš Seljak} 
  \affiliation{University of California, Berkeley, Department of Physics, Berkeley, CA} 
 
\date{\today} 

\begin{abstract}
Galaxy-galaxy strong gravitational lenses have become a popular probe of dark matter (DM) by providing a window into structure formation on the smallest scales. In particular, the convergence power spectrum of subhalos within lensing galaxies has been suggested as a promising observable to study DM. However, the distances involved in strong-lensing systems are vast, and we expect the relevant volume to contain line-of-sight (LOS) halos that are not associated with the main lens. We develop a formalism to calculate the effect of LOS halos as an effective convergence power spectrum. The multi-lens plane equation couples the angular deflections of consecutive lens planes, but by assuming that the perturbations due to the LOS halos are small, we show that they can be projected onto the main-lens plane as effective subhalos. We test our formalism by simulating lensing systems using the full multi-plane lens equation and find excellent agreement. We show how the relative contribution of LOS halos and subhalos depends on the source and lens redshift, as well as the assumed halo and subhalo mass functions. For a fiducial system with fraction of DM halo mass in substructure $f_{\rm sub}=0.4\%$ for subhalo masses $[10^5-10^8]\rm{M}_{\odot}$, the interloper contribution to the power spectrum is at least several times greater than that of subhalos for source redshifts $z_s\gtrsim0.5$. Furthermore, it is likely that for the SLACS and BELLS lenses the interloper contribution dominates: $f_{\rm sub}\gtrsim2\%$ ($4\%$) is needed for subhalos to dominate in SLACS (BELLS), which is higher than current upper bounds on $f_{\rm sub}$ for our mass range. Since the halo mass function is better understood from first principles, the dominance of interlopers in galaxy-galaxy lenses with high-quality imaging can be seen as a significant advantage when translating this observable into a constraint on DM.
\end{abstract}

\maketitle
\section{Introduction} \label{sec:intro}

The particle nature of dark matter (DM) remains one of the biggest puzzles in modern cosmology. In the standard cosmological model, the Lambda Cold Dark Matter ($\Lambda$CDM) model, DM is assumed to be cold and non-dissipative, and only interacts via gravity, meaning that it does not interact with light, with baryons, or with itself. Structure formation based on the $\Lambda$CDM paradigm \cite{1981ApJ...250..423D,Blumenthal:1982mv,Blumenthal:1984bp,Davis:1985rj}
has been extremely successful at predicting and explaining many different observables in our universe \cite{BOSS,Pantheon,Planck2018}. Nevertheless, it remains untested on small (sub-galactic) scales. 

The reason for this is two-fold. On the theoretical side, making predictions on these scales is complicated by the fact that they are deeply in the non-linear regime, meaning that high-resolution simulations are necessary. Furthermore, baryonic physics cannot be ignored \cite{Brooks_Zolotov,Brooks_2013,Arraki,Onorbe_2015,Wetzel_2016,2016MNRAS.457.1931S,2017MNRAS.468.2283C,2017MNRAS.467.4383S,2017arXiv170103792G}, so in fact $N$-body simulations do not suffice and hydrodynamical ones are required. Not only are these simulations very computationally expensive, but how to model astrophysical phenomena accurately remains an open problem (see, e.g., Ref. \cite{Vogelsberger:2019ynw} for an overview of different approaches to modeling baryonic physics in cosmological hydrodynamical simulations). On the observational side we face another considerable challenge: the efficiency of galaxy formation decreases with decreasing halo mass  \cite{Fitts:2016usl,2017MNRAS.467.2019R}, meaning that small-scale dark matter halos are largely devoid of gas and stars, unlike their more massive counterparts. While we have been able to find some of these small-scale halos in the Local Group by the faint galaxies they host (e.g., Refs. \cite{2015ApJ...807...50B,Koposov:2015cua,Drlica-Wagner:2015ufc,Homma_2017}), we cannot rely on light to find them further away from the Milky Way (MW). This, and the fact that the contribution and impact of astrophysical processes is currently unclear, motivates pursuing a purely gravitational method to probe sub-galactic scales and consequently test the $\Lambda$CDM paradigm in this regime.

To date, the most prominent method used to probe small scales is strong gravitational lensing. The idea is that, while there is a large dark matter halo doing the lensing (which we will henceforth refer to as the main lens or host), additional, smaller halos can perturb the images generated by the main lens. By comparing the observed images (and their fluxes, in the case of lensed time-varying sources such as quasars or supernovae) to those that would be created solely by the main dark matter halo, we can infer the presence of additional dark matter clumps \cite{Mao,Nierenberg:2017vlg,Gilman:2017voy,grav_imaging1,spat_res_spec1,Hezaveh_powerspec,pcatlens,Birrer:2017rpp,Brewer:2015yya,dark_census,powerspec1,mining_substructure,DiazRivero:2019hxf}. This method has been successfully used to find small clumps in several different lensing systems and place some constraints on the particle nature of dark matter \cite{spat_res_spec2,detection_2010_mnras,vegetti_nature,bells_2018,2014MNRAS.442.2017V,Vegetti:2018dly}.

While this method was proposed at the turn of the century \cite{Mao}, until recently most analyses of strong gravitational lenses in this context have been done under the assumption that the additional clumps lie within the dark matter halo of the main lens. These types of clumps are commonly referred to as subhalos or substructures. However, it has been claimed that a large (in fact most likely larger) contribution to the perturbations in strong lenses comes from free dark matter clumps along the line of sight (LOS) \cite{Li_los}. These halos are commonly referred to as LOS halos or interlopers. Their contribution to lensing observables is an active area of study \cite{daloisio11, McCully:2016yfe,Despali_los,Gilman:2019vca}. 

Evidently, it is crucial to take the LOS contribution into account before making any claim about dark matter; otherwise, we risk wrongfully falsifying or reinforcing the standard $\Lambda$CDM scenario. Recent analyses of strong gravitational lenses have begun to take it into account when placing constraints on $\Lambda$CDM \cite{Vegetti:2018dly,Gilman:2019nap,Gilman:2019bdm,Hsueh:2019ynk,Wong:2019kwg,Rusu:2019xrq,Chen:2019ejq}. If, as expected, the contribution of interlopers really is greater than that of substructure, this could be good news for the ability of lensing observations to constrain the properties of dark matter: while subhalos are subject to messy, ill-understood processes as they travel through the main lens halo (such as tidal disruption), by virtue of being in the field, LOS interlopers are much less subject to environmental effects that might cause them to be disrupted.

Let us present a toy example that illustrates both the benefit gained from having the LOS contribution supersede that of subhalos and also how failing to take into account the LOS contribution can bias any inference about the particle nature of dark matter. Let us assume that the cold dark matter (CDM) paradigm really is the true dark matter model in our universe. In this scenario, the subhalo mass function rises steeply at the low-mass end and we expect a very high number of subhalos. However, if by traveling within their host's halo a large number of them are tidally disrupted, effectively disappearing, the observable number of subhalos might be a lot smaller than the expected number of subhalos under the CDM assumption. If the subhalo contribution is dominant, so we only consider subhalos, we might wrongfully falsify CDM if we do not observe a certain number of subhalos in a given mass range: for example, we may attribute the lack of halos to warm or self-interacting dark matter. If the LOS contribution really is dominant, then the lack of detection of halos in a given mass range is a much more faithful reflection of the fact that there may be some exotic dark matter physics reducing the number of halos with respect to the CDM expectation.

In this paper, we focus on the LOS contribution to the convergence power spectrum. This observable has been analyzed extensively in the context of the subhalo contribution \cite{Hezaveh_powerspec,powerspec1,powerspec2,Brennan} (it has interchangeably been referred to as the substructure power spectrum) and identified as a powerful statistical method to constrain dark matter from strong gravitational lens images. It is a particularly valuable observable because it ties mass scales (what dark matter theories provide) to length scales (deflection angles on the lens plane). The amplitude, shape, and slope of the power spectrum all contain valuable information that can be tied back to dark matter theories \cite{powerspec1,powerspec2}. However, because previous analyses have neglected the LOS contribution, some features that have been deemed significant in past works may not be if the interlopers are included in the analysis. On the other hand, the dominance of the interloper contribution could facilitate deriving constraints on DM from the convergence power spectrum. 

This paper is organized as follows. 
In \S\ref{sec:analytic}, we quantify the LOS contribution to the convergence power spectrum analytically by deriving an effective convergence for the LOS halos. In \S\ref{sec:numeric}, we quantify the same contribution numerically by simulating a multi-plane lens system populated by LOS halos, then solving the multi-plane lens equation without any approximation. We discuss our findings and conclude in \S\ref{sec:conclusions}.

We shall refer to the halo that dominates the strong lensing as the \textit{main lens}, the LOS halos as \textit{interlopers} and the halos within the main lens as \textit{subhalos} throughout the rest of this paper. When we are agnostic to whether a perturbation is due to subhalos or interlopers we refer to them as \textit{perturbers}. Since we incorporate the LOS contribution to the power spectrum formalism, we will exclusively refer to this observable as the (effective) convergence power spectrum, instead of the substructure power spectrum, throughout the remainder of this paper. We assume flat $\Lambda$CDM cosmology when calculating distances and the halo mass function.

\section{Analytical Calculation} \label{sec:analytic}

In the case where perturbations to strong-lens images are assumed to be caused by subhalos, the relative length scales in the problem are the physical size of the main lens along the line of sight and the distance traveled by light rays from the source to the observer. Obviously, the former is orders of magnitude smaller than the latter. Therefore, all the mass that is doing the lensing can be thought of as being on a single thin-lens plane (aptly called the thin-lens approximation). The convergence power spectrum calculation in this case is relatively straightforward since the convergence field is well defined. The addition of interlopers complicates the calculation since there is no well-defined convergence for a case with multiple consecutive thin-lens planes where each one deflects the light rays before they go onto the next plane. The angular deflections are not only added as vector fields but also are coupled to each other. 

In this section we circumvent these problems by defining an effective convergence for a special case with a massive main lens coupled to low-mass interlopers. We will first go over some fundamentals of multi-plane lensing before deriving this effective convergence and, ultimately, arriving at expressions for the convergence power spectrum in the presence of interlopers in front of and behind the main lens.

\subsection{Multi-plane Lens Equation}

\begin{figure}[ht]
    \centering
    \includegraphics[width=\linewidth]{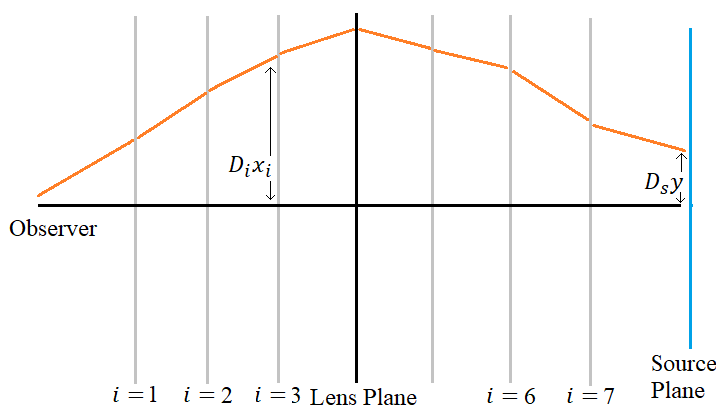}
    \caption{\footnotesize{A simplified 2D sketch of the strong-lens system with interlopers. This is an example of a multi-lens plane system with $N=7$ and $l=4$. The orange line depicts the path that a single light ray travels from the source to the observer. The vertical gray lines correspond to the interloper planes, the vertical black line corresponds to the main-lens plane, and the vertical blue line to the source plane. $D_ix_i$ is the physical distance between the point where the light ray intersects the $i$th plane and the main axis that connects the center of the source plane to the observer. $D_s y = D_{N+1}x_{N+1}$ is the physical distance between the origin of the light ray and the center of the source plane.}}
    \label{fig:lensing_sketch}
\end{figure}

We model the main lens and interlopers as $N$ consecutive thin-lens planes at redshifts $z_i$, where $i=1,\,2,...,l,...,\,N$ and $i>j$ implies $z_i>z_j$ (see Fig. \ref{fig:lensing_sketch}). The main-lens plane is indexed by $l$ and the source plane by $s = N+1$. 

Let us start with the multi-plane lens equation,
\begin{equation}\label{lens_eq}
    \vec{y}  = \vec{x}_1 - \sum^N_{i=1}\vec{\alpha}_i(\vec{x}_i),
\end{equation}
\noindent where $\vec{y}$, $\vec{x}_1$ $\in \mathbb{R}^2$ are the positions on the source plane and image plane, respectively. $\vec x_i$ is the angular position and $\vec{\alpha}_i$ the \textit{deflection angle}  of the light ray at the $i$th lens plane. Recall that, in strong gravitational lensing, the deflection angle is given by

\begin{equation}\label{alpha_eq}
    \vec{\alpha}_i(\vec x_i) = \dfrac{1}{\pi} \int_{\mathbb{R}^2}^{ } d^2 \vec x \dfrac{\vec{x}_i-\vec{x}}{|\vec{x}_i-\vec{x}|^2} \kappa_i(\vec{x}).    
\end{equation}
\noindent $\kappa_i$ is the \textit{convergence} of lens plane $i$, defined as the \textit{projected mass density} $\Sigma_i$ weighted by the \textit{critical surface density} $\Sigma_{\mathrm{cr},i}$,

\begin{align}\label{kappa_def}
    \kappa_i(\vec x) \equiv \dfrac{\Sigma_i(D_i\vec x)}{\Sigma_{\mathrm{cr},i}}, \\
    \Sigma_{\mathrm{cr},i}\equiv \dfrac{c^2 D_s}{4\pi G D_i D_{is}},
\end{align}
\noindent where $c$ is the speed of light and $G$ is the gravitational constant. $D_i$ and $D_{ij}$ are the angular diameter distances from the observer to the lens plane $i$ and from lens plane $i$ to lens plane $j$, respectively. For each lens plane, the derivative of the deflection angle can be written as 
\begin{equation}
    \dfrac{\partial\vec\alpha_i(\vec x_i)}{\partial\vec x_i} = \begin{pmatrix}
    \kappa_i + \gamma_{1,i} & \gamma_{2,i}\\[1ex]
    \gamma_{2,i}&\kappa_i - \gamma_{1,i}&
    \end{pmatrix}(\vec x_i),
\end{equation}
which is a symmetric matrix because the deflection angle at each lens plane is the gradient of the lensing potential, which is a scalar function \cite{lensing_book}. Here $\gamma_1$ and $\gamma_2$ are shear terms that determine the amount that the image is elliptically distorted.

The intermediate lens-plane positions can be obtained by the following recursive equation:
\begin{equation}\label{recursive_eq}
    \vec{x}_j = \vec{x}_1 - \sum^{j-1}_{i=1}\beta_{ij} \vec{\alpha}_i(\vec{x}_i),\quad \mathrm{where}  \quad \beta_{ij} \equiv \dfrac{D_{ij}D_s}{D_jD_{is}}. \\
\end{equation}
\noindent Note that for $j=s$, $\vec x_s = \vec y$, and $\beta_{is} = 1$, we recover Eq. \eqref{lens_eq}.

\subsection{Effective Convergence for Interlopers}\label{effective_conv}

We want to define a single \textit{effective convergence} that gives rise to images that are indistinguishable from those from a system with a main lens and interlopers. 
In general, angular deflections of multiple lens planes cannot be recreated by a single lens plane, so we will need to make some approximations in order to do so. In \S \ref{curl_error}, we will show that the error introduced by these approximations is minimal.

We consider interlopers that are multiple orders of magnitude less massive than the main lens (an interloper with mass comparable to the main lens would distort the images enough to be detected and modeled directly). In strong-lensing systems, images are formed near the the Einstein radius $\theta_E$ of the main lens. We therefore use the Critical Sheet Born (CSB) approximation \cite{CSB}, where the intermediate lens-plane positions are assumed to be
\begin{equation}\label{CSB_eq}
    \vec x_i = \left\{
        \begin{array}{ll}
            \vec x_1 & \quad z_i < z_l \\[1ex]
            \left(1-\beta_{li}\right)\vec x_1 & \quad z_i > z_l.
        \end{array}
    \right.
\end{equation}

\noindent The underlying assumption is that the main lens does most of the lensing and the interlopers only slightly perturb the image.
More specifically, first it is assumed that the light follows a path that is only deflected by the main lens. Then, the gravitational pull of an interloper on the light is integrated over this path as a perturbation. These perturbations are added to the angular deflections caused only by the main lens to get the final angular deflections.

We take the derivative of the multi-plane lens equation (Eq. \ref{lens_eq}),
\begin{align}\label{jacob_lens}
    \dfrac{\pd \vec y}{\pd \vec x_1} &= I - \underbrace{\sum_{i=1}^{l-1}\dfrac{\partial\vec\alpha_i(\vec x_i)}{\partial\vec x_1}}_\text{foreground} \nonumber \\
    &\underbrace{-\dfrac{\partial\vec\alpha_l(\vec x_l)}{\partial\vec x_1}}_\text{main-lens coupling}-\underbrace{\sum_{i=l+1}^{s-1}\dfrac{\partial\vec\alpha_i(\vec x_i)}{\partial\vec x_1}}_\text{background},
\end{align}
and define the effective convergence as
\begin{equation}\label{eff_converg}
    \dfrac{\partial\vec\alpha(\vec x_1)}{\partial\vec x_1} = \begin{pmatrix}
    \kappa_\mathrm{eff} + \gamma_{1,\mathrm{eff}} & \gamma_{2,\mathrm{eff}}\\[1ex]
    \gamma_{2,\mathrm{eff}}&\kappa_\mathrm{eff} - \gamma_{1,\mathrm{eff}}&
    \end{pmatrix}(\vec x_1),
\end{equation}
where
\begin{equation}
    \vec \alpha(\vec x_1) \equiv \sum^{s-1}_{i=1} \vec \alpha_i(\vec x_i)
\end{equation}
is the total deflection angle. In Eq. \eqref{jacob_lens} we decompose the sum over lens planes of Eq. \eqref{lens_eq} into three separate terms: one for the foreground interlopers, one for the coupling to the main lens, and one for the background interlopers. Due to the recursive nature of the multi-plane lensing equation, and thus the different effect that interlopers have whether they are in front of or behind the main lens, these terms will each lead to different effective convergences.
In the remainder of this subsection we consider each term separately.

\subsubsection{Foreground Interlopers}\label{fore_eff}

We derive the effective convergence for foreground interlopers by rewriting the second term on the RHS of Eq. \eqref{jacob_lens} using the CSB approximation (Eq. \ref{CSB_eq}):
\begin{align}
    \sum_{i=1}^{l-1}\dfrac{\partial\vec\alpha_i(\vec x_i)}{\partial\vec x_1} &= \sum_{i=1}^{l-1}\dfrac{\partial\vec\alpha_i(\vec x_1)}{\partial\vec x_1} \nonumber \\ 
    &= \sum_{i=1}^{l-1}
    \begin{pmatrix}
    \kappa_i + \gamma_{1,i} & \gamma_{2,i}\\[1ex]
    \gamma_{2,i}&\kappa_i - \gamma_{1,i}&
    \end{pmatrix}(\vec x_1).
\end{align}
\noindent Thus, the effective convergence for foreground halos is simply the sum of each interloper's convergence up to the main lens:
\begin{equation}
    \kappa_\mathrm{eff,fg}(\vec x_1) =  \sum_{i=1}^{l-1} \kappa_i(\vec x_1) = \sum_{i=1}^{l-1} \dfrac{\Sigma_i(D_i \vec x_1)}{\Sigma_{\mathrm{cr},i}}.
\end{equation}
\noindent By taking the continuum limit, we can write this as an integral over comoving distance $\chi$,
\begin{equation}\label{fg_cont_int}
    \kappa_\mathrm{eff,fg}(\vec x_1) = \int^{\chi_l}_0 d\chi\, \dfrac{a(\chi)\bar\rho_m(\chi)\delta(\chi,\vec x_1)}{\Sigma_\mathrm{cr}(\chi)},
\end{equation}
\noindent where $a$ is the scale factor, $\bar\rho_m$ is the average matter density of the universe, $\delta$ is the overdensity, and $\chi_l$ is the comoving distance to the main-lens plane. We can rewrite this more concisely using the definition of the average matter density,
\begin{equation}
    \bar{\rho}_{\rm m} \equiv \dfrac{3H_0^2}{8\pi G}\dfrac{\Omega_{0,\rm{m}}}{a^3},
\end{equation}
where $H_0$ is the Hubble constant and $\Omega_{0,\rm{m}}$ the matter density parameter:
\begin{equation}
    \kappa_\mathrm{eff,fg}(\vec x_1) = C \int^{\chi_s}_0 d\chi\,W_\mathrm{fg}(\chi)\,\delta(\chi,\vec x_1),
\end{equation}
with $C = 3\Omega_{0,\rm{m}} H^2_0/2c^2$ and 
\begin{equation}\label{select1}
    W_\mathrm{fg}(\chi) = 
    \left\{
        \begin{array}{ll}
            \dfrac{\chi}{a(\chi)}\left(1-\dfrac{\chi}{\chi_s}\right) & \quad \chi \leq \chi_l \\[3ex]
            0& \quad \chi > \chi_l.
        \end{array}
    \right.
\end{equation}
Note that $W_\mathrm{fg}(\chi)$ plays the role of a selection function for the foreground interlopers.

\subsubsection{Main-Lens Coupling}

The third term on the RHS of Eq. \eqref{jacob_lens} corresponds to the coupling between the foreground interlopers and the main lens. This term arises because a small angular deflection by a foreground interloper not only shifts the image, but also shifts the location where the light ray intersects with the main-lens plane, consequently altering the deflection caused by the main lens.

The angular position at the lens plane can be found evaluating Eq. \eqref{recursive_eq} at $j=l$:
\begin{align}
    \dfrac{\pd \vec x_l}{\pd \vec x_1} &= I - \sum^{l-1}_{i=1}\beta_{il}\dfrac{\pd \vec \alpha_i(\vec x_1)}{\pd \vec x_1} \nonumber \\
    &= I - \sum^{l-1}_{i=1}\beta_{il}
    \begin{pmatrix}
    \kappa_i + \gamma_{1,i} & \gamma_{2,i}\\[1ex]
    \gamma_{2,i}&\kappa_i - \gamma_{1,i}&
    \end{pmatrix}(\vec x_1).
\end{align}
\noindent We can thus see that the effective convergence corresponding to this coupling term is a weighted sum of the foreground interlopers:
\begin{equation}
    \kappa_\mathrm{eff,cp}(\vec x_1) = -\sum^{l-1}_{i=1}\beta_{il}\kappa_i(\vec x_1).
\end{equation}
\noindent Following the same procedure as in \S\ref{fore_eff}, we take the continuum limit and write this as an integral over the comoving distance,
\begin{equation}
    \kappa_\mathrm{eff,cp}(\vec x_1) = C \int^{\chi_s}_0 d\chi\,W_\mathrm{cp}(\chi)\,\delta(\chi,\vec x_1),
\end{equation}
where the selection function is now given by
\begin{equation}\label{select2}
    W_\mathrm{cp}(\chi) = 
    \left\{
        \begin{array}{ll}
         -\dfrac{\chi\beta_{\chi l}}{a(\chi)}\left(1-\dfrac{\chi}{\chi_s}\right) & \quad \chi \leq \chi_l \\[3ex]
            0& \quad \chi > \chi_l.
        \end{array}
    \right.
\end{equation}

\subsubsection{Background Interlopers}

Finally, we derive an effective convergence for background interlopers by rewriting the last term on the RHS of Eq. \eqref{jacob_lens},
\begin{align}
    \sum_{i=l+1}^{s-1}\dfrac{\partial\vec\alpha_i(\vec x_i)}{\partial\vec x_1} &= \sum_{i=l+1}^{s-1}\dfrac{\pd \vec x_i}{\pd \vec x_1}\dfrac{\partial\vec\alpha_i(\vec x_i)}{\partial\vec x_i}\nonumber \\
    &= \sum_{i=l+1}^{s-1}(1-\beta_{li})\dfrac{\partial\vec\alpha_i(\left(1- \beta_{li}\right)\vec x_1)}{\partial\vec x_i},
\end{align}
where on the last line we have used the CSB approximation. The effective convergence corresponding to the background interlopers is then
\begin{equation}
    \kappa_\mathrm{eff,bg} = \sum^{s-1}_{i=l+1}(1-\beta_{li})\kappa_i((1-\beta_{li})\vec x_1).
\end{equation}
\noindent In the continuum limit, we get
\begin{align}
    \kappa_\mathrm{eff,bg}(\vec x_1) &= C\int^{\chi_s}_0 d\chi\,W_\mathrm{bg}(\chi)\,\delta(\chi,(1-\beta_{l\chi})\vec x_1),
\end{align}
where the selection function is
\begin{equation}\label{select3}
    W_\mathrm{bg}(\chi) = 
    \left\{
        \begin{array}{ll}
            0 & \quad \chi \leq \chi_l \\[3ex]
            \dfrac{\chi(1-\beta_{l\chi})}{a(\chi)}\left(1-\dfrac{\chi}{\chi_s}\right)& \quad \chi > \chi_l .
        \end{array}
    \right.
\end{equation}

\bigskip 

\subsubsection{Interlopers as effective subhalos}\label{eff_subhalo} 

Combining these results, we see that for a strong-lensing system with some foreground and background perturbers we can write a single effective convergence that characterizes the effect of the interlopers as
\begin{align}\label{eq:k_eff}
    \kappa_\mathrm{eff}(\vec x) &= \underbrace{\sum^{l-1}_{i=1} (1-\beta_{il})\kappa_i(\vec{x})}_\text{foreground + coupling} \nonumber \\
    &+ \underbrace{\sum^N_{i=l+1} \left(1-\beta_{li}\right)\kappa_i((1-\beta_{li})\vec{x})}_\text{background}.
\end{align}
\noindent We can write this in the continuum limit as
\begin{equation}\label{keff_integ}
    \kappa_\mathrm{eff}(\vec x) = C\int^{\chi_s}_0d\chi\, W_\mathrm{I}(\chi) \delta(\chi,g(\chi)\vec x_1),
\end{equation}
where
\begin{equation}\label{eq:total_select}
    W_\mathrm{I} \equiv W_\mathrm{fg} + W_\mathrm{cp} + W_\mathrm{bg} = \frac{f(\chi) D_{\chi s} \chi^2}{D_\chi D_s},
\end{equation} 
and $f(\chi)$ and $g(\chi)$ are piecewise functions of the comoving distance:
\begin{equation}\label{eq:fdef}
    f(\chi)=\left\{
        \begin{array}{ll}
            1-\beta_{\chi l} & \quad \chi \leq \chi_l \\
            1-\beta_{l \chi} & \quad \chi > \chi_l
        \end{array}
    \right.
\end{equation}
\begin{equation}\label{eq:gdef}
    g(\chi)=\left\{
        \begin{array}{ll}
            1 & \quad \chi \leq \chi_l \\
            1-\beta_{l \chi} & \quad \chi > \chi_l.
        \end{array}
    \right.
\end{equation}

We can think of Eq. \eqref{eq:k_eff} as a projection that takes interlopers at some plane $i$ and projects them onto the lens plane $l$ with an effective convergence. Under this approximation, we calculate the deflection of the interlopers assuming that the light ray travels a path that is only deflected by the main lens. Since $\beta_{il}$ and $\beta_{li}$ go to $0$ as the distance between the planes $i$ and $l$ goes to $0$, the interlopers that are sufficiently close to the main lens are unchanged by this projection. Furthermore, since $\beta_{il}$ goes to $1$ as the distance between plane $i$ and the observer goes to $0$, and $\beta_{li}$ goes to $1$ as the distance between plane $i$ and the source goes to $0$, the interlopers that are sufficiently close to the observer and the source become insignificant after projection. At intermediate positions, where neither of these approximations hold, one could imagine that the error introduced by this projection could be significant. We discuss this further and quantify the error in \S \ref{curl_error}, showing that it is in fact an excellent approximation even in this intermediate regime.

For the remainder of this paper, when we have to specify a density profile for a halo we will do so with a truncated NFW profile \cite{tnfw} (tNFW). The form of this profile will be shown explicitly later on in Eq. \eqref{eq:tnfw_prof}; for now, the relevant aspect of this profile is that it is fully determined by two parameters, the scale radius $r_{\rm s}$ and the dimensionless truncation parameter $\tau$, defined in terms of the truncation radius $r_{\rm t}$: $\tau \equiv r_{\rm s}/r_{\rm t}$. 

\begin{figure}[ht!]
    \centering
    \includegraphics[width=\linewidth]{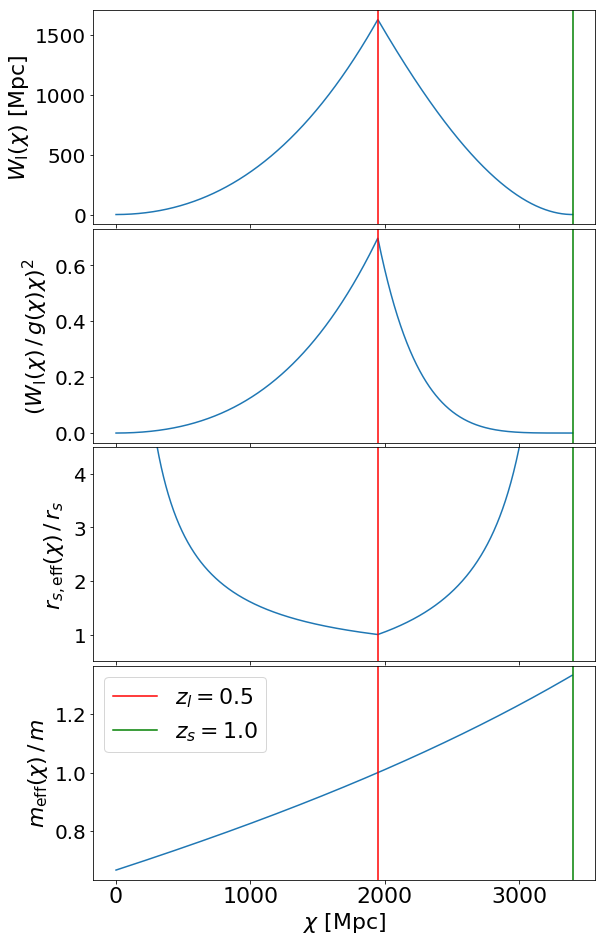}
    \caption{\footnotesize{The comoving distance dependence of (from the top) \textbf{1:} the selection function $W_\mathrm{I}$ which weighs the overdensities at different distances to get the effective convergence in Eq. \eqref{keff_integ}, \textbf{2:} $(W_\mathrm{I}/g(\chi)\chi)^2$ which weighs the 3D power to get the 2D power spectrum in Eq. \eqref{eq:power2Dinteg}, \textbf{3:} the ratio of effective scale radius to the scale radius of the interlopers (Eq. \ref{eq:rseff}), \textbf{4:} the ratio of effective mass to the mass of the interlopers (Eq. \ref{eq:meff}). These functions also depend on the source and lens distances. Here, $z_l = 0.5$ and $z_s = 1.0$ for the lens and source redshifts, respectively. Their comoving distances are shown as red and green vertical lines.}}
    \label{fig:useful_functions}
\end{figure}

Eq. \eqref{eq:k_eff} implies that an interloper with mass $m$ and convergence $\kappa$ at comoving distance $\chi$ has an effective convergence
\begin{equation}\label{eq:effective_subhalo}
    \kappa_{\chi,\mathrm{eff}}(\vec x\,;\,m,r_{\rm s},\tau) = f(\chi)\, \kappa(g(\chi)\vec x\,;\,m,r_{\rm s},\tau).
\end{equation}
Therefore, we can think of the interlopers as subhalos on the main-lens plane with a modified scale radius and mass. To obtain these scaling relations, we express the convergence in terms of the projected mass density (Eq. \ref{kappa_def}), 
\begin{equation}\label{sigma_transform}
    \dfrac{\Sigma_{\chi,\mathrm{eff}}(D_l \vec x\,;\,m,r_{\rm s},\tau)}{\Sigma_{\mathrm{cr},l}} = f(\chi)\dfrac{\Sigma(g(\chi) D_\chi \vec x\,;\,m,r_{\rm s},\tau)}{\Sigma_{\mathrm{cr},\chi}},
\end{equation}
and make use of the following rules:
\begin{equation}\label{scaling_mass}
    \epsilon \Sigma(\vec r\,;\, m, r_{\rm s},\tau) = \Sigma(\vec r\,;\, \epsilon m, r_{\rm s},\tau)
\end{equation}
\begin{equation}\label{scaling_radius}
    \Sigma(\eta \vec r\,;\, m, r_{\rm s},\tau) = \Sigma( \vec r\,;\, \frac{m}{\eta^2}, \frac{r_{\rm s}}{\eta},\tau),
\end{equation}
where $\epsilon$ and $\eta$ are scaling constants. These are derived in Appendix \ref{app:a}. With these in hand we can rewrite Eq. \eqref{sigma_transform} as
\begin{equation}\label{eq:effective_proj_mass}
    \Sigma_{\chi,\mathrm{eff}}(D_l \vec x\,;\,m,r_{\rm s},\tau)=\Sigma(D_l \vec x\,;\,m_\mathrm{eff},r_{\mathrm{s,eff}},\tau),
\end{equation}
where
\begin{equation}\label{eq:rseff}
    r_{\mathrm{s,eff}}(\chi) = \dfrac{D_l}{g(\chi)D_\chi}r_{\rm s}
\end{equation}
and
\begin{equation}\label{eq:meff}
    m_\mathrm{eff}(\chi) = f(\chi)\dfrac{\Sigma_{\mathrm{cr},l}}{\Sigma_{\mathrm{cr},\chi}} \left(\dfrac{D_l}{g(\chi)D_\chi}\right)^2m
\end{equation}
are the effective scale radius and the effective mass for interlopers, respectively (shown in Fig. \ref{fig:useful_functions}). The effective scale radius is larger than the true scale radius for both foreground and the background interlopers. The effective mass, on the other hand, is smaller for foreground and larger for background interlopers.

\subsection{Power Spectrum of Interlopers}\label{pow_spec_cal}

\begin{figure}[ht]
    \centering
    \includegraphics[width=\linewidth]{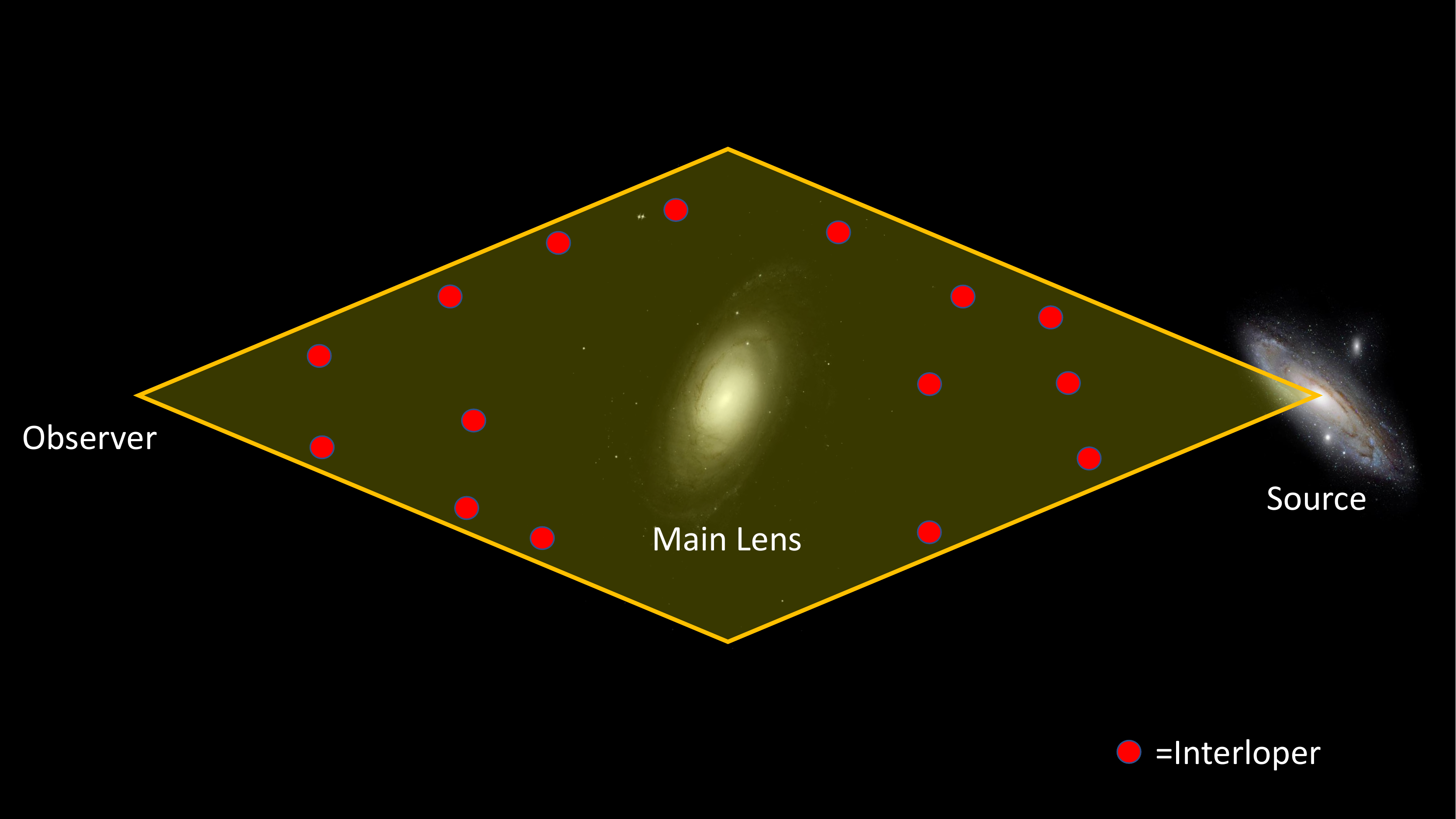}
    \caption{\footnotesize{Double cone (Eq. \ref{eq:cone_area}) volume integrated over for the analytical calculation of the effective convergence power spectrum due to interlopers (Eq. \ref{eq:effsubpower}).}}
    \label{fig:double_cone}
\end{figure}

In this section, we calculate the power spectrum of the interlopers using two different approaches. In the first approach we use the Limber approximation \cite{1953ApJ...117..134L}, which neglects the Fourier modes of the $3$D matter distribution parallel to the line of sight. For this, we will express the $2$D line-of-sight density as an integral over the $3$D density with a window function that weighs the integral over distance, as we did in \S \ref{effective_conv}. The second approach consists of replacing the interlopers with effective subhalos that live on the lens plane, using Eq. \eqref{eq:effective_subhalo}, and calculating the Fourier transform of their $2$D two-point correlation function. The interlopers that are projected onto the lens plane will be within a volume with the shape of a double cone, shown in Fig. \ref{fig:double_cone}. 

While the former approach might be more familiar to cosmologists due to its similarity to the weak lensing literature, the latter is more akin to previous works on the statistics of dark matter substructure in strong gravitational lenses. In particular, this calculation closely follows that of Ref. \cite{powerspec1} for the subhalo power spectrum. The main difference is the addition of the comoving distance $\chi$ as a parameter that an interloper has as an effective subhalo. We will see that ultimately both approaches are in agreement.

\subsubsection{Power Spectrum of the Interlopers using the Limber Approximation}\label{limber}

For a 2D projected density that can be written in the form \cite{dodelson:2003}
\begin{equation}
    \delta_\mathrm{2D}(\vec x) = \int^{\chi_s}_0 d\chi\,W(\chi)\delta_\mathrm{3D}(\chi,g(\chi)\chi \vec x),
\end{equation}
where $\delta_\mathrm{3D}$ is the 3D matter overdensity, $W(\chi)$ is the selection function that weighs distances, and $g(\chi)$ is a function that represents how the 3D density is projected down to 2D, the 2D power spectrum can be written in terms of the 3D power spectrum as
\begin{equation}\label{eq:power2Dinteg}
    P_\mathrm{2D}(\vec l) = \int^{\chi_s}_0 d\chi\,\left(\dfrac{W(\chi)}{g(\chi)\chi}\right)^2 P_\mathrm{3D}\left(\chi,\dfrac{\vec l}{g(\chi)\chi}\right),
\end{equation}
\noindent where $\vec l$ is the the Fourier conjugate to the angular position $\vec x$.

Combining the selection functions in Eqs. \eqref{select1}, \eqref{select2}, and \eqref{select3}, and noticing that $\vec k = \vec l/D_l$ at the lens plane, we can write the interloper power spectrum as
\begin{equation}\label{2D_pow_int}
    P_{\rm I}(\vec k) =C^2 D^2_l\int^{\chi_s}_0 d\chi\,\dfrac{W^2_\mathrm{I}(\chi)}{g^2(\chi)\chi^2} P_\mathrm{3D}\left(\chi,\dfrac{\vec k D_l}{g(\chi)\chi}\right).
\end{equation}

Our goal is to write this power spectrum as a function of the density profiles and ensemble properties of interlopers. Therefore, we will expand the 3D matter power spectrum in order to express it in terms of these quantities. 

Let us write the 2-point correlation function of the matter overdensities,
\begin{equation}\label{eq:3D_corr}
    \varepsilon(\vec r) = \frac{1}{V}\int d^3 \vec s\, \delta(\vec s) \delta(\vec s + \vec r),
\end{equation}
\noindent where $\vec s$ and $\vec r$ are positions in comoving coordinates, and $V$ is the comoving volume over which these correlations are integrated. Writing the overdensities as the sum of overdensities of each individual interloper, $\delta(\vec x) = \sum_i \delta_i(\vec x)$, we can rewrite the 2-point correlation function as the sum of the correlation of each interloper with itself and the correlation of each interloper with other interlopers,
\begin{eqnarray}\label{eq:haloterms}
    \varepsilon(\vec r) &=&\underbrace{\frac{1}{V} \sum_i \int d^3 \vec s\, \delta_i(\vec s) \delta_i(\vec s + \vec r)}_{\mathrm{1-halo\,\,term}}  \nonumber\\
    &+&\underbrace{\frac{1}{V} \sum_i \sum_{j\neq i} \int d^3 \vec s\, \delta_i(\vec s) \delta_j(\vec s + \vec r)}_{\mathrm{2-halo\,\,term}},
\end{eqnarray}
where the summation is now over interlopers rather than lens planes.

We assume that the interloper positions are uncorrelated, which makes the 2-halo term vanish. Therefore, we can write the 3D matter power spectrum as the Fourier transform of the 1-halo term,
\begin{eqnarray}\label{eq:P3dk}
    P_\mathrm{3D}(\vec k) &\equiv& \mathcal{F}[\varepsilon](\vec k)  \\
    &=& \dfrac{1}{V}\int d^3 \vec r\, \exp[-i\vec k \cdot \vec r]\, \nonumber \\ 
    &\times& \sum_i\int d^3 \vec s\, \delta_i(\vec s) \delta_i(\vec s + \vec r) \nonumber \\
    &=& \frac{1}{V} \sum_i \left|\mathcal{F}[\delta_i](\vec k)\right|^2.
\end{eqnarray}
For convenience, we denote the Fourier transform of a function as $\tilde \delta(\vec k) \equiv \mathcal{F}[\delta](\vec k)$ in the remainder of the text.

We now assume the interlopers to be tNFW halos, which have a spherically symmetric density profile (in physical coordinates) given by
\begin{equation} \label{eq:tnfw_prof}
    \rho(r;\,m,r_{\rm s},\tau) = \frac{m}{4\pi r(r+r_\mathrm{s})^2 u(\tau)}\bigg(\frac{\tau^2 r_\mathrm{s}^2}{r^2+\tau^2 r_\mathrm{s}^2}\bigg),
\end{equation}
where
\begin{equation}
    u(\tau) \equiv\frac{\tau^2}{(\tau^2+1)^2}\big[(\tau^2-1)\ln(\tau)+\tau\pi-(\tau^2+1)\big].
\end{equation}
The density profile can be cast in dimensionless form by defining $\xi = r/r_{\rm s}$ and
\begin{equation}\label{eq:dimless}
    \phi(\xi;\,\tau) \equiv \frac{1}{4\pi \xi (\xi+1)^2 u(\tau)}\bigg(\frac{\tau^2}{\xi^2+\tau^2}\bigg),
\end{equation}
such that
\begin{equation}
    \rho(r;\,m,r_{\rm s},\tau) = \frac{m}{r^3_{\rm s}}\phi\left(\dfrac{r}{r_{\rm s}};\,\tau\right).
\end{equation}
Near halos, the overdensity is much larger than  the average density $\bar \rho$, so we can write the overdensity due to a single tNFW interloper $i$ as
\begin{align}
    \delta_i(s,\chi;\, m_i,r_{{\rm s},i},\tau_i) &= \frac{\rho(a(\chi)s;\, m_i,r_{{\rm s},i},\tau_i) - \bar \rho(\chi)}{\bar\rho(\chi)}  \nonumber \\
    & \cong \frac{m_i}{r^3_{{\rm s},i}\bar\rho(\chi)} \phi\left(\frac{a(\chi)}{r_{{\rm s},i}} s;\,\tau_i\right),
\end{align}
and its Fourier transform as
\begin{equation}
    \tilde \delta(k,\chi;\,m_i,r_{{\rm s},i},\tau_i) = \frac{m_i}{\rho_0}\tilde \phi \left(\frac{r_{{\rm s},i}}{a(\chi)} k;\,\tau_i\right),
\end{equation}
where $\rho_0 = \bar \rho(\chi=0) a^3(\chi=0)$. Spherical symmetry of the tNFW profile implies $\tilde \delta(\vec k) = \tilde \delta(k)$, where $k \equiv |\vec k|$.

Substituting this into Eq. \eqref {eq:P3dk} we obtain an expression for the 3D power spectrum in terms of the density profile of interlopers:
\begin{equation}
    P_\mathrm{3D}(\chi,k) = \frac{1}{V}\sum_i \frac{m_i^2}{\rho^2_0}\left|\tilde \phi \left(\frac{r_{{\rm s},i}}{a(\chi)} k;\,\tau_i\right)\right|^2.
\end{equation}

Following the procedure in Ref. \cite{powerspec1}, we can convert this sum into an integral over the probability distribution of the interloper parameters,
\begin{align}
    P_\mathrm{3D}(\chi,k) &= \frac{N}{V}\int dm\,d^2\vec q\, \mathcal{P}(m,\vec q\,|\,\chi) \nonumber \\
    &\times \frac{m^2}{\rho^2_0}\left|\tilde \phi \left(\frac{r_{\rm s}}{a(\chi)}k;\,\tau\right)\right|^2,
\end{align}
where $\vec q = (r_s,\tau)$ are the intrinsic halo parameters, $N$ is the total number of interlopers in the ensemble, and $\mathcal P(m, \vec q\,|\,\chi)$ is the probability that an interloper has mass $m$ and intrinsic parameters $\vec q$, given that it is at comoving distance $\chi$. We now separate the probabilities,
\begin{equation}
    \mathcal{P}(m,\vec q\,|\chi) = \mathcal{P}(m\,|\,\chi) \mathcal{P}(\vec q\,|\,m,\chi),
\end{equation}
since $\mathcal P(A,B) = \mathcal P(A\,|\,B) \mathcal P(B)$ for any two propositions $A$ and $B$.  For the mass, we assume
\begin{equation}
    \mathcal P (m\,|\,\chi) = \frac{V}{N}n(m,\chi),
\end{equation}
where $n(m,\chi)$ is the Sheth-Tormen mass function \cite{sheth},
\begin{align}
    n(m,z) = -&B \dfrac{\rho_0}{m}\dfrac{2 g(z) \delta_{\rm c}}{\sigma^2_{ m}}\dfrac{d\sigma_{ m}}{dm}\left(\dfrac{2a}{\pi}\right)^{1/2} \nonumber\\
    &\times \left[1 + a^{-p}\left( \dfrac{g(z) \delta_{\rm c}}{\sigma_{ m}}\right)^{-2p}\right]\nonumber\\
    &\times \exp\left[-\dfrac{a(g(z)\delta_{\rm c})^2}{2 \sigma^2_{m}}\right],
\end{align}
with the free parameters set to $a = 0.707$ and $p = 0.3$ \cite{st_parameters}, where  $g(z)$ is the growth function, $B\equiv (1+(2^p\sqrt{\pi})^{-1}\Gamma(-p + 1/2))^{-1}$, $\sigma_m$ is the standard deviation of the matter fluctuations smoothed with a top-hat filter of size $\sim (m/\rho_0)^{1/3}$, and $\delta_{\rm c} = 1.686$ is the collapse threshold. This probability density is already normalized since $\int dm\, n(m,\chi) = N/V$. 

With this assumption, the 3D power spectrum has the form
\begin{align}
    P_\mathrm{3D}(\chi,k) &= \frac{1}{\rho^2_0}\int_{m_{\rm low}}^{m_{\rm high}} dm\, n(m,\chi) \,m^2  \nonumber \\
    &\times \int d^2\vec q\,\mathcal{P}(\vec q\,|\,m,\chi)\left|\tilde \phi \left(\frac{r_{\rm s}}{a(\chi)}k;\,\tau\right)\right|^2,
\end{align}
where $m_{\rm low}$ and $m_{\rm high}$ are the lower and upper bounds of the mass range of the interlopers.

Using Eq. \eqref{2D_pow_int} we can finally write the power spectrum of the effective convergence of the interlopers as a function of the density profile of interlopers:
\begin{align}\label{eq:limber_power_raw}
    P_{\rm I}(k) &=\left(\frac{4\pi G}{c^2}\right)^2 D^2_l\int^{\chi_s}_0 d\chi\,\dfrac{W^2_\mathrm{I}(\chi)}{g^2(\chi)\chi^2} \nonumber \\
    &\times\int dm\,n(m,\chi)\,m^2 \nonumber \\
    &\times \int d^2\vec q\,\mathcal{P}(\vec q\,|\,m,\chi)\left|\tilde \phi \left(\dfrac{D_lr_{\rm s}}{g(\chi)D_\chi} k;\,\tau\right)\right|^2.
\end{align}

\subsubsection{Power Spectrum of the Interlopers as Effective Subhalos}

We start with the 2-point correlation function of the effective convergence,
\begin{equation}\label{eq:2D_corr}
    \varepsilon_\mathrm{2D}(\vec r) = \frac{1}{A}\int d^2 \vec s\,\kappa_\mathrm{eff}(\vec s)\kappa_\mathrm{eff}(\vec s + \vec r),
\end{equation}
where $\vec r$ and $\vec s$ are physical coordinates on the lens plane, and $A$ is the lens plane area. The subscript 2D is added to differentiate this from the 3D 2-point correlation function of the matter overdensities in Eq. \eqref{eq:3D_corr}. Just like in the 3D case, we write the total effective convergence as the sum of the effective convergence of each interloper, $\kappa_\mathrm{eff}(\vec r) = \sum_i \kappa_{i,\mathrm{eff}}(\vec r)$. Again, we assume that the interloper positions are uncorrelated which makes the 2-halo term in Eq. \eqref{eq:haloterms} vanish and lets us write
\begin{equation}
    \varepsilon_\mathrm{2D}(\vec r) = \frac{1}{A}\sum_i\int d^2 \vec s\,\kappa_{i,\mathrm{eff}}(\vec s)\kappa_{i,\mathrm{eff}}(\vec s + \vec r),
\end{equation}
and consequently
\begin{align}\label{eq:2Dpowersum}
    P_\mathrm{2D}(k) &\equiv \tilde \varepsilon_\mathrm{2D}(k) = \frac{1}{A} \sum_i \left|\tilde \kappa_{i,\mathrm{eff}}( k)\right|^2,
\end{align}
where we again dropped the vector notation in $\vec k$ due to the radial symmetry of the effective convergence. 

Using Eqs. \eqref{eq:effective_proj_mass} and \eqref{eq:dimless}, we can write
\begin{align}
    \kappa_{i,\mathrm{eff}}(s) &= \frac{\Sigma(s; m_{\mathrm{eff},i},r_{\rm{ s,eff},\textit{i}},\tau_i)}{\Sigma_{\mathrm{cr},l}} \nonumber \\
    &= \dfrac{1 }{\Sigma_{\mathrm{cr},l}}\dfrac{m_{\mathrm{eff},i}}{r_{\rm{ s,eff},\textit{i}}^3}\int dz\;  \phi\left(\frac{\sqrt{s^2 + z^2}}{r_{\rm{ s,eff},\textit{i}}}\,;\,\tau_i\right).
\end{align}
In Appendix \ref{app:b}, we show that the Fourier transform of $\kappa_{i.\mathrm{eff}}$ can be expressed in terms of the Fourier transform of the dimensionless density profile $\phi$ and, consequently, we can rewrite Eq. \eqref{eq:2Dpowersum} as
\begin{equation}
    P_\mathrm{2D}(k) = \frac{1}{A}\sum_i \frac{m^2_{\mathrm{eff},i}}{\Sigma^2_{\mathrm{cr},l}}\left|\phi\left(r_{\rm{ s,eff},\textit{i}}k\,;\,\tau_i\right)\right|^2.
\end{equation}
\noindent Analogously to the procedure in \S\ref{limber}, we convert the sum into an integral over the probability distribution of the interloper parameters:
\begin{align}
    P_\mathrm{2D}(k) &= \frac{N}{A} \int d\chi\,dm\,d^2\vec q\, \mathcal{P}(\chi,m,\vec q)  \nonumber \\
    &\times \frac{m^2_\mathrm{eff}(\chi)}{\Sigma^2_{\mathrm{cr},l}}\left|\phi\left(r_{\rm s,\mathrm{eff}}(\chi)k\,;\,\tau\right)\right|^2,
\end{align}
where $\mathcal{P}(\chi,m,\vec q)$ is the probability of an interloper being at comoving distance $\chi$ and having mass $m$ and intrinsic parameters $\vec q$. We substitute the effective scale radius and mass from Eqs. \eqref{eq:rseff} and \eqref{eq:meff}, and again separate the probabilities $\mathcal{P}(\chi,m,\vec q) = \mathcal{P}(\vec q\,|\,m,\chi) \mathcal{P}(m,\chi)$, which gives
\begin{align}\label{eq:power_2D_second_to_last}
    P_\mathrm{2D}(k) &= \frac{N}{A} \int d\chi\left[f(\chi)\dfrac{1}{\Sigma_{\mathrm{cr},\chi}} \left(\dfrac{D_l}{g(\chi)D_\chi}\right)^2\right]^2   \nonumber \\
    &\times\int dm\,  \mathcal{P}(m,\chi)m^2 \nonumber\\
    &\times \int d^2\vec q\,\mathcal{P}(\vec q\,|\,m,\chi)\left|\phi\left(\dfrac{D_lr_{\rm s}}{g(\chi)D_\chi}k\,;\,\tau\right)\right|^2.
\end{align}

The probability that an interloper has mass $m$ and is at comoving distance $\chi$ is proportional to the halo mass function $n(m,\chi)$ and the cross section of the double cone with the lens plane as the base (Fig. \ref{fig:double_cone}). Thus, we can write\footnote{This is already normalized because $\int d\chi\,S(\chi)\int dm\,n(m,\chi) = N$.}
\begin{equation}\label{eq:prob_chi_m}
    \mathcal P(m,\chi) = \frac{S(\chi)}{N}n(m,\chi),
\end{equation}
where $S(\chi)$ is the cross section of the double cone in comoving units at comoving distance $\chi$ and is given by
\begin{equation}\label{eq:cross_sec}
    \frac{S(\chi)}{A} = \frac{\chi^2 }{D^2_l}g^2(\chi),
\end{equation}
where $A$ is the physical area of the main lens and $g(\chi)$ was defined in Eq. \eqref{eq:gdef} (see Appendix \ref{app:c} for a careful derivation).

Substituting Eqs. \eqref{eq:prob_chi_m} and \eqref{eq:cross_sec} into Eq. \eqref{eq:power_2D_second_to_last} gives
\begin{align}\label{eq:effsubpower}
    P_\mathrm{2D}(k) &= \left(\frac{4\pi G}{c^2}\right)^2 D^2_l\int d\chi\,  \left[\frac{f(\chi)D_{\chi s}\chi^2}{D_sD_\chi}\right]^2 \dfrac{1}{g^2(\chi)\chi^2}   \nonumber \\
    &\times\int dm\,  n(m,\chi)\,m^2  \nonumber\\
    &\times \int d^2\vec q\,\mathcal{P}(\vec q\,|\,m,\chi)\left|\phi\left(\dfrac{D_lr_{\rm s}}{g(\chi)D_\chi}k\,;\,\tau\right)\right|^2.
\end{align}

We see that the factor in square brackets is exactly the selection function in Eq. \eqref{eq:total_select}. Therefore, this equation is identical to Eq. \eqref{eq:limber_power_raw}. We conclude that calculating the power spectrum of the effective convergence on the lens plane after projecting the interlopers onto the lens plane as effective subhalos is equivalent to calculating it using the Limber approximation from the 3D matter power spectrum with the selection function derived in \S \ref{effective_conv}.

\subsection{Effective Convergence Power Spectrum for a Population of tNFW Perturbers}
\label{subsec:effpower}

In this section, we compare the convergence power spectrum of interlopers to that of subhalos, referring to both collectively as perturbers. We calculate both contributions to the convergence power spectrum for a fiducial system and show, independently of profile, how each contribution varies differently as a function of source and lens redshift.

Since the power spectrum depends on the perturber profile parameters, we now specify the probability distribution $\mathcal{P}(\vec q|m,\chi)$ of the intrinsic halo parameters $\vec q = (r_{\rm s}, \tau)$.
For both interlopers and subhalos, we assume the following form:
\begin{equation}\label{deltas}
    \mathcal{P}(\vec q\,|\,m,\chi) = \delta\left(r_{\rm s} - r_{\rm s}(m)\right)\delta\left(\tau - 20\right)
\end{equation}
\begin{equation}\label{rsfunc}
    r_{\rm s}(m) = r_{\rm s,0}\left[\frac{m}{m_0}\right]^\gamma,
\end{equation}
where $r_{\rm s,0}=0.1\,\mathrm{kpc}$, $m_0=10^6 M_\odot$, and $\gamma=1/3$ \cite{dark_census,subs_den_prof,halo_wmap}. These parameters are chosen so our results are directly comparable to the convergence power spectrum from only substructure in Ref. \cite{powerspec1}.

We carry out the integral over $\vec q$ in Eq. \eqref{eq:effsubpower}, which fixes $\tau = 20$ and $r_{\rm s} = r_{\rm s}(m)$ (given in Eqs. \ref{deltas} and \ref{rsfunc}). After these choices, the effective convergence power spectrum for interlopers is
\begin{align}\label{eq:final_power}
    &P_{\rm I}(k) =\left(\frac{4\pi G}{c^2}\right)^2 D^2_l\int^{\chi_s}_0 d\chi\,\frac{W^2_\mathrm{I}(\chi)}{g^2(\chi)\chi^2} \nonumber \\
    &\times\int^{m_\mathrm{high}}_{m_\mathrm{low}} dm\, n(m,\chi)m^2 \left|\tilde \phi\left(\frac{r_{\rm s}(m) D_l}{g(\chi)a(\chi)\chi}k;\,\tau\right)\right|^2,
\end{align}
where $n(m,\chi)$ is the Sheth-Tormen mass function \cite{sheth}. We use a perturber mass range from $m_\mathrm{low} = 10^5\,\msun$ to $m_\mathrm{high} = 10^8\,\msun$, since lower-mass perturbers contribute little to the total power and higher-mass perturbers can be modeled directly. 
For subhalos, the convergence power spectrum is \cite{powerspec1}
\begin{equation}\label{eq:powersub}
    P_{\rm S}(k) = \frac{1}{\Sigma_{\rm cr}^2}\int_{m_\mathrm{low}}^{m_\mathrm{high}}dm\, m^2 n_\mathrm{sub}(m)\abs{\tilde \phi(r_{\rm s}(m)k; \tau)}^2,
\end{equation}
where $n_\mathrm{sub}(m)$ is the number of subhalos per physical area per mass. We use (see Appendix \ref{app:nsub})
\begin{equation} \label{eq:nsub}
    n_\mathrm{sub}(m, z_l) = \frac{0.3\Sigma_{\mathrm{cr},0.5}f_{\mathrm{sub},0.5}(2+\beta)}{(m_\mathrm{high}^{2+\beta}-m_\mathrm{low}^{2+\beta})} \frac{(1+z_l)^{5/2}}{(1+0.5)^{5/2}} m^\beta ,
\end{equation}
where $\beta = -1.9$ and $f_{\mathrm{sub}, z_l}$ is the fraction of halo mass in substructure within the mass range at redshift $z_l$. The factor of $(1+z_l)^{5/2}$ accounts for the redshift evolution of the subhalo mass function as the subhalos travel within their host \cite{fsubredshift}.
A value of $\beta = -1.9\pm0.1$ is fairly well agreed-upon in the literature, both in observations \cite{okabe14} and simulations \cite{delucia04, madau08, bolyan10, gao12, wu13}, but $f_{\mathrm{sub}, z_l}$ is much less constrained. This is because the population of subhalos evolves as it travels within the host and is subject to tidal stripping.

There is no consensus of the extent to which tidal stripping happens $-$ both with and without baryons $-$ as a function of redshift and host mass.
Different $N$-body simulations have found $f_{\mathrm{sub}, 0}$ on the order of $10^{-3}$ to $10^{-2}$ using host halo masses of $\sim 10^{12}\,\msun$  \cite{aquarius08, ethos16, powerspec2, diemand07}, and we expect baryons to decrease these values.
Observations for similar lens redshifts and masses seem to be consistent with a wide range of possible values. Ref. \cite{2014MNRAS.442.2017V} found that for a sample of SLACS galaxies of similar masses (mass within Einstein radius $\sim 10^{11.4}\,\msun$ \cite{auger09}) and lens redshifts ($z \sim 0.2$), $f_{\mathrm{sub}, 0.2} = 0.0076_{-.0052}^{+.0208}$ for subhalos in $[4\E{6}-4\E{9}]\,\msun$. Ref. \cite{bells_2018} found that for BELLS lenses, the upper bound on $f_{\mathrm{sub},0.5}$ is $7\%$ with an upper subhalo mass bound of $10^{11}$ $\msun$. This constraint includes both interlopers and subhalos. Because our upper mass bound is 3 orders of magnitude below the one cited, we expect the upper bound on $f_{\rm sub,0.5}$ from BELLS to be significantly below that for our mass range.

Taking into account the considerable uncertainty in these observations, and the wide range of plausible values extracted from simulations, we settle on a fiducial value of $f_{\mathrm{sub}, 0.5} = 4\E{-3}$ for our mass range $[10^5 - 10^8] \, \msun$ and host redshift ($z_l = 0.5$), for typical galaxy-scale lenses.
We will nevertheless discuss in detail the dependence of our results on $f_{\mathrm{sub}, 0.5}$ below.

Fig. \ref{fig:compare_power} displays the convergence power spectrum due only to interlopers (blue), due only to subhalos (green), and due to both (red) for a fiducial lensing system with $z_l = 0.5$, $z_s=1$, and mass functions as described above. The numerical results in this figure will be described in detail in \S \ref{sec:numeric}.
For this lensing system, the power spectrum amplitude due to interlopers is $7.4$ times larger than that of subhalos, meaning the former would be the dominant contribution to any measured signal.

A signal known to be dominated by interlopers would be especially useful for constraining the low-mass end of the halo mass function, which is considered a key way of distinguishing between vanilla CDM and exotic dark matter scenarios that can lead to low-mass cutoffs. This is both because interlopers are simpler to model, as they are generally not subject to the same degree of astrophysics and tidal effects as subhalos, and because the density of interlopers is much better understood. Indeed, while $f_\mathrm{sub,0.5}$ may range between several orders of magnitude, the two commonly-used mass functions that would affect the interloper amplitude, Sheth-Tormen \cite{sheth} and Press-Schechter \cite{press}, only differ by about a factor of two. Simulations agree with Sheth-Tormen to roughly a $10\%$ level \cite{warren06, tinker08}, and future observations could in principle measure the halo mass function to percent-level accuracy \cite{castro16}.

Refs. \cite{powerspec1,powerspec2} pointed to several features of the power spectrum that could be used to constrain the particle nature of dark matter, such as the slope at $k > 2$ kpc$^{-1}$; however in the remainder of this section we focus on the amplitude at small values of $k$, i.e. the $k \rightarrow 0 $ limit,\footnote{To be precise, our definition of the plateau only matches the $k \to 0$ limit of power when we neglect the 2-halo term from Eq. \eqref{eq:haloterms}, which would contribute an additional term to Eqs. \eqref{eq:final_power} and \eqref{eq:powersub}. However, this term is expected to be small (particularly in the presence of baryons) and only becomes relevant at $k \lesssim 10^{-1}\un{kpc}^{-1}$ \cite{powerspec2}, so it is safe to neglect here.}
which primarily provides information about the overall abundance of perturbers within a given mass range. We will refer to this regime as the \textit{plateau} (due to the fact that the power spectrum is constant on these scales), whose amplitude we define as $P_{0} \equiv \lim_{k\rightarrow 0} P(k)$. 
Expressions for the interloper and subhalo plateau, $P_{\rm I,0}$ and $P_{\rm S,0}$, are derived in Appendix \ref{app:plateau}.

We focus on the low-$k$ scales for two main reasons. First, they are the most readily observable ones. Second, they neatly illustrate the importance of taking into account the contribution of interlopers in order to use strong-lens measurements to draw conclusions about dark matter, without having to worry about the specific details of how the interlopers and subhalos are modeled (which affect the power spectrum at higher wavenumbers \cite{powerspec1}), since the amplitude depends only on their mass functions. Due to the fact that both the halo and subhalo mass functions evolve with redshift, $P_0$ depends on the geometry of the lensing system. At higher source redshifts, there are more interlopers along the line of sight, which produces a higher interloper power. The subhalo power spectrum depends on source and lens redshifts through $\Sigma_{\mathrm{cr},l}$ and $n_\mathrm{sub}$, resulting in a somewhat different redshift dependence.

\begin{figure*}
    \centering
    \includegraphics[width=\linewidth]{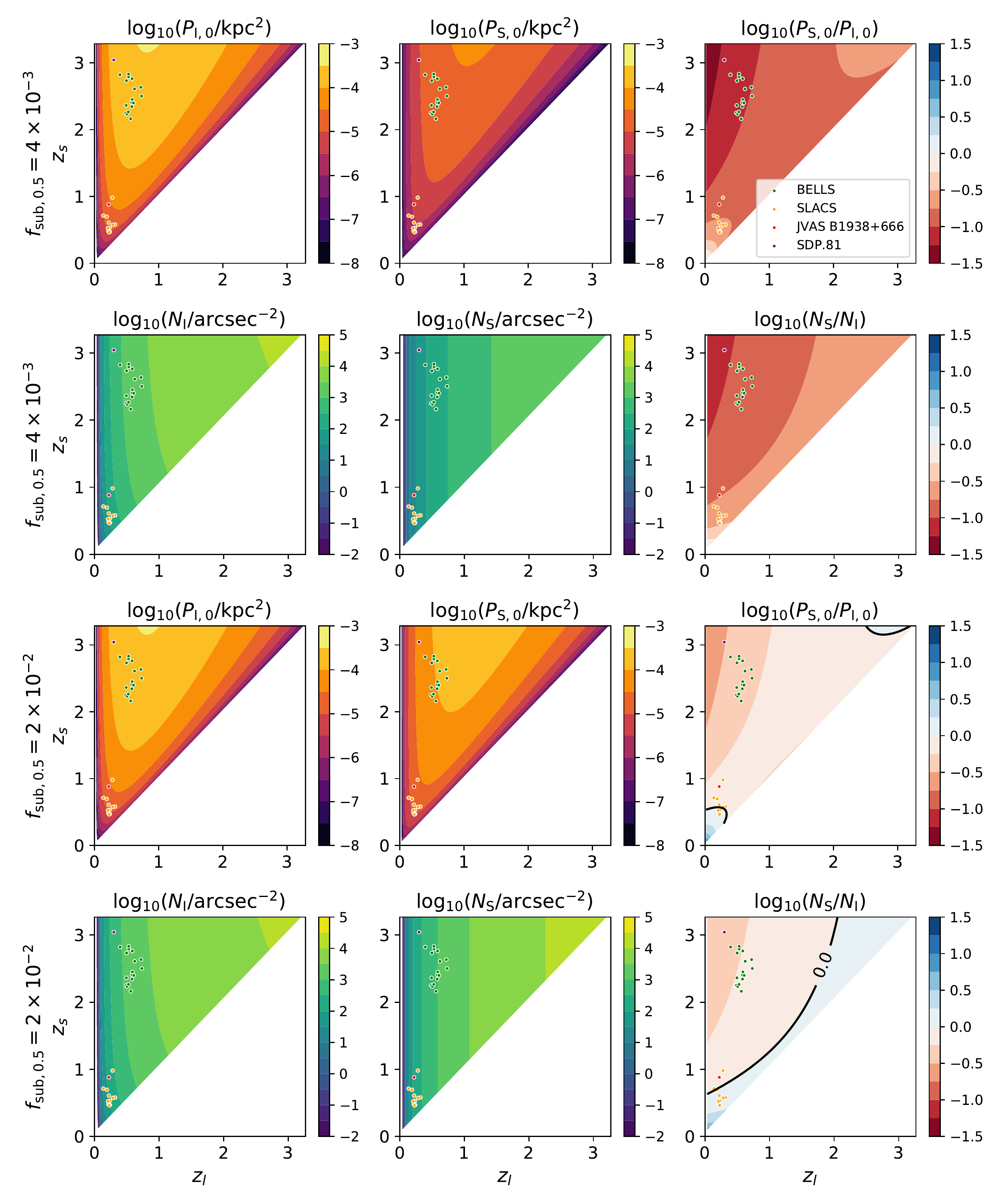}
    \caption{\footnotesize{\emph{Rows 1 and 3:} Plateau, i.e. $k\to 0$ limit, of the convergence power spectrum of perturbers in the range $[10^5\text{-}10^8]\,\msun$, for interlopers (\emph{left}), subhalos (\emph{center}), and the ratio of subhalos to interlopers (\emph{right}). \emph{Rows 2 and 4:} Number of perturbers per arcsec$^2$, for perturbers in mass range $[10^5\text{-}10^8]\,\msun$, for interlopers (\emph{left}), subhalos (\emph{center}), and the ratio (\emph{right}). }
    \emph{Top two rows:} Fiducial subhalo normalization, $f_{\mathrm{sub},0.5} = 4\times 10^{-3}$. \emph{Bottom two rows:} Higher subhalo normalization, $f_{\mathrm{sub}, 0.5} = 2\times 10^{-2}$, which gives a roughly equal subhalo and interloper contribution for SLACS systems. The line of equal contribution is shown in black.
    The dots represent some of the galaxy-galaxy lensing systems that have been studied in the literature \cite{bells_2018,Vegetti:2018dly,vegetti_nature,Hezaveh:2016ltk}. See \url{https://arthur-tsang.github.io/interloper_widget.html} for an interactive version of the power spectrum plateau plots, with adjustable $f_\mathrm{sub,0.5}$ and halo mass function.}
    \label{fig:plateau_comp}
\end{figure*}

The first row of Fig. \ref{fig:plateau_comp} shows the results for the power spectrum plateau due to interlopers (left), subhalos (center), and the ratio between the two (right), for our fiducial choice of $f_{\mathrm{sub}, 0.5}$. As expected, the relative interloper contribution generally increases with increasing source redshift, so interlopers contribute a greater fraction to the total power spectrum for the higher source redshift BELLS systems compared to SLACS: interlopers dominate over subhalos by a factor of a few for SLACS and by just over an order of magnitude for BELLS. The third row shows the same three panels but for a higher value of $f_{\mathrm{sub}, 0.5} = 0.02$. We can see that, for this value of the fraction of dark matter in substructure, the subhalo and interloper contributions for SLACS become roughly equal.

To better understand this turnover, and the relative contribution of perturbers more generally, we plot the number of perturbers per solid angle, as well as the ratio of subhalos to interlopers, in the second and fourth rows of Fig. \ref{fig:plateau_comp}. 
We see that the redshift dependence of these numbers is quite different from that of the plateaus,\footnote{In particular, the number of interlopers has a strong dependence on lens redshift, which comes from converting into angular dimensions.} but the ratios share a similar pattern.
They are slightly different because a factor of $m^2 n(m)$ goes into the integral for the plateau. Compared to the plateau ratios at a given redshift, the number density ratios tend to be slightly larger, which means that to a rough approximation, we can think of the plateau turnover as the place where there are slightly more subhalos than interlopers in the field of view.

Because these results are strongly dependent on the choice of subhalo and halo mass functions, we provide an interactive version of the power spectrum plots at \url{https://arthur-tsang.github.io/interloper_widget.html}, where the reader can adjust the value of $f_\mathrm{sub,0.5}$ as well as the halo mass function (Sheth-Tormen or Press-Schechter) to see how the results are affected. For SLACS lenses up to $f_{\rm sub,0.5} \sim 2\%$, the interloper contribution dominates. For larger values of $f_{\rm sub,0.5}$, however, the subhalo contribution takes over (albeit by less than an order of magnitude). For the BELLS lenses, a higher value of $f_{\rm sub,0.5} \sim 4\%$ is necessary for the subhalos to dominate due to their higher source redshifts.

\section{Numerical Calculation} \label{sec:numeric}

To verify our analytic results and, in particular, the validity of our approximations, we independently calculate the (effective) convergence power spectrum numerically. The main concern is the error introduced by approximating the interlopers as effective subhalos at the redshift of the main lens.\footnote{Note that other works have made this approximation as well \cite{Li_los, Despali_los}.} To understand the error introduced by this approximation, we calculate the convergence power spectrum from a lensing system simulated using the full multi-plane lens equation. 

In short, our procedure is to first use ray-tracing to generate a map of the total angular deflection $\vec \alpha
(\vec x_1)\equiv \sum^N_{i=1}\vec{\alpha}_i(\vec{x}_i)$, and then calculate
\begin{equation}\label{eq:effective_convergence}
    \kappa_\mathrm{eff,total} \equiv \dfrac 1 2 \nabla \cdot \vec \alpha,
\end{equation}
which is the multi-plane equivalent of the traditional convergence for single-plane lensing \cite{Gilman:2019vca}. To single out the effective convergence of the interlopers (or subhalos), we subtract the convergence of the main lens:
\begin{equation}\label{eq:eff_conv_for_int_and_sub}
    \kappa_\mathrm{eff} = \kappa_\mathrm{eff,total} - \kappa_l.
\end{equation}

\noindent We then convert $\kappa_\mathrm{eff}$ into a power spectrum using a 2D Fourier transform (squared) \cite{powerspec2}. Since we are interested in the monopole term, we perform an azimuthal average. 

In \S \ref{lensing_sim} we describe the lensing system we simulate, while in \S \ref{sec:num_ecps} we detail the procedure used to obtain the effective convergence power spectrum from it.

\subsection{A Simulated Strong-Lensing System with Interlopers and Subhalos}\label{lensing_sim}

We set the lens at $z_{l} = 0.5$ and the source at $z_{s} = 1.0$, and we assume a flat $\Lambda$CDM cosmology with $h=0.675$ and $\Omega_{\rm m} = 0.316$ \cite{Planck2018}.
The lens is a singular isothermal sphere (SIS) with an Einstein radius of $\theta_{\rm E} = 7\un{arcsec}$, which corresponds to a virial mass of $\sim 10^{14}\,\msun$.
This is larger than the typical $\sim [0.1-1] \un{arcsec}$ or $\sim [10^{12}-10^{13}]\,\msun$ of galaxy-galaxy lenses \cite{bells_i, slacs2}, although note that some systems may go up to $10^{14.5}\,\msun$ \cite{robertson20}. We use a somewhat larger lens in order to probe a broader range of wavenumbers; nevertheless, our results are fully applicable to galaxy-scale lenses because we use values of $f_{\rm sub,0.5}$ consistent with typical galaxy lenses, instead of scaling it up for a larger halos mass. Furthermore, the power spectrum is otherwise independent of lens size in our formalism. 
Note that because we calculate the power spectrum from deflection angles rather than from an observed image, the only relevant property of the source is its redshift. 

We randomly populate interlopers on 100 planes, evenly spaced in redshift between the source and observer.\footnote{Note that the result is unchanged if we increase the number of planes.}
The number of interlopers on each plane is chosen as a Poisson random variable whose expected value is the number of interlopers between adjacent redshift planes according to the Sheth-Tormen mass function \cite{sheth} (same as in \S\ref{pow_spec_cal}). The interloper masses are randomly chosen between $[10^5-10^8]\,\msun$, again according to Sheth-Tormen. Their positions within each plane are uncorrelated and uniformly distributed within the double pyramid of visible structure (analogous to the double cone in Fig. \ref{fig:double_cone}, but now instead we use a double pyramid, since we simulate a square field of view). The interloper profiles are modeled as in \S \ref{subsec:effpower}. While the true profile of interlopers may be slightly different, note that the low-$k$ limit of the power spectrum is sensitive to the abundance of interlopers and not to the intrinsic profile parameters. In addition to the interlopers, we add a negative mass sheet to each redshift plane to cancel out the net mass of the interlopers, which captures the fact that the underdense regions along the line of sight effectively lens as negative masses.

For completeness, we also simulate a lensing system that only has subhalos, and one that has both subhalos and interlopers. To populate the lens with subhalos, we assume that subhalos are uncorrelated, uniformly spatially distributed, and follow the mass function of Eq. \eqref{eq:nsub}. Note that it is possible to simulate the two-subhalo term by modifying the spatial distribution, however, in a realistic lensing galaxy with baryons we expect this term to be subdominant \cite{powerspec2}.

\subsection{Obtaining the Convergence Power Spectrum from a Simulation}\label{sec:num_ecps}

We calculate the total deflection vector $\vec \alpha(\vec x_1)$ using the full multi-plane lens equation for the simulated lensing systems using \texttt{lenstronomy} \cite{lenstronomy}, which is a publicly available \texttt{Python} package. 
We then calculate $\kappa_\mathrm{eff}$ by taking the divergence of $\vec \alpha$ using the five-point stencil method (see Appendix \ref{app:d}) to limit the numerical error. 

We run two different simulations, both with $(500\times 500)$ pixels, but with different fields of view: $(1.6\times 1.6)\un{arcsec}$ (small) and $(16\times 16)\un{arcsec}$ (large). This is to sample a wide range of wavenumbers that would otherwise require a much larger number of pixels and would thus be computationally intractable. These two different fields of view require different treatments to be processed into the convergence power spectrum. This is because when all parts of the $\kappa_\mathrm{eff}$ map are statistically equivalent, the Fourier transform squared of $\kappa_\mathrm{eff}$ is the two-dimensional interloper power spectrum.
However, our analytic approximation from \S\ref{sec:analytic} only applies near the Einstein radius (since the derivation relied on the CSB approximation).

\begin{figure}
    \centering
    \includegraphics[width=\linewidth]{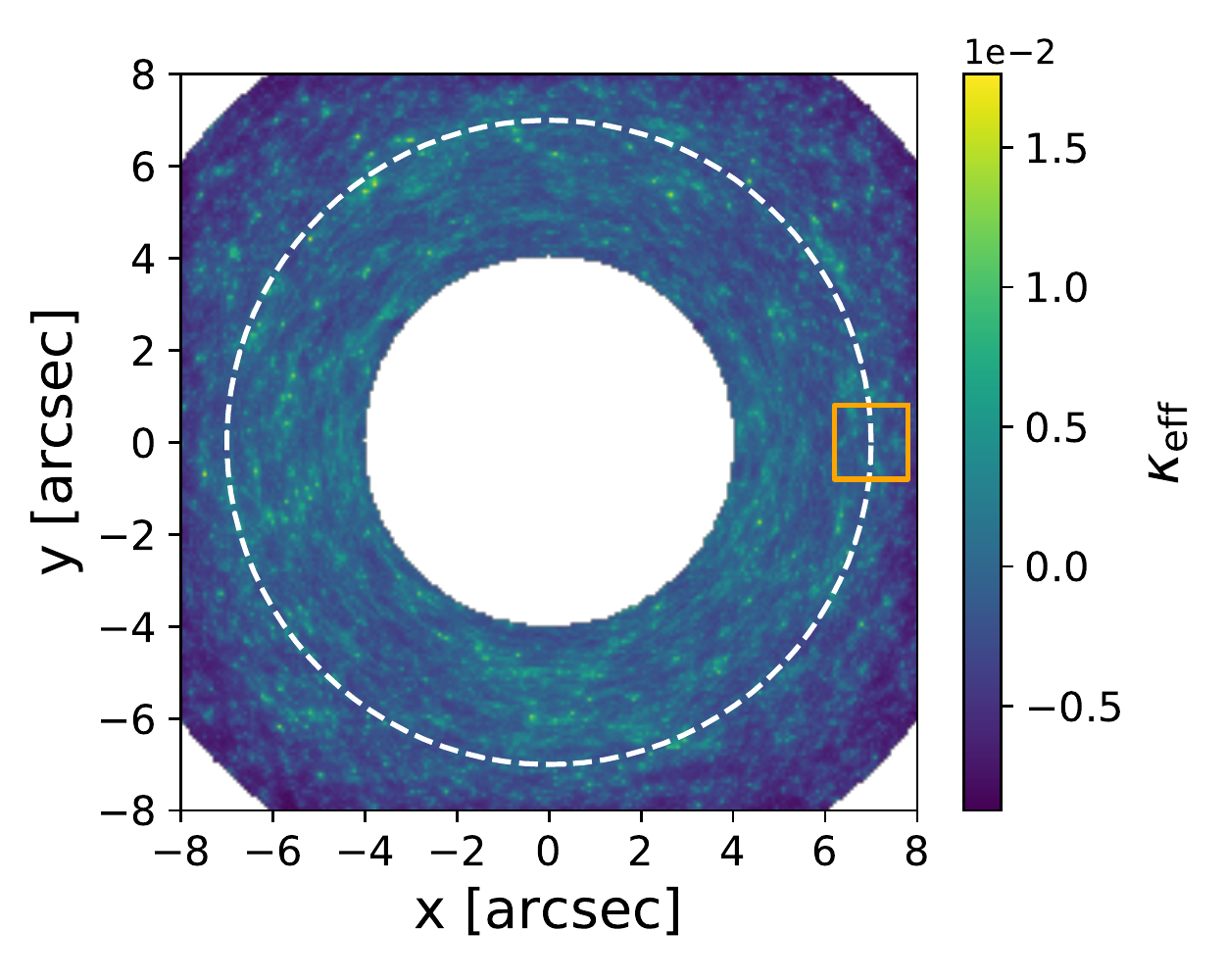}
    \caption{\footnotesize{Illustration of the effective convergence field $\kappa_\mathrm{eff}$ used for the full multi-plane lensing calculation with $z_l = 0.5$ and $z_s = 1.0$. The annular mask can be seen for the large field of view (16 arcsec). The small field of view (1.6 arcsec) is represented with an orange square on the right. The white dashed line shows the Einstein radius.}} 
    \label{fig:mask}
\end{figure}

Due to this, for the large field of view we filter $\kappa_\mathrm{eff}$ with an annular mask centered on the main lens, setting $\kappa_\mathrm{eff} = 0$ outside the mask (see Fig. \ref{fig:mask}).\footnote{Using a mask affects the normalization of the Fourier transform, so to correct for it we divide the power by the fraction of the image covered by the mask.} For the small field of view, we center the image on a point on the Einstein ring in order to remain in a regime where the CSB approximation is valid, so we can compare with our analytic results. 

Having to impose a mask for the large field of view has several limitations. First, while it is not desirable to use points too far from the Einstein ring (the CSB approximation gets progressively worse with increasing distance), a narrow mask does not estimate well the lowest-$k$ modes since they correspond to sizes larger than the annulus width. Furthermore, the Fourier transform of the mask can itself give rise to unphysical oscillations. Ultimately, we opted for a mask width of $\pm(3/7)\theta_{\rm E}$. An annulus of this width is sufficient to smooth out the oscillations and probe relatively low-$k$ modes. Furthermore, because it is quite wide, it allows us to be conservative when comparing it to the analytical results: we know that as the mask becomes wider, the validity of the CSB approximation decreases. 

\subsubsection{Comparison to Analytical Results}

\begin{figure}[ht]
\centering
\includegraphics[width=\linewidth]{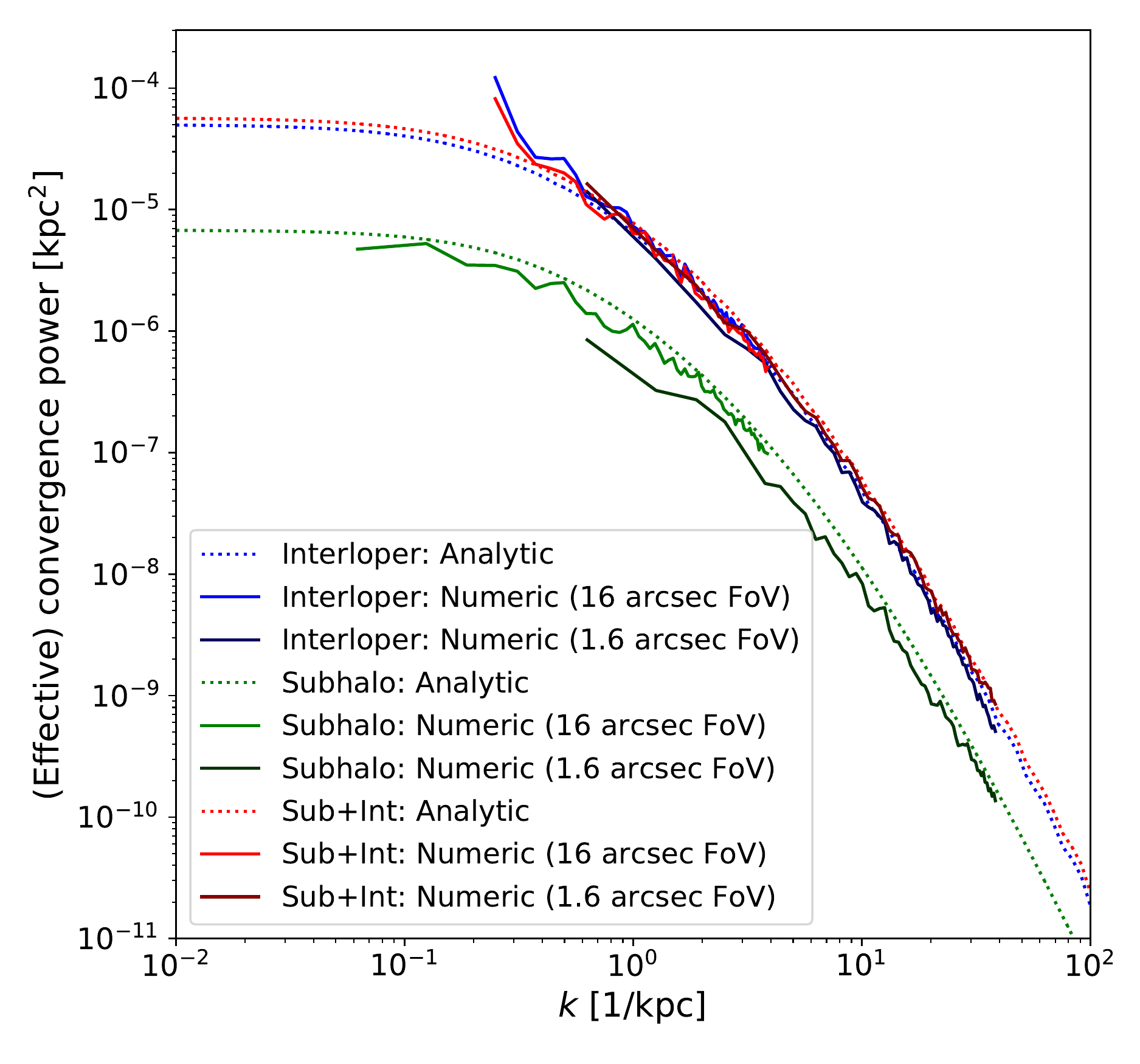}
\caption{\footnotesize{Power spectra for interlopers (blue), subhalos (green), and both (red). We show analytic results (dotted) and numerical results (solid) for the simulations with field of view (FoV) of 1.6 arcsec (darker) and 16 arcsec (lighter). All numerical curves are cut off at high $k$ due to loss of power near a pixel size, and the masked simulations are cut off at low $k$, corresponding to the mask width.}}
\label{fig:compare_power}
\end{figure}

In Fig. \ref{fig:compare_power}, we compare the effective convergence power spectrum obtained following the procedure outlined above to the analytical predictions derived in \S\ref{subsec:effpower}. We show the contribution due solely to interlopers (blue), solely to subhalos (green), and their combination (red).\footnote{We used the same subhalo and interloper population characteristics, as described in Eqs. \eqref{deltas} to \eqref{eq:nsub}.} The analytical results are shown as dotted lines, and the numerical results are shown as solid lines. We see that the two independent estimates of the effective convergence power spectrum show excellent agreement,\footnote{Note that, for images without masks (subhalo only and all small field of view images), the minimum $k$ we plot is $2\pi/L$, where $L$ is the width of the image. For the images with masks (interloper and combined for large field of view), the minimum $k$ is $2\pi/L'$, where $L' = L/4$ is the width of the annulus mask (see Fig. \ref{fig:mask}).} even though the annular mask used was quite wide. This agreement shows that treating the interlopers as effective subhalos using our framework introduces a very small error compared to the full ray tracing results, even in a regime far from the Einstein radius. Furthermore, we note that the small difference between the analytical and numerical results is much smaller than even the most optimistic expected error bars from near-future surveys \cite{beyond_subhalos}.

\subsubsection{Quantifying the Error in the Numerical Effective Convergence}\label{curl_error}

The advantage of calculating the effective convergence as in Eq. \eqref{eq:effective_convergence} is that we do not have to make any approximations, such as the CSB approximation that was necessary for the analytical derivation in \S\ref{sec:analytic}. However, this single-plane effective convergence, and the one derived in \S\ref{sec:analytic}, do not reproduce the deflection angles exactly. To see why this is the case let us separate the deflection angles into a curl-free and a divergence-free component:
\begin{equation}
    \vec \alpha = \vec \alpha_{\mathrm{div}} + \vec \alpha_{\mathrm{curl}} 
\end{equation}
\begin{equation}
    \nabla \times \vec\alpha_{\mathrm{div}} = 0 \And \nabla \cdot \vec\alpha_{\mathrm{curl}} = 0.
\end{equation}
We call $\vec \alpha_{\mathrm{div}}$ the divergence component and $\vec \alpha_{\mathrm{curl}}$ the curl component. 

In single-plane lensing, $\vec \alpha_\mathrm{curl}$ vanishes because the deflection angle can be written as the gradient of the lensing potential $\Psi$ \cite{lensing_book}:
\begin{equation}\label{eq:curlless}
    \vec \alpha(\vec x) = \nabla_x \Psi(\vec x) \quad \Rightarrow \quad \nabla \times \vec \alpha = 0.
\end{equation}

However, in the multi-plane lensing case, the coupling between the successive lens planes introduces a curl component \cite{Gilman:2019nap}. In our calculation it is a nuisance since it indicates that the single-plane effective convergence is not fully capturing the multi-plane lensing of the interlopers. In terms of the total deflection angle, the two components can be written as (see Appendix \ref{app:d})
\begin{equation}
    \vec \alpha_{\mathrm{div}}(\vec x) = \dfrac{1}{\pi} \int d^2x'\, \dfrac{\vec{x}-\vec{x}'}{|\vec{x} - \vec{x}'|^2}\left[\dfrac{1}{2}\nabla_{\vec{x}'} \cdot \vec{\alpha}\right]
\end{equation}
\begin{equation}
    \vec \alpha_{\mathrm{curl}}(\vec x) = \hat{z} \times \dfrac{1}{\pi} \int d^2x'\, \dfrac{\vec{x}-\vec{x}'}{|\vec{x} - \vec{x}'|^2}\left[\dfrac{1}{2}\nabla_{\vec{x}'} \times \vec{\alpha}\right].
\end{equation}
Here $\hat z$ is the unit vector that is orthogonal to the lens-plane and pointing towards the observer. We see that what we defined as $\kappa_\mathrm{eff}$ in Eq. \eqref{eq:effective_convergence} sources the divergence component, i.e.
\begin{equation}\label{eq:kappadiv}
    \kappa_\mathrm{div} = \kappa_\mathrm{eff} \equiv \dfrac{1}{2}\nabla\cdot \vec \alpha,
\end{equation}
\noindent and 
\begin{equation}\label{eq:kappacurl}
    \kappa_\mathrm{curl} \equiv \dfrac{1}{2}\nabla\times\vec \alpha
\end{equation}
sources the curl component. So we can compare $\kappa_\mathrm{curl}$ to $\kappa_\mathrm{eff}$ in order to gain an understanding of the error introduced by treating interlopers as effective subhalos.

\begin{figure*}[ht]
    \centering
    \includegraphics[width=\linewidth]{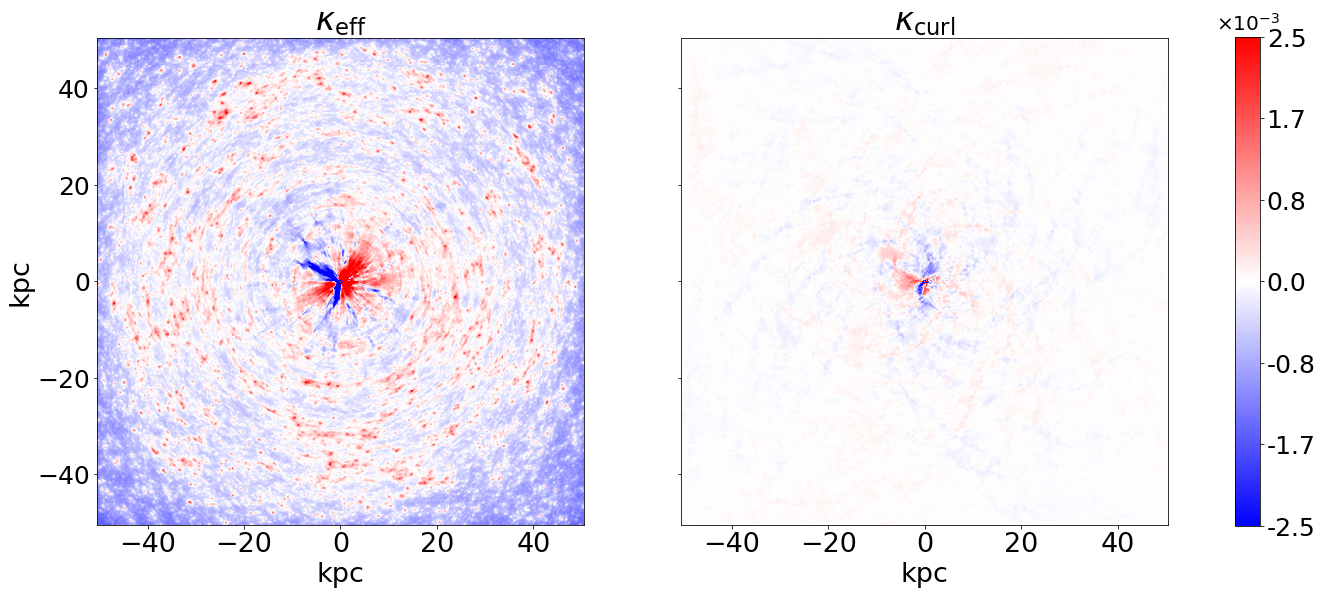}
    \caption{\footnotesize{Two types of effective convergences calculated from the divergence and the curl of angular deflections in a simulated strong-lensing system with interlopers and subhalos. \textit{Left:} Effective convergence of the interlopers + subhalos, defined in Eq. \eqref{eq:eff_conv_for_int_and_sub}, which sources the divergence component of the angular deflections. The central dipole is caused by the coupling between the uneven distribution of interlopers and subhalos and the main lens. We can safely ignore the dipole as we are interested in a small annulus around the Einstein ring where the strong-lensing images are produced. \textit{Right:} $\kappa_\mathrm{curl}$, defined in Eq. \eqref{eq:kappacurl}, which sources the curl component of the angular deflections. The interlopers that are far from the lens plane, either towards the observer or towards the source, contribute more to the curl component.}}
    \label{fig:compare.png}
\end{figure*}

We see in Fig. \ref{fig:compare.png} that for our simulation, $\kappa_\mathrm{curl} \ll \kappa_\mathrm{eff}$,\footnote{We know from Eq. \eqref{eq:curlless} that the curl component has to be zero when there are no interlopers, i.e. when the lensing is caused by mass on a single lens plane. In Appendix \ref{app:d} we show that the curl component that we measure is not a numerical artifact but indeed a result of the coupling of the lensing effect of multiple lens planes at various redshifts.} especially near the Einstein radius.
We thus conclude that the curl component of the angular deflection is also much smaller than the divergence component, meaning that the coupling between the interlopers and the main lens is small enough to justify the projection of interlopers as effective subhalos in the main lens. To test whether the $\kappa_\mathrm{curl}$ in Fig. \ref{fig:compare.png} is simply a numerical artifact, we simulate a system with only a main lens and subhalos (no interlopers) in Appendix \ref{app:d}, and show in Fig. \ref{fig:numeric_curl} that the numerical error is more than two orders of magnitude smaller than the curl observed in Fig. \ref{fig:compare.png}, showing that the curl term sourced by the interlopers is physical.

\section{Conclusions}\label{sec:conclusions}

Strong gravitational lensing has long provided some of the most decisive evidence for the existence of dark matter in our universe. Within the past couple of decades, the use of gravitational lensing for dark matter science has expanded considerably, and significant theoretical and observational advances have turned it into one of the most promising probes of the nature of dark matter. In particular, there is great interest in using strongly lensed images to constrain the very low-mass end of the halo mass function ($\lesssim 10
^8$ $\msun$), since this regime can distinguish between vanilla CDM and more exotic models, such as warm dark matter or self-interacting dark matter, that can lead to low-mass cutoffs. 

The canonical approach followed has been to try to directly detect individual dark matter clumps, such as in gravitational imaging \cite{grav_imaging1}. This approach has claimed detections of substructures with masses as small as a few times $10^8$ $\msun$ \cite{spat_res_spec2,detection_2010_mnras,vegetti_nature,bells_2018,2014MNRAS.442.2017V,Vegetti:2018dly}, but reliably reaching lower masses in galaxy-galaxy lenses with this approach has remained elusive.

The idea of using the subhalo convergence power spectrum \cite{Hezaveh_powerspec,powerspec1,powerspec2,Brennan} was developed as a statistical detection method to obtain population-level constraints without having to individually resolve clumps. Unlike in direct detection efforts where by construction the sensitivity is maximal for the most massive clump close to the lensed images, in a power spectrum approach the higher number of lower-mass halos can actually make the sensitivity peak for the mass range of $10^7-10^8$ $\msun$ for a CDM population of subhalos, and still maintain some sensitivity at lower masses \cite{powerspec2}. 

However, while perturbations due to line-of-sight halos have been studied in the context of lensed galaxies \cite{Li_los,Despali_los} and lensed quasars \cite{McCully:2016yfe,Gilman:2019vca}, and has become standard practice in direct detection pipelines \cite{Vegetti:2018dly,Gilman:2019nap,Gilman:2019bdm,Hsueh:2019ynk}, its contribution to the convergence power spectrum had not yet been quantified. In this work, we set out to do so. 

We show that it is possible to define an effective convergence for multi-plane lensing systems with a dominant main lens coupled to lower-mass interlopers. One can think of this effective convergence as mapping an interloper at any point along the line of sight onto the lens plane as an effective subhalo with a modified mass and scale radius.\footnote{The scale radius is the relevant lensing length scale in our density profile of choice (truncated NFW). Other density profiles would see an analogous rescaling of relevant intrinsic parameters.} It is then possible to analytically calculate its power spectrum, incorporating the relative effect of interlopers at different redshifts as a lensing kernel. We find that the interlopers that are closest to the main-lens plane have the largest contribution to the power spectrum, while those close to the observer or source are negligible.

Because the halo and subhalo mass functions evolve with redshift, and in fact there are still considerable unknowns with respect to the subhalo mass function's normalization, we do not expect that a global statement can be made about the importance of one versus the other. Instead, we opt to thoroughly quantify each contribution as a function of both source and lens redshift, and for different choices of mass function and subhalo mass function normalizations. We show specifically what we expect the ratio of power between subhalos and interlopers to be for galaxy-galaxy lensing systems for which we currently have high-resolution imaging (Fig. \ref{fig:plateau_comp}).

For our fiducial choice of $f_{\rm sub,0.5} = 4 \times 10^{-3}$, we find that for both the BELLS and SLACS lenses the interloper contribution dominates, albeit to different extents. Due to the higher redshift of the BELLS sources, the interloper contribution is larger by over an order of magnitude, while the lower redshift of the SLACS sources lead the interloper plateau to only be larger by a factor of a few. As we increase or decrease $f_{\rm sub,0.5}$, each group of lenses is affected differently: for SLACS, the subhalos dominate for as little as $f_{\rm sub,0.5} \gtrsim 2\%$, while for BELLS they do so for $f_{\rm sub,0.5} \gtrsim 4\%$. This can be intuitively understood: with increasing source redshift, the LOS volume increases, overwhelming the subhalo signal.

Let us put these bounds into context by discussing the expected values of $f_{\rm sub,0.5}$ in these systems.
As we discussed in \S \ref{subsec:effpower}, there is considerable uncertainty both on the simulation side and on the observational side. Nevertheless, even with this uncertainty, it seems unlikely that $f_{\rm sub,0.2}$ and $f_{\rm sub,0.5}$ would reach these values, especially with an upper subhalo mass of $10^8$ $\msun$. Ref. \cite{2014MNRAS.442.2017V} found that for SLACS lenses, the upper bound on $f_{\rm sub,0.2}$ was about 2.7\% with an upper mass bound of $4 \times 10^9$ $\msun$. Ref. \cite{bells_2018} found that for BELLS lenses, the upper bound on $f_{\mathrm{sub},0.5}$ is $7\%$ with an upper mass bound of $10^{11}$ $\msun$, which corresponds to $f_{\mathrm{sub},0.5}<2.3\%$ for an upper mass bound of $10^{8}$ $\msun$. Thus, we conclude that it is likely for the interloper contribution to dominate in these two ensembles of lenses.

One worry about this approach might be that treating the interlopers as effective subhalos is overly simplistic since it neglects the recursive nature of the multi-plane lens equation, which couples the deflection angles of successive lens planes. To study this, we tested the analytical calculation with mock lensing simulations obtained by doing ray tracing with the multi-lens plane equation (without any approximations). We find that the power spectrum from the simulations matches the analytical prediction extremely well. 

Furthermore, we note that defining the effective convergence as the divergence of the deflection angle does not capture the divergence-free part (what we call the curl component in \S \ref{curl_error}) of the angular deflections sourced by interlopers. For our analysis, which projects the interlopers onto the main lens as effective subhalos, the curl part quantifies the error in doing such a projection since a single-plane effective convergence cannot create such a term (See Appendix \ref{app:d}). We show in Fig. \ref{fig:compare.png} that the curl term for low-mass interlopers is very small compared to the divergence term, meaning that the projection introduces minimal error in calculating the angular deflections for multi-plane lensing. 
We point out that, for more massive perturbers, the fact that multiple lens planes source a curl term suggests a novel way of identifying multi-plane lensing and therefore distinguish interlopers from subhalos, by for example measuring a B-mode power spectrum. 

Our results on the importance of incorporating interlopers into the analysis of strong-lensing systems are broadly consistent with Refs. \cite{Li_los} and \cite{Despali_los}, which did so in the context of direct detection efforts. An important point to keep in mind is that the mass ranges and subhalo mass function normalizations they consider are different to ours; their perturber mass range spans $\sim 10^6\,\msun-10^{11}\,\msun$, so converting to our definition of $f_{\rm sub,0.5}$ gives a lower value than what they cite. Ref. \cite{Li_los} considered a single lensing configuration ($z_{l} = 0.2$ and $z_{s} = 1$) and found that the number of interlopers is roughly four times higher than that of subhalos. Ref. \cite{Despali_los} found that the number of line-of-sight perturbers is comparable to subhalo perturbers for low-redshift lenses (e.g. SLACS lenses) and dominant over subhalo perturbers for high-redshift lenses (e.g. BELLS lenses). The difference between our results and these can be understood because of the different mass range: although they have a smaller subhalo mass function normalization, the extra subhalos at larger masses make the subhalo contribution comparable for SLACS instead of subdominant, as is the case in our fiducial results.\footnote{While these two references developed the notion of using interlopers as effective subhalos, we note that our projection prescription is inherently different. 
For example, Ref. \cite{Despali_los} relied on first projecting the interloper positions onto the lens plane by ensuring they remained on the same line of sight, and then varied their mass to minimize the residual of the angular deflections between the projected interloper and a subhalo of a given mass. The downside of this projection prescription is that the minimization is done for an unobservable lensing quantity.}

As the LOS contribution has gained recognition as an integral ingredient in lens modeling, new systems have been analyzed, and older systems reanalyzed, taking it into account. To date, a single real lens has been analyzed through a power spectrum approach. Ref. \cite{bayer_power} placed an upper bound on the convergence power spectrum due to subhalos using lens system SDSS J0252+0039 ($z_l = 0.280$, $z_s=0.982$)\footnote{This system had been analyzed with gravitational imaging in Ref. \cite{2014MNRAS.442.2017V} and no evidence of a substructure above the mass-detection threshold was found.}. Their upper bound on the power spectrum was significantly higher than the expected amplitude due to subhalos in a CDM scenario, but interestingly, according to our results, for this redshift combination we expect the line-of-sight contribution to dominate. 

The fact that under many lensing configurations and reasonable subhalo population assumptions the interloper contribution dominates the signal can be good news for the capacity of strong gravitational lenses to constrain the particle nature of dark matter. The amplitude of the convergence power spectrum can essentially be tied back to a mass function (halo mass function for interlopers, subhalo mass function for subhalos), so in order to translate a power spectrum amplitude to a dark matter theory, it is paramount to understand the relevant mass function(s). The subhalo mass function is very hard to pin down. It depends on the host mass and inevitably evolves with redshift. How it is affected by subhalos traveling within the host's dark matter halo, as well as due to any baryons in the host, remains an open problem: neither theory nor simulations have yet converged on a satisfying answer to these questions. In comparison, halos that are only subject to large-scale tidal fields have relatively calmer lives, and their evolution is better understood. Therefore, having a window into the smallest dark matter scales in the universe without relying on subhalos can make gravitational lensing a much more powerful (and reliable) probe of dark matter.

Furthermore, we note that the advantages of statistical detection efforts compared to direct detection ones that were introduced in the subhalo context, namely taking advantage of the much more numerous population of low-mass halos that are individually below the detection threshold, are undeniably advantageous in the interloper context as well. For instance, the number of interlopers that are massive enough for direct detection was shown to be roughly unity for the $17$ BELLS lenses shown in Fig. \ref{fig:plateau_comp} \cite{bells_2018}, which prevents lack of detections in the ensemble of lenses to be used to rule out CDM. Since the lower-mass interlopers are expected to be much more abundant, if the power spectrum can be measured 
(which Refs. \cite{Hezaveh_powerspec,beyond_subhalos} claim can be done using near future observations), the lack of power at high redshift lenses can more decisively rule out the CDM scenario.

There is considerable momentum being harnessed by gravitational lensing as a cosmological probe. Much progress has been made over the course of the last decade regarding how to model these systems. Furthermore, in the last couple of years new methods that harness the image recognition power of machine learning methods have started being developed to accelerate the indirect and direct detection of perturbers in strong-lens images \cite{mining_substructure,Alexander:2019puy,DiazRivero:2019hxf,Varma:2020kbq}. Between these advances in detecting perturbers in optical imaging data, and the fact that we expect thousands of new high-quality optical imaging strong-lens systems to become available in the near future \cite{LSST_lenses,2014MNRAS.439.3392P,Collett_2015},
we expect to have a treasure trove of data for dark matter science soon. In order for strong gravitational lensing to establish itself as a premier way of constraining dark matter, however, we need to ensure that the mapping from observations to theory is done correctly, which undoubtedly involves accounting for the line-of-sight contribution.

\section*{Acknowledgments}

We thank Simon Birrer for useful discussions. We would also like to thank Irwin Shapiro for pointing us to a typo in Eq. \eqref{eq:typo} in a previous version of this manuscript.
CD was partially supported by Department of Energy (DOE) grant DE-SC0020223.

\appendix

\section{Scaling Relations for tNFW Projected Density}\label{app:a}

In this Appendix we derive the scaling laws for the surface density that we used in \S\ref{eff_subhalo}. Let us start with Eq. \eqref{scaling_mass}, which is trivial because scaling the surface density everywhere by a constant amount scales the total mass by the same amount. Eq. \eqref{scaling_radius} can be derived by first denoting
\begin{equation}\label{sigma_compare}
    \Sigma(\vec r\,;\, m', r'_{\rm s},\tau)=\Sigma(\eta \vec r\,;\, m, r_{\rm s},\tau).
\end{equation}
The new mass $m'$ can then be found by integrating over the whole 2D plane:
\begin{align}
    m' &= \int d^2\vec r \,\,\Sigma(\vec r\,;\, m', r'_{\rm s},\tau') \\
       &= \frac{1}{\eta^2}\int d^2(\eta \vec r) \,\,\Sigma(\eta \vec r\,;\, m, r_{\rm s},\tau)\\
       &= \frac{m}{\eta^2}.
\end{align}
The new scale radius can be found by setting $|\vec r| = r'_{\rm s}$ in Eq. \eqref{sigma_compare}, which gives the relation
\begin{equation}
    r'_{\rm s} = \frac{r_{\rm s}}{\eta}. 
\end{equation}

\section{2D Fourier Transform of a Projection}\label{app:b}

In this Appendix we derive a useful relationship between the Fourier transform of the dimensionless tNFW density profile $\phi$ (defined in Eq. \ref{eq:dimless}) and its projection. 

Let us write the effective convergence as
\begin{equation}
    \kappa_{i,\mathrm{eff}}(s) = \dfrac{1 }{\Sigma_{\mathrm{cr},l}}\dfrac{m_{\mathrm{eff},i}}{r^2_{s,\mathrm{eff},i}}\Theta\left(\dfrac{s}{r_{s,\mathrm{eff},i}}\,;\,\tau_i\right),   
\end{equation}
where $\Theta$ is defined in terms of $\phi$ as
\begin{equation}
    \Theta(t\,;\,\tau) \equiv \int dw \, \phi(\sqrt{t^2 + w^2}\,;\,\tau)
\end{equation}
and its Fourier transform is
\begin{equation}\label{eq:kappa_four}
    \tilde \kappa_{i,\mathrm{eff}}(k) = \dfrac{m_{\mathrm{eff},i} }{\Sigma_{\mathrm{cr},l}}\tilde\Theta\left(r_{s,\mathrm{eff},i}k\,;\,\tau_i\right).   
\end{equation}
If we write the Fourier transform of $\Theta$ explicitly, we can relate it to the Fourier transform of $\phi$.  
Let us start with
\begin{align}
    \tilde \Theta(k\,;\,\tau) &= \int d^2 \vec t\, \exp[-i\vec k \cdot \vec t\,]\Theta(t\,;\,\tau) \nonumber \\
    &= \int d^2 \vec t\,dw\, \exp[-i\vec k \cdot \vec t\,]\phi(\sqrt{t^2 + w^2}\,;\,\tau).
\end{align}
We can combine $(\vec t, w)$ into a 3D vector $\vec r$. We can also replace $\vec k \cdot \vec t$ with $\vec k \cdot \vec r$ since $\vec k$ is perpendicular to the $z$-axis. We then obtain
\begin{align}\label{eq:2D_3D}
    \tilde \Theta(k\,;\,\tau) &= \int d^3 \vec r\,\exp[-i\vec k \cdot \vec r\,]\phi(r\,;\,\tau) \nonumber \\
    &= \tilde \phi(k\,;\,\tau).
\end{align}
Therefore, we can write the Fourier transform of the effective convergence of each interloper $i$ as
\begin{equation}\label{eq:kappa_four2}
    \tilde \kappa_{i,\mathrm{eff}}(k) = \dfrac{m_{\mathrm{eff},i} }{\Sigma_{\mathrm{cr},l}}\tilde\phi\left(r_{s,\mathrm{eff},i}k\,;\,\tau_i\right).   
\end{equation}

\section{The Area of the Double Cone}\label{app:c}

The radius of the double cone (Fig. \ref{fig:double_cone}) increases linearly from 0 to its maximum at the lens with comoving distance from the observer $\chi$, and it decreases linearly from its maximum at the lens to 0 with the comoving distance from the lens $\chi-\chi_l$. With this in mind, we can write the follow relation for the radius of the double cone $R(\chi)$:
\begin{equation}\label{eq:raw_cone_area}
    \frac{R(\chi)}{R(\chi_l)} = \left\{
        \begin{array}{ll}
            a_1(\chi + b_1) & \quad \chi \leq \chi_l \\[3ex]
            a_2(\chi + b_2) & \quad \chi > \chi_l,
        \end{array}
    \right.
\end{equation}
where $a_1,a_2,b_1,$ and $b_2$ are the linearity constants which we will obtain from the boundary values. We know that
\begin{align}
    \frac{R(0)}{R(\chi_l)} = 0 &\qrq b_1 = 0 \\
    \frac{R(\chi_l)}{R(\chi_l)} = 1 &\qrq a_{1,2}(\chi_l+b_{1,2}) = 1  \\
    \frac{R(\chi_s)}{R(\chi_l)} = 0 &\qrq a_2(\chi_s + b_2) = 0. \label{eq:typo}
\end{align}
Solving for the constants we get
\begin{equation}\label{eq:cone_area}
    \frac{R(\chi)}{R(\chi_l)} = \left\{
        \begin{array}{ll}
            \dfrac{\chi}{\chi_l} & \quad \chi \leq \chi_l \\[3ex]
            \dfrac{\chi_s - \chi}{\chi_s - \chi_l} & \quad \chi > \chi_l.
        \end{array}
    \right.
\end{equation}
Then, the normalized area is
\begin{align}\label{eq:cone_area}
    \frac{S(\chi)}{A/a^2(\chi_l)} &= \left\{
        \begin{array}{ll}
            \left(\dfrac{\chi}{\chi_l}\right)^2 & \quad \chi \leq \chi_l \\[3ex]
            \left(\dfrac{\chi_s - \chi}{\chi_s - \chi_l}\right)^2 & \quad \chi > \chi_l
        \end{array}
    \right. \\
    &=\frac{\chi^2 }{\chi^2_l}g^2(\chi)
\end{align}
where $A$ is the physical area of the main lens.

\section{Number density of subhalos}
\label{app:nsub}

Here we derive Eq. \eqref{eq:nsub}, which gives the number of subhalos per area as a function of mass and lens redshift. In particular, the expression is shown in terms of the fraction of halo mass in substructure, $f_{{\rm sub},z_l}$, which is commonly used in the literature as a proxy for the subhalo mass function normalization.

At any redshift, the total mass in subhalos within the mass range [$10^5 - 10^8$] $\msun$ can be written as
\begin{equation}\label{eq:total_num}
    M_\mathrm{sub} = A \bar\kappa_\mathrm{sub}\Sigma_{\mathrm{cr},}
\end{equation}
where $A$ is the area of the lens, and $\bar \kappa_\mathrm{sub}$ is the average convergence of the subhalos. We assume a moderately elliptical isothermal lens, so that the convergence of the main lens near the Einstein radius is $\kappa_l = 0.5$, \  and roughly $40\%$ of matter within the Einstein radius comes from stars \cite{auger09}. Therefore, we can write
\begin{equation}
    f_{\mathrm{sub},z_l} = \frac{2}{0.6}\bar \kappa_\mathrm{sub},
\end{equation}
which we can substitute into Eq. \eqref{eq:total_num} and get
\begin{equation}\label{eq:fsubeq}
    M_\mathrm{sub} = \frac{0.6}{2}A f_{\mathrm{sub},z_l}\Sigma_{\mathrm{cr},z_l}.
\end{equation}

Our goal is to express the normalization of $n_\mathrm{sub}$ evaluated at $z_l = 0.5$, denoted $F$, in terms of $f_\mathrm{sub,0.5}$ evaluated at that same redshift. So we write
\begin{equation}\label{eq:simple_nsub}
    n_\mathrm{sub}(m,z_l=0.5) = F m^{\beta}.
\end{equation}
The total mass in subhalos can be calculated from $n_\mathrm{sub}$:
\begin{align}
    M_\mathrm{sub} &= \int dA\int^{m_\mathrm{high}}_{m_\mathrm{low}} dm \, m\, n_\mathrm{sub}(m,0.5) \nonumber \\
     &= AF\int^{m_\mathrm{high}}_{m_\mathrm{low}} dm \, m^{\beta+1}  \nonumber \\
     &= AF \frac{m^{2+\beta}_\mathrm{high}-m^{2+\beta}_\mathrm{low}}{2+\beta}.
\end{align}
Combined with Eq. \eqref{eq:fsubeq}, we get
\begin{equation}
    F = \Sigma_{\mathrm{cr},0.5}\frac{0.6f_{\mathrm{sub},0.5}}{2}\frac{2+\beta}{m_\mathrm{high}^{2+\beta}-m_\mathrm{low}^{2+\beta}}.
\end{equation}

Plugging this into Eq. \eqref{eq:simple_nsub} we obtain the final expression for the number density of subhalos:
\begin{equation}
    n_\mathrm{sub}(m, z_l=0.5) = \frac{0.3\Sigma_{\mathrm{cr},0.5}f_{\mathrm{sub},0.5}(2+\beta)}{(m_\mathrm{high}^{2+\beta}-m_\mathrm{low}^{2+\beta})} m^\beta.
\end{equation}

The redshift dependence of the halo number density per comoving area is found to be $(1+z_l)^{1/2}$ \cite{fsubredshift}. The number density per physical area will then depend on lens redshift as $(1+z_l)^{5/2}$ where the extra factor of $2$ comes from the scale factor. We now have the final expression for the subhalo mass function:
\begin{equation}
    n_\mathrm{sub}(m, z_l) = n_\mathrm{sub}(m, z_l=0.5)\frac{(1+z_l)^{5/2}}{(1+0.5)^{5/2}}.
\end{equation}

\section{$k \rightarrow 0$ Limit of the Perturber Power Spectrum}
\label{app:plateau}

For interlopers, we first calculate the Fourier transform of the profile $\phi$ (defined in Eq. \ref{eq:dimless}) in the $k\rightarrow 0$ limit. Letting $\zeta \equiv (D_l r_s/ g(\chi) D_\chi)k$, we can write

\begin{align} \label{eq:philimit}
    \lim_{\zeta\rightarrow 0}\tilde \phi(\zeta\,;\,\tau) &= \int^{\infty}_0 4\pi\xi^2\,d \xi\,\lim_{\zeta\rightarrow 0}\left[\frac{\sin(\zeta\xi)}{\zeta\xi}\right]\,\phi(\xi\,;\,\tau) \nonumber \\
    &= \int^{\infty}_0 4\pi\xi^2\,d \xi\,\phi(\xi\,;\,\tau) 
    =1.
\end{align}
This means that in the $k\rightarrow0$ limit, Eq. \eqref{eq:limber_power_raw} gives
\begin{align}\label{eq:limber_plateau}
    P_{\rm I,0} \equiv P_{\rm I}(k\rightarrow0) &=\left(\frac{4\pi G}{c^2}\right)^2 D^2_l\int^{\chi_s}_0 d\chi\,\dfrac{W^2_\mathrm{I}(\chi)}{g^2(\chi)\chi^2} \nonumber \\
    &\times\int^{m_\mathrm{high}}_{m_\mathrm{low}} dm\,n(m,\chi)\,m^2,
\end{align}
where $W_\mathrm{I}(\chi)$ and $g(\chi)$ depend on $z_l$ and $z_s$. The intrinsic halo parameters completely drop out because $\int d^2\vec q \,\mathcal{P}(\vec q\,|\,m,\chi) = 1$.

From Eq. \eqref{eq:limber_plateau}, there are two limits where the interloper power spectrum (in Fig. \ref{fig:plateau_comp}) goes to zero. In one limit, $z_l \rightarrow 0$, the power goes to zero due to the factor of $D_l^2$. In the other limit, $z_l \rightarrow z_s$, the power goes to zero because $f(\chi) \to 0$ for all $\chi$, and $W_I$ in the integrand contains a factor of $f(\chi)$.

For subhalos, we take the $k \to 0$ limit by applying Eq. \eqref{eq:philimit} to Eq. \eqref{eq:powersub}, which gives
\begin{equation}
    P_{\rm S,0} = \frac{1}{\Sigma_{\rm cr}^2}\int_{m_\mathrm{low}}^{m_\mathrm{high}}dm\, m^2 n_\mathrm{sub}(m).
\end{equation}

\section{Curl and Divergence Components}
\label{app:d}
\paragraph{\textbf{Helmholtz Decomposition in 2D:}}
We can express a 2D vector function $\vec \alpha(\vec x)$ in a volume $V$ from its divergence $\nabla\cdot \vec \alpha$, curl $\nabla \times \vec \alpha$, and its values on the boundary $\partial V$. We use the fact that $\frac{1}{2\pi}\ln |\vec x - \vec x'|$ is the Green's function for the Laplacian in 2D, i.e. we can write
\begin{equation}
    \delta_\mathrm{2D}(\vec x - \vec x') = \frac{1}{2\pi} \nabla^2 \ln |\vec x - \vec x'|,
\end{equation}
where $\delta_\mathrm{2D}$ is the 2D delta function and $\nabla^2$ is the Laplace operator that acts on $\vec{x}$. 

We have
\begin{align}
    \vec \alpha(\vec x) &= \int_V dV'\, \vec \alpha(\vec x') \delta(\vec x - \vec x') \\
    & = \frac{1}{2\pi}\nabla^2\int_V dV'\, \vec \alpha(\vec x') \ln |\vec x - \vec x'|.
\end{align}
Using the identities
\begin{align}
    \nabla^2 \vec q &= \nabla(\nabla \cdot \vec q) - \nabla \times (\nabla \times \vec q) \\
    \vec q \cdot \nabla \phi &= -\phi(\nabla \cdot \vec q) + \nabla\cdot(\phi\vec q)\\
    \vec q \times \nabla \phi&= \phi(\nabla \times \vec q) - \nabla \times (\phi \vec q),
\end{align}
we obtain
\begin{align}
    \vec \alpha(\vec x) &= \frac{1}{2\pi} \nabla\left(\nabla \cdot \int_V dV'\, \vec \alpha(\vec x')\ln |\vec x' - \vec x|\right) \nonumber \\
    &-\frac{1}{2\pi}\nabla\times\left(\nabla \times \int_V dV'\, \vec \alpha(\vec x')\ln |\vec x' - \vec x|\right) \\
    \vec \alpha(\vec x)&= -\frac{1}{2\pi} \nabla\left(\int_V dV'\, \vec \alpha(\vec x')\cdot \nabla ' \ln |\vec x' - \vec x|\right) \nonumber \\
    &-\frac{1}{2\pi}\nabla\times\left(\int_V dV'\, \vec \alpha(\vec x')\times \nabla'\ln |\vec x' - \vec x|\right) \\
    \vec \alpha(\vec x)&= \frac{1}{2\pi}\nabla\left(\int_V dV'\, \ln|\vec x -\vec x'|\nabla'\cdot \vec \alpha(\vec x')\right) \nonumber \\
    &- \frac{1}{2\pi}\nabla\left(\int_V dV'\, \nabla'\cdot \left[\ln|\vec x -\vec x'|\vec \alpha(\vec x')\right]\right) \nonumber \\
    &+\frac{1}{2\pi}\nabla\times\left(\int_V dV'\, \ln|\vec x -\vec x'|\nabla'\times \vec \alpha(\vec x')\right) \nonumber \\
    &- \frac{1}{2\pi}\nabla\times\left(\int_V dV'\, \nabla'\times \left[\ln|\vec x -\vec x'|\vec \alpha(\vec x')\right]\right),
\end{align}
where $\nabla'$ acts on $\vec x'$. 
Now we use the divergence theorem to write
\begin{align}
    \vec \alpha(\vec x)&= \frac{1}{2\pi}\nabla\left(\int_V dV'\, \ln|\vec x -\vec x'|\nabla'\cdot \vec \alpha(\vec x')\right) \nonumber \\
    &- \frac{1}{2\pi}\nabla\left(\oint_{\partial V} dS'\, \hat n'\cdot \left[\ln|\vec x -\vec x'|\vec \alpha(\vec x')\right]\right) \nonumber \\
    &-\frac{1}{2\pi}\nabla\times\left(\int_V dV'\, \ln|\vec x -\vec x'|\nabla'\times \vec \alpha(\vec x')\right) \nonumber \\
    &+\frac{1}{2\pi}\nabla\times\left(\oint_{\partial V} dS'\, \hat n'\times \left[\ln|\vec x -\vec x'|\vec \alpha(\vec x')\right]\right),   
\end{align}
where $\hat n'$ is the unit vector normal to the boundary $\partial V$. If $\vec \alpha$ vanishes faster than $1/|\vec x'|\ln |\vec x'| $, the boundary terms vanish as we make $V$ infinitely large. This allows us to write
\begin{align}
    \vec \alpha(\vec x)&= \frac{1}{2\pi}\nabla\left(\int_V dV'\, \ln|\vec x -\vec x'|\nabla'\cdot \vec \alpha(\vec x')\right) \nonumber \\
    &-\frac{1}{2\pi}\nabla\times\left(\int_V dV'\, \ln|\vec x -\vec x'|\nabla'\times \vec \alpha(\vec x')\right) \\
    &=\frac{1}{\pi}\int_V dV'\,\left\{\nabla \cdot \ln|\vec x -\vec x'|\right\}\left[\frac{1}{2}\nabla'\cdot \vec \alpha(\vec x')\right] \nonumber \\
    &-\frac{1}{\pi}\int_V dV'\,\left\{\nabla \times \ln|\vec x -\vec x'|\right\}\left[\frac{1}{2}\nabla'\times \vec \alpha(\vec x')\right] \\
    &=\frac{1}{\pi}\int_V dV'\,\left\{\frac{\vec x - \vec x'}{|\vec x  - \vec x'|^2}\right\}\underbrace{\left[\frac{1}{2}\nabla'\cdot \vec \alpha(\vec x')\right]}_{\textstyle \kappa_\mathrm{eff}} \nonumber \\
    &-\frac{1}{\pi}\int_V dV'\,\left\{\hat x_3\times \frac{\vec x - \vec x'}{|\vec x - \vec x'|^2}\right\}\underbrace{\left[\frac{1}{2}\nabla'\times \vec \alpha(\vec x')\right]}_{\textstyle \kappa_\mathrm{curl}},
\end{align}
where $\hat x_3 \equiv \hat x_1 \times \hat x_2$ is the unit vector that is perpendicular to the 2D plane on which $\vec \alpha$ lives. $\hat x_1$ and $\hat x_2$ are the unit vectors of the 2D plane.
\newline
\paragraph{\textbf{Five-Point Stencil:}}
For a function $f:\mathbb{R}\rightarrow \mathbb{R}$, the first derivative can be approximated by \cite{sauer2013numerical},
\begin{align}
    f'(x) =& \mathrm{5pt}[f](x\,;\,h) +\frac{h^4}{30}f^{(5)}(x) + \mathcal{O}(h^5),
\end{align}
where
\begin{align}
    \mathrm{5pt}[f](x\,;\,h) &\equiv \frac{2}{3}\left[f(x+h) - f(x-h)\right] \nonumber \\
    - &\frac{1}{12}\left[f(x+2h) - f(x-2h)\right].
\end{align}

The error scales with the fourth power of the discrete interval size  $h$ and the 5th derivative of the function. Defining $\vec x \equiv (x_1, x_2)$, $\alpha_1(x) \equiv \hat x_1 \cdot \vec \alpha(x,x_2)$, and $\alpha_2(x) \equiv \hat x_2 \cdot \vec \alpha(x_1,x)$, the divergence of $\vec \alpha$ can be calculated as
\begin{align}
    \nabla \cdot \vec \alpha \cong \mathrm{5pt}[\alpha_1](x_1\,;\,h) + \mathrm{5pt}[\alpha_2](x_2\,;\,h),
\end{align}
where the leading error term is
\begin{equation}
    \frac{h^4}{30}\left(\alpha^{(5)}_1(x_1)+\alpha^{(5)}_2(x_2)\right).
\end{equation}
\newline
\paragraph{\textbf{Numerical Artifacts:}}
As with the divergence, we calculate the curl using the five-point stencil. Defining $\vec x \equiv (x_1, x_2)$, $\alpha_1(x) \equiv \hat x_1 \cdot \vec \alpha(x_1,x)$, and $\alpha_2(x) \equiv \hat x_2 \cdot \vec \alpha(x,x_2)$, the curl of $\vec \alpha$ is calculated as
\begin{align}\label{eq:curl_numeric}
    \nabla \times \vec \alpha \cong \mathrm{5pt}[\alpha_1](x_2\,;\,h) - \mathrm{5pt}[\alpha_2](x_1\,;\,h),
\end{align}
where the leading error term is
\begin{equation}
    \frac{h^4}{30}\left(\alpha^{(5)}_1(x_2)-\alpha^{(5)}_2(x_1)\right).
\end{equation}

\begin{figure}[ht]
\centering
\includegraphics[width=\linewidth]{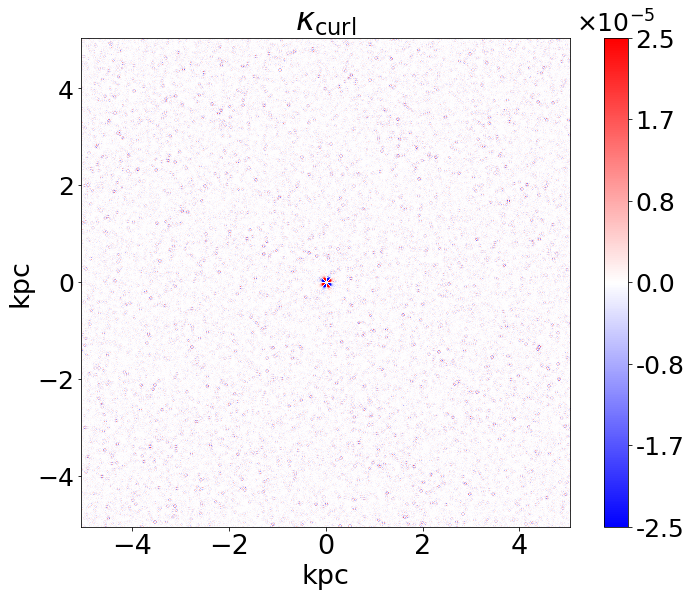}
\caption{\footnotesize{$\kappa_\mathrm{curl}$ of a simulated lensing system described in \S \ref{lensing_sim} without the interlopers. Without interlopers there is nothing that can source a curl component. This figure shows the numerical error in calculating the curl using discrete pixel values.}}
\label{fig:numeric_curl}
\end{figure}

To study how this numerical effect appears, we simulate a lensing system with a main lens and subhalos (i.e. no interlopers). From Eq. \eqref{eq:curlless}, we know that we should have $\kappa_\mathrm{curl} = 0$. Therefore, any non-zero value we get after calculating the curl using Eq. \eqref{eq:curl_numeric} will be a numerical artifact. In Fig. \ref{fig:numeric_curl}, we see that this numerical effect is only present at the centers of subhalos, as well as the center of the main lens, where the central cusp has a large 5th derivative, which increases the error. Nevertheless, it is more that 2 orders of magnitude smaller than the curl that we calculate in Fig. \ref{fig:compare.png}.

\counterwithin{figure}{section}

\newpage
\bibliography{biblio}

\end{document}

%% file: preamble.tex
\usepackage{amsthm}
\usepackage{mathtools}
\usepackage{physics}
\usepackage{xcolor}
\usepackage{graphicx}
\usepackage[left=23mm,right=13mm,top=35mm,columnsep=15pt]{geometry} 
\usepackage{adjustbox}
\usepackage{placeins}
\usepackage[T1]{fontenc}
\usepackage{lipsum}
\usepackage{csquotes}

%% file: main.bbl
\begin{thebibliography}{92}%
\makeatletter
\providecommand \@ifxundefined [1]{%
 \@ifx{#1\undefined}
}%
\providecommand \@ifnum [1]{%
 \ifnum #1\expandafter \@firstoftwo
 \else \expandafter \@secondoftwo
 \fi
}%
\providecommand \@ifx [1]{%
 \ifx #1\expandafter \@firstoftwo
 \else \expandafter \@secondoftwo
 \fi
}%
\providecommand \natexlab [1]{#1}%
\providecommand \enquote  [1]{``#1''}%
\providecommand \bibnamefont  [1]{#1}%
\providecommand \bibfnamefont [1]{#1}%
\providecommand \citenamefont [1]{#1}%
\providecommand \href@noop [0]{\@secondoftwo}%
\providecommand \href [0]{\begingroup \@sanitize@url \@href}%
\providecommand \@href[1]{\@@startlink{#1}\@@href}%
\providecommand \@@href[1]{\endgroup#1\@@endlink}%
\providecommand \@sanitize@url [0]{\catcode `\\12\catcode `\$12\catcode
  `\&12\catcode `\#12\catcode `\^12\catcode `\_12\catcode `\%12\relax}%
\providecommand \@@startlink[1]{}%
\providecommand \@@endlink[0]{}%
\providecommand \url  [0]{\begingroup\@sanitize@url \@url }%
\providecommand \@url [1]{\endgroup\@href {#1}{\urlprefix }}%
\providecommand \urlprefix  [0]{URL }%
\providecommand \Eprint [0]{\href }%
\providecommand \doibase [0]{http://dx.doi.org/}%
\providecommand \selectlanguage [0]{\@gobble}%
\providecommand \bibinfo  [0]{\@secondoftwo}%
\providecommand \bibfield  [0]{\@secondoftwo}%
\providecommand \translation [1]{[#1]}%
\providecommand \BibitemOpen [0]{}%
\providecommand \bibitemStop [0]{}%
\providecommand \bibitemNoStop [0]{.\EOS\space}%
\providecommand \EOS [0]{\spacefactor3000\relax}%
\providecommand \BibitemShut  [1]{\csname bibitem#1\endcsname}%
\let\auto@bib@innerbib\@empty
\bibitem [{\citenamefont {{Davis}}\ \emph {et~al.}(1981)\citenamefont
  {{Davis}}, \citenamefont {{Lecar}}, \citenamefont {{Pryor}},\ and\
  \citenamefont {{Witten}}}]{1981ApJ...250..423D}%
  \BibitemOpen
  \bibfield  {author} {\bibinfo {author} {\bibfnamefont {M.}~\bibnamefont
  {{Davis}}}, \bibinfo {author} {\bibfnamefont {M.}~\bibnamefont {{Lecar}}},
  \bibinfo {author} {\bibfnamefont {C.}~\bibnamefont {{Pryor}}}, \ and\
  \bibinfo {author} {\bibfnamefont {E.}~\bibnamefont {{Witten}}},\ }\href
  {\doibase 10.1086/159390} {\bibfield  {journal} {\bibinfo  {journal} {\apj}\
  }\textbf {\bibinfo {volume} {250}},\ \bibinfo {pages} {423} (\bibinfo {year}
  {1981})}\BibitemShut {NoStop}%
\bibitem [{\citenamefont {Blumenthal}\ \emph {et~al.}(1982)\citenamefont
  {Blumenthal}, \citenamefont {Pagels},\ and\ \citenamefont
  {Primack}}]{Blumenthal:1982mv}%
  \BibitemOpen
  \bibfield  {author} {\bibinfo {author} {\bibfnamefont {G.~R.}\ \bibnamefont
  {Blumenthal}}, \bibinfo {author} {\bibfnamefont {H.}~\bibnamefont {Pagels}},
  \ and\ \bibinfo {author} {\bibfnamefont {J.~R.}\ \bibnamefont {Primack}},\
  }\href {\doibase 10.1038/299037a0} {\bibfield  {journal} {\bibinfo  {journal}
  {Nature}\ }\textbf {\bibinfo {volume} {299}},\ \bibinfo {pages} {37}
  (\bibinfo {year} {1982})}\BibitemShut {NoStop}%
\bibitem [{\citenamefont {Blumenthal}\ \emph {et~al.}(1984)\citenamefont
  {Blumenthal}, \citenamefont {Faber}, \citenamefont {Primack},\ and\
  \citenamefont {Rees}}]{Blumenthal:1984bp}%
  \BibitemOpen
  \bibfield  {author} {\bibinfo {author} {\bibfnamefont {G.~R.}\ \bibnamefont
  {Blumenthal}}, \bibinfo {author} {\bibfnamefont {S.}~\bibnamefont {Faber}},
  \bibinfo {author} {\bibfnamefont {J.~R.}\ \bibnamefont {Primack}}, \ and\
  \bibinfo {author} {\bibfnamefont {M.~J.}\ \bibnamefont {Rees}},\ }\href
  {\doibase 10.1038/311517a0} {\bibfield  {journal} {\bibinfo  {journal}
  {Nature}\ }\textbf {\bibinfo {volume} {311}},\ \bibinfo {pages} {517}
  (\bibinfo {year} {1984})}\BibitemShut {NoStop}%
\bibitem [{\citenamefont {Davis}\ \emph {et~al.}(1985)\citenamefont {Davis},
  \citenamefont {Efstathiou}, \citenamefont {Frenk},\ and\ \citenamefont
  {White}}]{Davis:1985rj}%
  \BibitemOpen
  \bibfield  {author} {\bibinfo {author} {\bibfnamefont {M.}~\bibnamefont
  {Davis}}, \bibinfo {author} {\bibfnamefont {G.}~\bibnamefont {Efstathiou}},
  \bibinfo {author} {\bibfnamefont {C.~S.}\ \bibnamefont {Frenk}}, \ and\
  \bibinfo {author} {\bibfnamefont {S.~D.}\ \bibnamefont {White}},\ }\href
  {\doibase 10.1086/163168} {\bibfield  {journal} {\bibinfo  {journal} {\apj}\
  }\textbf {\bibinfo {volume} {292}},\ \bibinfo {pages} {371} (\bibinfo {year}
  {1985})}\BibitemShut {NoStop}%
\bibitem [{\citenamefont {Alam}\ \emph {et~al.}(2017)\citenamefont {Alam} \emph
  {et~al.}}]{BOSS}%
  \BibitemOpen
  \bibfield  {author} {\bibinfo {author} {\bibfnamefont {S.}~\bibnamefont
  {Alam}} \emph {et~al.} (\bibinfo {collaboration} {BOSS}),\ }\href {\doibase
  10.1093/mnras/stx721} {\bibfield  {journal} {\bibinfo  {journal} {Mon. Not.
  Roy. Astron. Soc.}\ }\textbf {\bibinfo {volume} {470}},\ \bibinfo {pages}
  {2617} (\bibinfo {year} {2017})},\ \Eprint {http://arxiv.org/abs/1607.03155}
  {arXiv:1607.03155 [astro-ph.CO]} \BibitemShut {NoStop}%
\bibitem [{\citenamefont {Scolnic}\ \emph {et~al.}(2018)\citenamefont {Scolnic}
  \emph {et~al.}}]{Pantheon}%
  \BibitemOpen
  \bibfield  {author} {\bibinfo {author} {\bibfnamefont {D.~M.}\ \bibnamefont
  {Scolnic}} \emph {et~al.},\ }\href {\doibase 10.3847/1538-4357/aab9bb}
  {\bibfield  {journal} {\bibinfo  {journal} {Astrophys. J.}\ }\textbf
  {\bibinfo {volume} {859}},\ \bibinfo {pages} {101} (\bibinfo {year}
  {2018})},\ \Eprint {http://arxiv.org/abs/1710.00845} {arXiv:1710.00845
  [astro-ph.CO]} \BibitemShut {NoStop}%
\bibitem [{\citenamefont {Aghanim}\ \emph {et~al.}(2018)\citenamefont {Aghanim}
  \emph {et~al.}}]{Planck2018}%
  \BibitemOpen
  \bibfield  {author} {\bibinfo {author} {\bibfnamefont {N.}~\bibnamefont
  {Aghanim}} \emph {et~al.} (\bibinfo {collaboration} {Planck}),\ }\href@noop
  {} {\  (\bibinfo {year} {2018})},\ \Eprint {http://arxiv.org/abs/1807.06209}
  {arXiv:1807.06209 [astro-ph.CO]} \BibitemShut {NoStop}%
\bibitem [{\citenamefont {{Brooks}}\ and\ \citenamefont
  {{Zolotov}}(2014)}]{Brooks_Zolotov}%
  \BibitemOpen
  \bibfield  {author} {\bibinfo {author} {\bibfnamefont {A.~M.}\ \bibnamefont
  {{Brooks}}}\ and\ \bibinfo {author} {\bibfnamefont {A.}~\bibnamefont
  {{Zolotov}}},\ }\href {\doibase 10.1088/0004-637X/786/2/87} {\bibfield
  {journal} {\bibinfo  {journal} {apj}\ }\textbf {\bibinfo {volume} {786}},\
  \bibinfo {eid} {87} (\bibinfo {year} {2014})},\ \Eprint
  {http://arxiv.org/abs/1207.2468} {arXiv:1207.2468} \BibitemShut {NoStop}%
\bibitem [{\citenamefont {{Brooks}}\ \emph {et~al.}(2013)\citenamefont
  {{Brooks}}, \citenamefont {{Kuhlen}}, \citenamefont {{Zolotov}},\ and\
  \citenamefont {{Hooper}}}]{Brooks_2013}%
  \BibitemOpen
  \bibfield  {author} {\bibinfo {author} {\bibfnamefont {A.~M.}\ \bibnamefont
  {{Brooks}}}, \bibinfo {author} {\bibfnamefont {M.}~\bibnamefont {{Kuhlen}}},
  \bibinfo {author} {\bibfnamefont {A.}~\bibnamefont {{Zolotov}}}, \ and\
  \bibinfo {author} {\bibfnamefont {D.}~\bibnamefont {{Hooper}}},\ }\href
  {\doibase 10.1088/0004-637X/765/1/22} {\bibfield  {journal} {\bibinfo
  {journal} {\apj}\ }\textbf {\bibinfo {volume} {765}},\ \bibinfo {eid} {22}
  (\bibinfo {year} {2013})},\ \Eprint {http://arxiv.org/abs/1209.5394}
  {arXiv:1209.5394} \BibitemShut {NoStop}%
\bibitem [{\citenamefont {{Arraki}}\ \emph {et~al.}(2014)\citenamefont
  {{Arraki}}, \citenamefont {{Klypin}}, \citenamefont {{More}},\ and\
  \citenamefont {{Trujillo-Gomez}}}]{Arraki}%
  \BibitemOpen
  \bibfield  {author} {\bibinfo {author} {\bibfnamefont {K.~S.}\ \bibnamefont
  {{Arraki}}}, \bibinfo {author} {\bibfnamefont {A.}~\bibnamefont {{Klypin}}},
  \bibinfo {author} {\bibfnamefont {S.}~\bibnamefont {{More}}}, \ and\ \bibinfo
  {author} {\bibfnamefont {S.}~\bibnamefont {{Trujillo-Gomez}}},\ }\href
  {\doibase 10.1093/mnras/stt2279} {\bibfield  {journal} {\bibinfo  {journal}
  {\mnras}\ }\textbf {\bibinfo {volume} {438}},\ \bibinfo {pages} {1466}
  (\bibinfo {year} {2014})},\ \Eprint {http://arxiv.org/abs/1212.6651}
  {arXiv:1212.6651} \BibitemShut {NoStop}%
\bibitem [{\citenamefont {O{\~n}orbe}\ \emph {et~al.}(2015)\citenamefont
  {O{\~n}orbe}, \citenamefont {Boylan-Kolchin}, \citenamefont {Bullock},
  \citenamefont {Hopkins}, \citenamefont {Ker{\v e}s}, \citenamefont
  {Faucher-Gigu{\`e}re}, \citenamefont {Quataert},\ and\ \citenamefont
  {Murray}}]{Onorbe_2015}%
  \BibitemOpen
  \bibfield  {author} {\bibinfo {author} {\bibfnamefont {J.}~\bibnamefont
  {O{\~n}orbe}}, \bibinfo {author} {\bibfnamefont {M.}~\bibnamefont
  {Boylan-Kolchin}}, \bibinfo {author} {\bibfnamefont {J.~S.}\ \bibnamefont
  {Bullock}}, \bibinfo {author} {\bibfnamefont {P.~F.}\ \bibnamefont
  {Hopkins}}, \bibinfo {author} {\bibfnamefont {D.}~\bibnamefont {Ker{\v e}s}},
  \bibinfo {author} {\bibfnamefont {C.-A.}\ \bibnamefont
  {Faucher-Gigu{\`e}re}}, \bibinfo {author} {\bibfnamefont {E.}~\bibnamefont
  {Quataert}}, \ and\ \bibinfo {author} {\bibfnamefont {N.}~\bibnamefont
  {Murray}},\ }\href {\doibase 10.1093/mnras/stv2072} {\bibfield  {journal}
  {\bibinfo  {journal} {Mon. Not. Roy. Astron. Soc.}\ }\textbf {\bibinfo
  {volume} {454}},\ \bibinfo {pages} {2092} (\bibinfo {year} {2015})},\ \Eprint
  {http://arxiv.org/abs/1502.02036} {arXiv:1502.02036 [astro-ph.GA]}
  \BibitemShut {NoStop}%
\bibitem [{\citenamefont {{Wetzel}}\ \emph {et~al.}(2016)\citenamefont
  {{Wetzel}}, \citenamefont {{Hopkins}}, \citenamefont {{Kim}}, \citenamefont
  {{Faucher-Gigu{\`e}re}}, \citenamefont {{Kere{\v s}}},\ and\ \citenamefont
  {{Quataert}}}]{Wetzel_2016}%
  \BibitemOpen
  \bibfield  {author} {\bibinfo {author} {\bibfnamefont {A.~R.}\ \bibnamefont
  {{Wetzel}}}, \bibinfo {author} {\bibfnamefont {P.~F.}\ \bibnamefont
  {{Hopkins}}}, \bibinfo {author} {\bibfnamefont {J.-h.}\ \bibnamefont
  {{Kim}}}, \bibinfo {author} {\bibfnamefont {C.-A.}\ \bibnamefont
  {{Faucher-Gigu{\`e}re}}}, \bibinfo {author} {\bibfnamefont {D.}~\bibnamefont
  {{Kere{\v s}}}}, \ and\ \bibinfo {author} {\bibfnamefont {E.}~\bibnamefont
  {{Quataert}}},\ }\href {\doibase 10.3847/2041-8205/827/2/L23} {\bibfield
  {journal} {\bibinfo  {journal} {\apjl}\ }\textbf {\bibinfo {volume} {827}},\
  \bibinfo {eid} {L23} (\bibinfo {year} {2016})},\ \Eprint
  {http://arxiv.org/abs/1602.05957} {arXiv:1602.05957} \BibitemShut {NoStop}%
\bibitem [{\citenamefont {{Sawala}}\ \emph {et~al.}(2016)\citenamefont
  {{Sawala}}, \citenamefont {{Frenk}}, \citenamefont {{Fattahi}}, \citenamefont
  {{Navarro}}, \citenamefont {{Bower}}, \citenamefont {{Crain}}, \citenamefont
  {{Dalla Vecchia}}, \citenamefont {{Furlong}}, \citenamefont {{Helly}},
  \citenamefont {{Jenkins}}, \citenamefont {{Oman}}, \citenamefont
  {{Schaller}}, \citenamefont {{Schaye}}, \citenamefont {{Theuns}},
  \citenamefont {{Trayford}},\ and\ \citenamefont
  {{White}}}]{2016MNRAS.457.1931S}%
  \BibitemOpen
  \bibfield  {author} {\bibinfo {author} {\bibfnamefont {T.}~\bibnamefont
  {{Sawala}}}, \bibinfo {author} {\bibfnamefont {C.~S.}\ \bibnamefont
  {{Frenk}}}, \bibinfo {author} {\bibfnamefont {A.}~\bibnamefont {{Fattahi}}},
  \bibinfo {author} {\bibfnamefont {J.~F.}\ \bibnamefont {{Navarro}}}, \bibinfo
  {author} {\bibfnamefont {R.~G.}\ \bibnamefont {{Bower}}}, \bibinfo {author}
  {\bibfnamefont {R.~A.}\ \bibnamefont {{Crain}}}, \bibinfo {author}
  {\bibfnamefont {C.}~\bibnamefont {{Dalla Vecchia}}}, \bibinfo {author}
  {\bibfnamefont {M.}~\bibnamefont {{Furlong}}}, \bibinfo {author}
  {\bibfnamefont {J.~C.}\ \bibnamefont {{Helly}}}, \bibinfo {author}
  {\bibfnamefont {A.}~\bibnamefont {{Jenkins}}}, \bibinfo {author}
  {\bibfnamefont {K.~A.}\ \bibnamefont {{Oman}}}, \bibinfo {author}
  {\bibfnamefont {M.}~\bibnamefont {{Schaller}}}, \bibinfo {author}
  {\bibfnamefont {J.}~\bibnamefont {{Schaye}}}, \bibinfo {author}
  {\bibfnamefont {T.}~\bibnamefont {{Theuns}}}, \bibinfo {author}
  {\bibfnamefont {J.}~\bibnamefont {{Trayford}}}, \ and\ \bibinfo {author}
  {\bibfnamefont {S.~D.~M.}\ \bibnamefont {{White}}},\ }\href {\doibase
  10.1093/mnras/stw145} {\bibfield  {journal} {\bibinfo  {journal} {\mnras}\
  }\textbf {\bibinfo {volume} {457}},\ \bibinfo {pages} {1931} (\bibinfo {year}
  {2016})},\ \Eprint {http://arxiv.org/abs/1511.01098} {arXiv:1511.01098}
  \BibitemShut {NoStop}%
\bibitem [{\citenamefont {{Creasey}}\ \emph {et~al.}(2017)\citenamefont
  {{Creasey}}, \citenamefont {{Sameie}}, \citenamefont {{Sales}}, \citenamefont
  {{Yu}}, \citenamefont {{Vogelsberger}},\ and\ \citenamefont
  {{Zavala}}}]{2017MNRAS.468.2283C}%
  \BibitemOpen
  \bibfield  {author} {\bibinfo {author} {\bibfnamefont {P.}~\bibnamefont
  {{Creasey}}}, \bibinfo {author} {\bibfnamefont {O.}~\bibnamefont {{Sameie}}},
  \bibinfo {author} {\bibfnamefont {L.~V.}\ \bibnamefont {{Sales}}}, \bibinfo
  {author} {\bibfnamefont {H.-B.}\ \bibnamefont {{Yu}}}, \bibinfo {author}
  {\bibfnamefont {M.}~\bibnamefont {{Vogelsberger}}}, \ and\ \bibinfo {author}
  {\bibfnamefont {J.}~\bibnamefont {{Zavala}}},\ }\href {\doibase
  10.1093/mnras/stx522} {\bibfield  {journal} {\bibinfo  {journal} {\mnras}\
  }\textbf {\bibinfo {volume} {468}},\ \bibinfo {pages} {2283} (\bibinfo {year}
  {2017})},\ \Eprint {http://arxiv.org/abs/1612.03903} {arXiv:1612.03903}
  \BibitemShut {NoStop}%
\bibitem [{\citenamefont {{Sawala}}\ \emph {et~al.}(2017)\citenamefont
  {{Sawala}}, \citenamefont {{Pihajoki}}, \citenamefont {{Johansson}},
  \citenamefont {{Frenk}}, \citenamefont {{Navarro}}, \citenamefont {{Oman}},\
  and\ \citenamefont {{White}}}]{2017MNRAS.467.4383S}%
  \BibitemOpen
  \bibfield  {author} {\bibinfo {author} {\bibfnamefont {T.}~\bibnamefont
  {{Sawala}}}, \bibinfo {author} {\bibfnamefont {P.}~\bibnamefont
  {{Pihajoki}}}, \bibinfo {author} {\bibfnamefont {P.~H.}\ \bibnamefont
  {{Johansson}}}, \bibinfo {author} {\bibfnamefont {C.~S.}\ \bibnamefont
  {{Frenk}}}, \bibinfo {author} {\bibfnamefont {J.~F.}\ \bibnamefont
  {{Navarro}}}, \bibinfo {author} {\bibfnamefont {K.~A.}\ \bibnamefont
  {{Oman}}}, \ and\ \bibinfo {author} {\bibfnamefont {S.~D.~M.}\ \bibnamefont
  {{White}}},\ }\href {\doibase 10.1093/mnras/stx360} {\bibfield  {journal}
  {\bibinfo  {journal} {\mnras}\ }\textbf {\bibinfo {volume} {467}},\ \bibinfo
  {pages} {4383} (\bibinfo {year} {2017})},\ \Eprint
  {http://arxiv.org/abs/1609.01718} {arXiv:1609.01718} \BibitemShut {NoStop}%
\bibitem [{\citenamefont {{Garrison-Kimmel}}\ \emph {et~al.}(2017)\citenamefont
  {{Garrison-Kimmel}}, \citenamefont {{Wetzel}}, \citenamefont {{Bullock}},
  \citenamefont {{Hopkins}}, \citenamefont {{Boylan-Kolchin}}, \citenamefont
  {{Faucher-Giguere}}, \citenamefont {{Keres}}, \citenamefont {{Quataert}},
  \citenamefont {{Sanderson}}, \citenamefont {{Graus}},\ and\ \citenamefont
  {{Kelley}}}]{2017arXiv170103792G}%
  \BibitemOpen
  \bibfield  {author} {\bibinfo {author} {\bibfnamefont {S.}~\bibnamefont
  {{Garrison-Kimmel}}}, \bibinfo {author} {\bibfnamefont {A.~R.}\ \bibnamefont
  {{Wetzel}}}, \bibinfo {author} {\bibfnamefont {J.~S.}\ \bibnamefont
  {{Bullock}}}, \bibinfo {author} {\bibfnamefont {P.~F.}\ \bibnamefont
  {{Hopkins}}}, \bibinfo {author} {\bibfnamefont {M.}~\bibnamefont
  {{Boylan-Kolchin}}}, \bibinfo {author} {\bibfnamefont {C.-A.}\ \bibnamefont
  {{Faucher-Giguere}}}, \bibinfo {author} {\bibfnamefont {D.}~\bibnamefont
  {{Keres}}}, \bibinfo {author} {\bibfnamefont {E.}~\bibnamefont {{Quataert}}},
  \bibinfo {author} {\bibfnamefont {R.~E.}\ \bibnamefont {{Sanderson}}},
  \bibinfo {author} {\bibfnamefont {A.~S.}\ \bibnamefont {{Graus}}}, \ and\
  \bibinfo {author} {\bibfnamefont {T.}~\bibnamefont {{Kelley}}},\ }\href@noop
  {} {\bibfield  {journal} {\bibinfo  {journal} {ArXiv e-prints}\ } (\bibinfo
  {year} {2017})},\ \Eprint {http://arxiv.org/abs/1701.03792}
  {arXiv:1701.03792} \BibitemShut {NoStop}%
\bibitem [{\citenamefont {Vogelsberger}\ \emph {et~al.}(2020)\citenamefont
  {Vogelsberger}, \citenamefont {Marinacci}, \citenamefont {Torrey},\ and\
  \citenamefont {Puchwein}}]{Vogelsberger:2019ynw}%
  \BibitemOpen
  \bibfield  {author} {\bibinfo {author} {\bibfnamefont {M.}~\bibnamefont
  {Vogelsberger}}, \bibinfo {author} {\bibfnamefont {F.}~\bibnamefont
  {Marinacci}}, \bibinfo {author} {\bibfnamefont {P.}~\bibnamefont {Torrey}}, \
  and\ \bibinfo {author} {\bibfnamefont {E.}~\bibnamefont {Puchwein}},\ }\href
  {\doibase 10.1038/s42254-019-0127-2} {\bibfield  {journal} {\bibinfo
  {journal} {Nature Rev. Phys.}\ }\textbf {\bibinfo {volume} {2}},\ \bibinfo
  {pages} {42} (\bibinfo {year} {2020})},\ \Eprint
  {http://arxiv.org/abs/1909.07976} {arXiv:1909.07976 [astro-ph.GA]}
  \BibitemShut {NoStop}%
\bibitem [{\citenamefont {Fitts}\ \emph {et~al.}(2016)\citenamefont {Fitts}
  \emph {et~al.}}]{Fitts:2016usl}%
  \BibitemOpen
  \bibfield  {author} {\bibinfo {author} {\bibfnamefont {A.}~\bibnamefont
  {Fitts}} \emph {et~al.},\ }\href@noop {} {\  (\bibinfo {year} {2016})},\
  \Eprint {http://arxiv.org/abs/1611.02281} {arXiv:1611.02281 [astro-ph.GA]}
  \BibitemShut {NoStop}%
\bibitem [{\citenamefont {{Read}}\ \emph {et~al.}(2017)\citenamefont {{Read}},
  \citenamefont {{Iorio}}, \citenamefont {{Agertz}},\ and\ \citenamefont
  {{Fraternali}}}]{2017MNRAS.467.2019R}%
  \BibitemOpen
  \bibfield  {author} {\bibinfo {author} {\bibfnamefont {J.~I.}\ \bibnamefont
  {{Read}}}, \bibinfo {author} {\bibfnamefont {G.}~\bibnamefont {{Iorio}}},
  \bibinfo {author} {\bibfnamefont {O.}~\bibnamefont {{Agertz}}}, \ and\
  \bibinfo {author} {\bibfnamefont {F.}~\bibnamefont {{Fraternali}}},\ }\href
  {\doibase 10.1093/mnras/stx147} {\bibfield  {journal} {\bibinfo  {journal}
  {\mnras}\ }\textbf {\bibinfo {volume} {467}},\ \bibinfo {pages} {2019}
  (\bibinfo {year} {2017})},\ \Eprint {http://arxiv.org/abs/1607.03127}
  {arXiv:1607.03127} \BibitemShut {NoStop}%
\bibitem [{\citenamefont {{Bechtol}}\ \emph {et~al.}(2015)\citenamefont
  {{Bechtol}}, \citenamefont {{Drlica-Wagner}}, \citenamefont {{Balbinot}},
  \citenamefont {{Pieres}}, \citenamefont {{Simon}}, \citenamefont {{Yanny}},
  \citenamefont {{Santiago}}, \citenamefont {{Wechsler}}, \citenamefont
  {{Frieman}}, \citenamefont {{Walker}}, \citenamefont {{Williams}},
  \citenamefont {{Rozo}}, \citenamefont {{Rykoff}}, \citenamefont {{Queiroz}},
  \citenamefont {{Luque}}, \citenamefont {{Benoit-L{\'e}vy}}, \citenamefont
  {{Tucker}}, \citenamefont {{Sevilla}}, \citenamefont {{Gruendl}},
  \citenamefont {{da Costa}}, \citenamefont {{Fausti Neto}}, \citenamefont
  {{Maia}}, \citenamefont {{Abbott}}, \citenamefont {{Allam}}, \citenamefont
  {{Armstrong}}, \citenamefont {{Bauer}}, \citenamefont {{Bernstein}},
  \citenamefont {{Bernstein}}, \citenamefont {{Bertin}}, \citenamefont
  {{Brooks}}, \citenamefont {{Buckley-Geer}}, \citenamefont {{Burke}},
  \citenamefont {{Carnero Rosell}}, \citenamefont {{Castander}}, \citenamefont
  {{Covarrubias}}, \citenamefont {{D'Andrea}}, \citenamefont {{DePoy}},
  \citenamefont {{Desai}}, \citenamefont {{Diehl}}, \citenamefont {{Eifler}},
  \citenamefont {{Estrada}}, \citenamefont {{Evrard}}, \citenamefont
  {{Fernandez}}, \citenamefont {{Finley}}, \citenamefont {{Flaugher}},
  \citenamefont {{Gaztanaga}}, \citenamefont {{Gerdes}}, \citenamefont
  {{Girardi}}, \citenamefont {{Gladders}}, \citenamefont {{Gruen}},
  \citenamefont {{Gutierrez}}, \citenamefont {{Hao}}, \citenamefont
  {{Honscheid}}, \citenamefont {{Jain}}, \citenamefont {{James}}, \citenamefont
  {{Kent}}, \citenamefont {{Kron}}, \citenamefont {{Kuehn}}, \citenamefont
  {{Kuropatkin}}, \citenamefont {{Lahav}}, \citenamefont {{Li}}, \citenamefont
  {{Lin}}, \citenamefont {{Makler}}, \citenamefont {{March}}, \citenamefont
  {{Marshall}}, \citenamefont {{Martini}}, \citenamefont {{Merritt}},
  \citenamefont {{Miller}}, \citenamefont {{Miquel}}, \citenamefont {{Mohr}},
  \citenamefont {{Neilsen}}, \citenamefont {{Nichol}}, \citenamefont {{Nord}},
  \citenamefont {{Ogando}}, \citenamefont {{Peoples}}, \citenamefont
  {{Petravick}}, \citenamefont {{Plazas}}, \citenamefont {{Romer}},
  \citenamefont {{Roodman}}, \citenamefont {{Sako}}, \citenamefont {{Sanchez}},
  \citenamefont {{Scarpine}}, \citenamefont {{Schubnell}}, \citenamefont
  {{Smith}}, \citenamefont {{Soares-Santos}}, \citenamefont {{Sobreira}},
  \citenamefont {{Suchyta}}, \citenamefont {{Swanson}}, \citenamefont
  {{Tarle}}, \citenamefont {{Thaler}}, \citenamefont {{Thomas}}, \citenamefont
  {{Wester}}, \citenamefont {{Zuntz}},\ and\ \citenamefont {{DES
  Collaboration}}}]{2015ApJ...807...50B}%
  \BibitemOpen
  \bibfield  {author} {\bibinfo {author} {\bibfnamefont {K.}~\bibnamefont
  {{Bechtol}}}, \bibinfo {author} {\bibfnamefont {A.}~\bibnamefont
  {{Drlica-Wagner}}}, \bibinfo {author} {\bibfnamefont {E.}~\bibnamefont
  {{Balbinot}}}, \bibinfo {author} {\bibfnamefont {A.}~\bibnamefont
  {{Pieres}}}, \bibinfo {author} {\bibfnamefont {J.~D.}\ \bibnamefont
  {{Simon}}}, \bibinfo {author} {\bibfnamefont {B.}~\bibnamefont {{Yanny}}},
  \bibinfo {author} {\bibfnamefont {B.}~\bibnamefont {{Santiago}}}, \bibinfo
  {author} {\bibfnamefont {R.~H.}\ \bibnamefont {{Wechsler}}}, \bibinfo
  {author} {\bibfnamefont {J.}~\bibnamefont {{Frieman}}}, \bibinfo {author}
  {\bibfnamefont {A.~R.}\ \bibnamefont {{Walker}}}, \bibinfo {author}
  {\bibfnamefont {P.}~\bibnamefont {{Williams}}}, \bibinfo {author}
  {\bibfnamefont {E.}~\bibnamefont {{Rozo}}}, \bibinfo {author} {\bibfnamefont
  {E.~S.}\ \bibnamefont {{Rykoff}}}, \bibinfo {author} {\bibfnamefont
  {A.}~\bibnamefont {{Queiroz}}}, \bibinfo {author} {\bibfnamefont
  {E.}~\bibnamefont {{Luque}}}, \bibinfo {author} {\bibfnamefont
  {A.}~\bibnamefont {{Benoit-L{\'e}vy}}}, \bibinfo {author} {\bibfnamefont
  {D.}~\bibnamefont {{Tucker}}}, \bibinfo {author} {\bibfnamefont
  {I.}~\bibnamefont {{Sevilla}}}, \bibinfo {author} {\bibfnamefont {R.~A.}\
  \bibnamefont {{Gruendl}}}, \bibinfo {author} {\bibfnamefont {L.~N.}\
  \bibnamefont {{da Costa}}}, \bibinfo {author} {\bibfnamefont
  {A.}~\bibnamefont {{Fausti Neto}}}, \bibinfo {author} {\bibfnamefont
  {M.~A.~G.}\ \bibnamefont {{Maia}}}, \bibinfo {author} {\bibfnamefont
  {T.}~\bibnamefont {{Abbott}}}, \bibinfo {author} {\bibfnamefont
  {S.}~\bibnamefont {{Allam}}}, \bibinfo {author} {\bibfnamefont
  {R.}~\bibnamefont {{Armstrong}}}, \bibinfo {author} {\bibfnamefont {A.~H.}\
  \bibnamefont {{Bauer}}}, \bibinfo {author} {\bibfnamefont {G.~M.}\
  \bibnamefont {{Bernstein}}}, \bibinfo {author} {\bibfnamefont {R.~A.}\
  \bibnamefont {{Bernstein}}}, \bibinfo {author} {\bibfnamefont
  {E.}~\bibnamefont {{Bertin}}}, \bibinfo {author} {\bibfnamefont
  {D.}~\bibnamefont {{Brooks}}}, \bibinfo {author} {\bibfnamefont
  {E.}~\bibnamefont {{Buckley-Geer}}}, \bibinfo {author} {\bibfnamefont
  {D.~L.}\ \bibnamefont {{Burke}}}, \bibinfo {author} {\bibfnamefont
  {A.}~\bibnamefont {{Carnero Rosell}}}, \bibinfo {author} {\bibfnamefont
  {F.~J.}\ \bibnamefont {{Castander}}}, \bibinfo {author} {\bibfnamefont
  {R.}~\bibnamefont {{Covarrubias}}}, \bibinfo {author} {\bibfnamefont {C.~B.}\
  \bibnamefont {{D'Andrea}}}, \bibinfo {author} {\bibfnamefont {D.~L.}\
  \bibnamefont {{DePoy}}}, \bibinfo {author} {\bibfnamefont {S.}~\bibnamefont
  {{Desai}}}, \bibinfo {author} {\bibfnamefont {H.~T.}\ \bibnamefont
  {{Diehl}}}, \bibinfo {author} {\bibfnamefont {T.~F.}\ \bibnamefont
  {{Eifler}}}, \bibinfo {author} {\bibfnamefont {J.}~\bibnamefont {{Estrada}}},
  \bibinfo {author} {\bibfnamefont {A.~E.}\ \bibnamefont {{Evrard}}}, \bibinfo
  {author} {\bibfnamefont {E.}~\bibnamefont {{Fernandez}}}, \bibinfo {author}
  {\bibfnamefont {D.~A.}\ \bibnamefont {{Finley}}}, \bibinfo {author}
  {\bibfnamefont {B.}~\bibnamefont {{Flaugher}}}, \bibinfo {author}
  {\bibfnamefont {E.}~\bibnamefont {{Gaztanaga}}}, \bibinfo {author}
  {\bibfnamefont {D.}~\bibnamefont {{Gerdes}}}, \bibinfo {author}
  {\bibfnamefont {L.}~\bibnamefont {{Girardi}}}, \bibinfo {author}
  {\bibfnamefont {M.}~\bibnamefont {{Gladders}}}, \bibinfo {author}
  {\bibfnamefont {D.}~\bibnamefont {{Gruen}}}, \bibinfo {author} {\bibfnamefont
  {G.}~\bibnamefont {{Gutierrez}}}, \bibinfo {author} {\bibfnamefont
  {J.}~\bibnamefont {{Hao}}}, \bibinfo {author} {\bibfnamefont
  {K.}~\bibnamefont {{Honscheid}}}, \bibinfo {author} {\bibfnamefont
  {B.}~\bibnamefont {{Jain}}}, \bibinfo {author} {\bibfnamefont
  {D.}~\bibnamefont {{James}}}, \bibinfo {author} {\bibfnamefont
  {S.}~\bibnamefont {{Kent}}}, \bibinfo {author} {\bibfnamefont
  {R.}~\bibnamefont {{Kron}}}, \bibinfo {author} {\bibfnamefont
  {K.}~\bibnamefont {{Kuehn}}}, \bibinfo {author} {\bibfnamefont
  {N.}~\bibnamefont {{Kuropatkin}}}, \bibinfo {author} {\bibfnamefont
  {O.}~\bibnamefont {{Lahav}}}, \bibinfo {author} {\bibfnamefont {T.~S.}\
  \bibnamefont {{Li}}}, \bibinfo {author} {\bibfnamefont {H.}~\bibnamefont
  {{Lin}}}, \bibinfo {author} {\bibfnamefont {M.}~\bibnamefont {{Makler}}},
  \bibinfo {author} {\bibfnamefont {M.}~\bibnamefont {{March}}}, \bibinfo
  {author} {\bibfnamefont {J.}~\bibnamefont {{Marshall}}}, \bibinfo {author}
  {\bibfnamefont {P.}~\bibnamefont {{Martini}}}, \bibinfo {author}
  {\bibfnamefont {K.~W.}\ \bibnamefont {{Merritt}}}, \bibinfo {author}
  {\bibfnamefont {C.}~\bibnamefont {{Miller}}}, \bibinfo {author}
  {\bibfnamefont {R.}~\bibnamefont {{Miquel}}}, \bibinfo {author}
  {\bibfnamefont {J.}~\bibnamefont {{Mohr}}}, \bibinfo {author} {\bibfnamefont
  {E.}~\bibnamefont {{Neilsen}}}, \bibinfo {author} {\bibfnamefont
  {R.}~\bibnamefont {{Nichol}}}, \bibinfo {author} {\bibfnamefont
  {B.}~\bibnamefont {{Nord}}}, \bibinfo {author} {\bibfnamefont
  {R.}~\bibnamefont {{Ogando}}}, \bibinfo {author} {\bibfnamefont
  {J.}~\bibnamefont {{Peoples}}}, \bibinfo {author} {\bibfnamefont
  {D.}~\bibnamefont {{Petravick}}}, \bibinfo {author} {\bibfnamefont {A.~A.}\
  \bibnamefont {{Plazas}}}, \bibinfo {author} {\bibfnamefont {A.~K.}\
  \bibnamefont {{Romer}}}, \bibinfo {author} {\bibfnamefont {A.}~\bibnamefont
  {{Roodman}}}, \bibinfo {author} {\bibfnamefont {M.}~\bibnamefont {{Sako}}},
  \bibinfo {author} {\bibfnamefont {E.}~\bibnamefont {{Sanchez}}}, \bibinfo
  {author} {\bibfnamefont {V.}~\bibnamefont {{Scarpine}}}, \bibinfo {author}
  {\bibfnamefont {M.}~\bibnamefont {{Schubnell}}}, \bibinfo {author}
  {\bibfnamefont {R.~C.}\ \bibnamefont {{Smith}}}, \bibinfo {author}
  {\bibfnamefont {M.}~\bibnamefont {{Soares-Santos}}}, \bibinfo {author}
  {\bibfnamefont {F.}~\bibnamefont {{Sobreira}}}, \bibinfo {author}
  {\bibfnamefont {E.}~\bibnamefont {{Suchyta}}}, \bibinfo {author}
  {\bibfnamefont {M.~E.~C.}\ \bibnamefont {{Swanson}}}, \bibinfo {author}
  {\bibfnamefont {G.}~\bibnamefont {{Tarle}}}, \bibinfo {author} {\bibfnamefont
  {J.}~\bibnamefont {{Thaler}}}, \bibinfo {author} {\bibfnamefont
  {D.}~\bibnamefont {{Thomas}}}, \bibinfo {author} {\bibfnamefont
  {W.}~\bibnamefont {{Wester}}}, \bibinfo {author} {\bibfnamefont
  {J.}~\bibnamefont {{Zuntz}}}, \ and\ \bibinfo {author} {\bibnamefont {{DES
  Collaboration}}},\ }\href {\doibase 10.1088/0004-637X/807/1/50} {\bibfield
  {journal} {\bibinfo  {journal} {\apj}\ }\textbf {\bibinfo {volume} {807}},\
  \bibinfo {eid} {50} (\bibinfo {year} {2015})},\ \Eprint
  {http://arxiv.org/abs/1503.02584} {arXiv:1503.02584 [astro-ph.GA]}
  \BibitemShut {NoStop}%
\bibitem [{\citenamefont {Koposov}\ \emph {et~al.}(2015)\citenamefont
  {Koposov}, \citenamefont {Belokurov}, \citenamefont {Torrealba},\ and\
  \citenamefont {Evans}}]{Koposov:2015cua}%
  \BibitemOpen
  \bibfield  {author} {\bibinfo {author} {\bibfnamefont {S.~E.}\ \bibnamefont
  {Koposov}}, \bibinfo {author} {\bibfnamefont {V.}~\bibnamefont {Belokurov}},
  \bibinfo {author} {\bibfnamefont {G.}~\bibnamefont {Torrealba}}, \ and\
  \bibinfo {author} {\bibfnamefont {N.~W.}\ \bibnamefont {Evans}},\ }\href
  {\doibase 10.1088/0004-637X/805/2/130} {\bibfield  {journal} {\bibinfo
  {journal} {Astrophys. J.}\ }\textbf {\bibinfo {volume} {805}},\ \bibinfo
  {pages} {130} (\bibinfo {year} {2015})},\ \Eprint
  {http://arxiv.org/abs/1503.02079} {arXiv:1503.02079 [astro-ph.GA]}
  \BibitemShut {NoStop}%
\bibitem [{\citenamefont {Drlica-Wagner}\ \emph {et~al.}(2015)\citenamefont
  {Drlica-Wagner} \emph {et~al.}}]{Drlica-Wagner:2015ufc}%
  \BibitemOpen
  \bibfield  {author} {\bibinfo {author} {\bibfnamefont {A.}~\bibnamefont
  {Drlica-Wagner}} \emph {et~al.} (\bibinfo {collaboration} {DES}),\ }\href
  {\doibase 10.1088/0004-637X/813/2/109} {\bibfield  {journal} {\bibinfo
  {journal} {Astrophys. J.}\ }\textbf {\bibinfo {volume} {813}},\ \bibinfo
  {pages} {109} (\bibinfo {year} {2015})},\ \Eprint
  {http://arxiv.org/abs/1508.03622} {arXiv:1508.03622 [astro-ph.GA]}
  \BibitemShut {NoStop}%
\bibitem [{\citenamefont {Homma}\ \emph {et~al.}(2017)\citenamefont {Homma},
  \citenamefont {Chiba}, \citenamefont {Okamoto}, \citenamefont {Komiyama},
  \citenamefont {Tanaka}, \citenamefont {Tanaka}, \citenamefont {Ishigaki},
  \citenamefont {Hayashi}, \citenamefont {Arimoto}, \citenamefont {Garmilla},\
  and\ \citenamefont {et~al.}}]{Homma_2017}%
  \BibitemOpen
  \bibfield  {author} {\bibinfo {author} {\bibfnamefont {D.}~\bibnamefont
  {Homma}}, \bibinfo {author} {\bibfnamefont {M.}~\bibnamefont {Chiba}},
  \bibinfo {author} {\bibfnamefont {S.}~\bibnamefont {Okamoto}}, \bibinfo
  {author} {\bibfnamefont {Y.}~\bibnamefont {Komiyama}}, \bibinfo {author}
  {\bibfnamefont {M.}~\bibnamefont {Tanaka}}, \bibinfo {author} {\bibfnamefont
  {M.}~\bibnamefont {Tanaka}}, \bibinfo {author} {\bibfnamefont {M.~N.}\
  \bibnamefont {Ishigaki}}, \bibinfo {author} {\bibfnamefont {K.}~\bibnamefont
  {Hayashi}}, \bibinfo {author} {\bibfnamefont {N.}~\bibnamefont {Arimoto}},
  \bibinfo {author} {\bibfnamefont {J.~A.}\ \bibnamefont {Garmilla}}, \ and\
  \bibinfo {author} {\bibnamefont {et~al.}},\ }\href {\doibase
  10.1093/pasj/psx050} {\bibfield  {journal} {\bibinfo  {journal} {Publications
  of the Astronomical Society of Japan}\ }\textbf {\bibinfo {volume} {70}}
  (\bibinfo {year} {2017}),\ 10.1093/pasj/psx050}\BibitemShut {NoStop}%
\bibitem [{\citenamefont {Mao}\ and\ \citenamefont {Schneider}(1998)}]{Mao}%
  \BibitemOpen
  \bibfield  {author} {\bibinfo {author} {\bibfnamefont {S.-d.}\ \bibnamefont
  {Mao}}\ and\ \bibinfo {author} {\bibfnamefont {P.}~\bibnamefont
  {Schneider}},\ }\href {\doibase 10.1046/j.1365-8711.1998.01319.x} {\bibfield
  {journal} {\bibinfo  {journal} {Mon. Not. Roy. Astron. Soc.}\ }\textbf
  {\bibinfo {volume} {295}},\ \bibinfo {pages} {587} (\bibinfo {year}
  {1998})},\ \Eprint {http://arxiv.org/abs/astro-ph/9707187}
  {arXiv:astro-ph/9707187 [astro-ph]} \BibitemShut {NoStop}%
\bibitem [{\citenamefont {Nierenberg}\ \emph {et~al.}(2017)\citenamefont
  {Nierenberg}, \citenamefont {Treu}, \citenamefont {Brammer}, \citenamefont
  {Peter}, \citenamefont {Fassnacht}, \citenamefont {Keeton}, \citenamefont
  {Kochanek}, \citenamefont {Schmidt}, \citenamefont {Sluse},\ and\
  \citenamefont {Wright}}]{Nierenberg:2017vlg}%
  \BibitemOpen
  \bibfield  {author} {\bibinfo {author} {\bibfnamefont {A.~M.}\ \bibnamefont
  {Nierenberg}}, \bibinfo {author} {\bibfnamefont {T.}~\bibnamefont {Treu}},
  \bibinfo {author} {\bibfnamefont {G.}~\bibnamefont {Brammer}}, \bibinfo
  {author} {\bibfnamefont {A.~H.~G.}\ \bibnamefont {Peter}}, \bibinfo {author}
  {\bibfnamefont {C.~D.}\ \bibnamefont {Fassnacht}}, \bibinfo {author}
  {\bibfnamefont {C.~R.}\ \bibnamefont {Keeton}}, \bibinfo {author}
  {\bibfnamefont {C.~S.}\ \bibnamefont {Kochanek}}, \bibinfo {author}
  {\bibfnamefont {K.~B.}\ \bibnamefont {Schmidt}}, \bibinfo {author}
  {\bibfnamefont {D.}~\bibnamefont {Sluse}}, \ and\ \bibinfo {author}
  {\bibfnamefont {S.~A.}\ \bibnamefont {Wright}},\ }\href {\doibase
  10.1093/mnras/stx1400} {\bibfield  {journal} {\bibinfo  {journal} {Mon. Not.
  Roy. Astron. Soc.}\ }\textbf {\bibinfo {volume} {471}},\ \bibinfo {pages}
  {2224} (\bibinfo {year} {2017})},\ \Eprint {http://arxiv.org/abs/1701.05188}
  {arXiv:1701.05188 [astro-ph.CO]} \BibitemShut {NoStop}%
\bibitem [{\citenamefont {Gilman}\ \emph {et~al.}(2018)\citenamefont {Gilman},
  \citenamefont {Birrer}, \citenamefont {Treu}, \citenamefont {Keeton},\ and\
  \citenamefont {Nierenberg}}]{Gilman:2017voy}%
  \BibitemOpen
  \bibfield  {author} {\bibinfo {author} {\bibfnamefont {D.}~\bibnamefont
  {Gilman}}, \bibinfo {author} {\bibfnamefont {S.}~\bibnamefont {Birrer}},
  \bibinfo {author} {\bibfnamefont {T.}~\bibnamefont {Treu}}, \bibinfo {author}
  {\bibfnamefont {C.~R.}\ \bibnamefont {Keeton}}, \ and\ \bibinfo {author}
  {\bibfnamefont {A.}~\bibnamefont {Nierenberg}},\ }\href {\doibase
  10.1093/mnras/sty2261} {\bibfield  {journal} {\bibinfo  {journal} {Mon. Not.
  Roy. Astron. Soc.}\ }\textbf {\bibinfo {volume} {481}},\ \bibinfo {pages}
  {819} (\bibinfo {year} {2018})},\ \Eprint {http://arxiv.org/abs/1712.04945}
  {arXiv:1712.04945 [astro-ph.CO]} \BibitemShut {NoStop}%
\bibitem [{\citenamefont {Koopmans}(2005)}]{grav_imaging1}%
  \BibitemOpen
  \bibfield  {author} {\bibinfo {author} {\bibfnamefont {L.~V.~E.}\
  \bibnamefont {Koopmans}},\ }\href {\doibase 10.1111/j.1365-2966.2005.09523.x}
  {\bibfield  {journal} {\bibinfo  {journal} {Mon. Not. Roy. Astron. Soc.}\
  }\textbf {\bibinfo {volume} {363}},\ \bibinfo {pages} {1136} (\bibinfo {year}
  {2005})},\ \Eprint {http://arxiv.org/abs/astro-ph/0501324}
  {arXiv:astro-ph/0501324 [astro-ph]} \BibitemShut {NoStop}%
\bibitem [{\citenamefont {Moustakas}\ and\ \citenamefont
  {Metcalf}(2003)}]{spat_res_spec1}%
  \BibitemOpen
  \bibfield  {author} {\bibinfo {author} {\bibfnamefont {L.~A.}\ \bibnamefont
  {Moustakas}}\ and\ \bibinfo {author} {\bibfnamefont {R.~B.}\ \bibnamefont
  {Metcalf}},\ }\href {\doibase 10.1046/j.1365-8711.2003.06055.x} {\bibfield
  {journal} {\bibinfo  {journal} {Mon. Not. Roy. Astron. Soc.}\ }\textbf
  {\bibinfo {volume} {339}},\ \bibinfo {pages} {607} (\bibinfo {year}
  {2003})},\ \Eprint {http://arxiv.org/abs/astro-ph/0206176}
  {arXiv:astro-ph/0206176 [astro-ph]} \BibitemShut {NoStop}%
\bibitem [{\citenamefont {Hezaveh}\ \emph
  {et~al.}(2016{\natexlab{a}})\citenamefont {Hezaveh}, \citenamefont {Dalal},
  \citenamefont {Holder}, \citenamefont {Kisner}, \citenamefont {Kuhlen},\ and\
  \citenamefont {Perreault~Levasseur}}]{Hezaveh_powerspec}%
  \BibitemOpen
  \bibfield  {author} {\bibinfo {author} {\bibfnamefont {Y.}~\bibnamefont
  {Hezaveh}}, \bibinfo {author} {\bibfnamefont {N.}~\bibnamefont {Dalal}},
  \bibinfo {author} {\bibfnamefont {G.}~\bibnamefont {Holder}}, \bibinfo
  {author} {\bibfnamefont {T.}~\bibnamefont {Kisner}}, \bibinfo {author}
  {\bibfnamefont {M.}~\bibnamefont {Kuhlen}}, \ and\ \bibinfo {author}
  {\bibfnamefont {L.}~\bibnamefont {Perreault~Levasseur}},\ }\href {\doibase
  10.1088/1475-7516/2016/11/048} {\bibfield  {journal} {\bibinfo  {journal}
  {JCAP}\ }\textbf {\bibinfo {volume} {1611}},\ \bibinfo {pages} {048}
  (\bibinfo {year} {2016}{\natexlab{a}})},\ \Eprint
  {http://arxiv.org/abs/1403.2720} {arXiv:1403.2720 [astro-ph.CO]} \BibitemShut
  {NoStop}%
\bibitem [{\citenamefont {Daylan}\ \emph {et~al.}(2018)\citenamefont {Daylan},
  \citenamefont {Cyr-Racine}, \citenamefont {Diaz~Rivero}, \citenamefont
  {Dvorkin},\ and\ \citenamefont {Finkbeiner}}]{pcatlens}%
  \BibitemOpen
  \bibfield  {author} {\bibinfo {author} {\bibfnamefont {T.}~\bibnamefont
  {Daylan}}, \bibinfo {author} {\bibfnamefont {F.-Y.}\ \bibnamefont
  {Cyr-Racine}}, \bibinfo {author} {\bibfnamefont {A.}~\bibnamefont
  {Diaz~Rivero}}, \bibinfo {author} {\bibfnamefont {C.}~\bibnamefont
  {Dvorkin}}, \ and\ \bibinfo {author} {\bibfnamefont {D.~P.}\ \bibnamefont
  {Finkbeiner}},\ }\href {\doibase 10.3847/1538-4357/aaaa1e} {\bibfield
  {journal} {\bibinfo  {journal} {Astrophys. J.}\ }\textbf {\bibinfo {volume}
  {854}},\ \bibinfo {pages} {141} (\bibinfo {year} {2018})},\ \Eprint
  {http://arxiv.org/abs/1706.06111} {arXiv:1706.06111 [astro-ph.CO]}
  \BibitemShut {NoStop}%
\bibitem [{\citenamefont {Birrer}\ \emph
  {et~al.}(2017{\natexlab{a}})\citenamefont {Birrer}, \citenamefont {Amara},\
  and\ \citenamefont {Refregier}}]{Birrer:2017rpp}%
  \BibitemOpen
  \bibfield  {author} {\bibinfo {author} {\bibfnamefont {S.}~\bibnamefont
  {Birrer}}, \bibinfo {author} {\bibfnamefont {A.}~\bibnamefont {Amara}}, \
  and\ \bibinfo {author} {\bibfnamefont {A.}~\bibnamefont {Refregier}},\ }\href
  {\doibase 10.1088/1475-7516/2017/05/037} {\bibfield  {journal} {\bibinfo
  {journal} {JCAP}\ }\textbf {\bibinfo {volume} {1705}},\ \bibinfo {pages}
  {037} (\bibinfo {year} {2017}{\natexlab{a}})},\ \Eprint
  {http://arxiv.org/abs/1702.00009} {arXiv:1702.00009 [astro-ph.CO]}
  \BibitemShut {NoStop}%
\bibitem [{\citenamefont {Brewer}\ \emph {et~al.}(2016)\citenamefont {Brewer},
  \citenamefont {Huijser},\ and\ \citenamefont {Lewis}}]{Brewer:2015yya}%
  \BibitemOpen
  \bibfield  {author} {\bibinfo {author} {\bibfnamefont {B.~J.}\ \bibnamefont
  {Brewer}}, \bibinfo {author} {\bibfnamefont {D.}~\bibnamefont {Huijser}}, \
  and\ \bibinfo {author} {\bibfnamefont {G.~F.}\ \bibnamefont {Lewis}},\ }\href
  {\doibase 10.1093/mnras/stv2370} {\bibfield  {journal} {\bibinfo  {journal}
  {Mon. Not. Roy. Astron. Soc.}\ }\textbf {\bibinfo {volume} {455}},\ \bibinfo
  {pages} {1819} (\bibinfo {year} {2016})},\ \Eprint
  {http://arxiv.org/abs/1508.00662} {arXiv:1508.00662 [astro-ph.IM]}
  \BibitemShut {NoStop}%
\bibitem [{\citenamefont {Cyr-Racine}\ \emph {et~al.}(2016)\citenamefont
  {Cyr-Racine}, \citenamefont {Moustakas}, \citenamefont {Keeton},
  \citenamefont {Sigurdson},\ and\ \citenamefont {Gilman}}]{dark_census}%
  \BibitemOpen
  \bibfield  {author} {\bibinfo {author} {\bibfnamefont {F.}~\bibnamefont
  {Cyr-Racine}}, \bibinfo {author} {\bibfnamefont {L.}~\bibnamefont
  {Moustakas}}, \bibinfo {author} {\bibfnamefont {C.}~\bibnamefont {Keeton}},
  \bibinfo {author} {\bibfnamefont {K.}~\bibnamefont {Sigurdson}}, \ and\
  \bibinfo {author} {\bibfnamefont {D.}~\bibnamefont {Gilman}},\ }\href
  {\doibase 10.1103/PhysRevD.94.043505} {\bibfield  {journal} {\bibinfo
  {journal} {Physical Review D}\ }\textbf {\bibinfo {volume} {94}} (\bibinfo
  {year} {2016}),\ 10.1103/PhysRevD.94.043505}\BibitemShut {NoStop}%
\bibitem [{\citenamefont {Diaz~Rivero}\ \emph {et~al.}(2018)\citenamefont
  {Diaz~Rivero}, \citenamefont {Cyr-Racine},\ and\ \citenamefont
  {Dvorkin}}]{powerspec1}%
  \BibitemOpen
  \bibfield  {author} {\bibinfo {author} {\bibfnamefont {A.}~\bibnamefont
  {Diaz~Rivero}}, \bibinfo {author} {\bibfnamefont {F.-Y.}\ \bibnamefont
  {Cyr-Racine}}, \ and\ \bibinfo {author} {\bibfnamefont {C.}~\bibnamefont
  {Dvorkin}},\ }\href {\doibase 10.1103/PhysRevD.97.023001} {\bibfield
  {journal} {\bibinfo  {journal} {Phys. Rev.}\ }\textbf {\bibinfo {volume}
  {D97}},\ \bibinfo {pages} {023001} (\bibinfo {year} {2018})},\ \Eprint
  {http://arxiv.org/abs/1707.04590} {arXiv:1707.04590 [astro-ph.CO]}
  \BibitemShut {NoStop}%
\bibitem [{\citenamefont {Brehmer}\ \emph {et~al.}(2019)\citenamefont
  {Brehmer}, \citenamefont {Mishra-Sharma}, \citenamefont {Hermans},
  \citenamefont {Louppe},\ and\ \citenamefont {Cranmer}}]{mining_substructure}%
  \BibitemOpen
  \bibfield  {author} {\bibinfo {author} {\bibfnamefont {J.}~\bibnamefont
  {Brehmer}}, \bibinfo {author} {\bibfnamefont {S.}~\bibnamefont
  {Mishra-Sharma}}, \bibinfo {author} {\bibfnamefont {J.}~\bibnamefont
  {Hermans}}, \bibinfo {author} {\bibfnamefont {G.}~\bibnamefont {Louppe}}, \
  and\ \bibinfo {author} {\bibfnamefont {K.}~\bibnamefont {Cranmer}},\
  }\href@noop {} {\  (\bibinfo {year} {2019})},\ \Eprint
  {http://arxiv.org/abs/1909.02005} {arXiv:1909.02005 [astro-ph.CO]}
  \BibitemShut {NoStop}%
\bibitem [{\citenamefont {Diaz~Rivero}\ and\ \citenamefont
  {Dvorkin}(2020)}]{DiazRivero:2019hxf}%
  \BibitemOpen
  \bibfield  {author} {\bibinfo {author} {\bibfnamefont {A.}~\bibnamefont
  {Diaz~Rivero}}\ and\ \bibinfo {author} {\bibfnamefont {C.}~\bibnamefont
  {Dvorkin}},\ }\href {\doibase 10.1103/PhysRevD.101.023515} {\bibfield
  {journal} {\bibinfo  {journal} {Phys. Rev.}\ }\textbf {\bibinfo {volume}
  {D101}},\ \bibinfo {pages} {023515} (\bibinfo {year} {2020})},\ \Eprint
  {http://arxiv.org/abs/1910.00015} {arXiv:1910.00015 [astro-ph.CO]}
  \BibitemShut {NoStop}%
\bibitem [{\citenamefont {{Hezaveh}}\ \emph {et~al.}(2013)\citenamefont
  {{Hezaveh}}, \citenamefont {{Dalal}}, \citenamefont {{Holder}}, \citenamefont
  {{Kuhlen}}, \citenamefont {{Marrone}}, \citenamefont {{Murray}},\ and\
  \citenamefont {{Vieira}}}]{spat_res_spec2}%
  \BibitemOpen
  \bibfield  {author} {\bibinfo {author} {\bibfnamefont {Y.}~\bibnamefont
  {{Hezaveh}}}, \bibinfo {author} {\bibfnamefont {N.}~\bibnamefont {{Dalal}}},
  \bibinfo {author} {\bibfnamefont {G.}~\bibnamefont {{Holder}}}, \bibinfo
  {author} {\bibfnamefont {M.}~\bibnamefont {{Kuhlen}}}, \bibinfo {author}
  {\bibfnamefont {D.}~\bibnamefont {{Marrone}}}, \bibinfo {author}
  {\bibfnamefont {N.}~\bibnamefont {{Murray}}}, \ and\ \bibinfo {author}
  {\bibfnamefont {J.}~\bibnamefont {{Vieira}}},\ }\href {\doibase
  10.1088/0004-637X/767/1/9} {\bibfield  {journal} {\bibinfo  {journal} {\apj}\
  }\textbf {\bibinfo {volume} {767}},\ \bibinfo {eid} {9} (\bibinfo {year}
  {2013})},\ \Eprint {http://arxiv.org/abs/1210.4562} {arXiv:1210.4562
  [astro-ph.CO]} \BibitemShut {NoStop}%
\bibitem [{\citenamefont {{Vegetti}}\ \emph {et~al.}(2010)\citenamefont
  {{Vegetti}}, \citenamefont {{Koopmans}}, \citenamefont {{Bolton}},
  \citenamefont {{Treu}},\ and\ \citenamefont
  {{Gavazzi}}}]{detection_2010_mnras}%
  \BibitemOpen
  \bibfield  {author} {\bibinfo {author} {\bibfnamefont {S.}~\bibnamefont
  {{Vegetti}}}, \bibinfo {author} {\bibfnamefont {L.~V.~E.}\ \bibnamefont
  {{Koopmans}}}, \bibinfo {author} {\bibfnamefont {A.}~\bibnamefont
  {{Bolton}}}, \bibinfo {author} {\bibfnamefont {T.}~\bibnamefont {{Treu}}}, \
  and\ \bibinfo {author} {\bibfnamefont {R.}~\bibnamefont {{Gavazzi}}},\ }\href
  {\doibase 10.1111/j.1365-2966.2010.16865.x} {\bibfield  {journal} {\bibinfo
  {journal} {\mnras}\ }\textbf {\bibinfo {volume} {408}},\ \bibinfo {pages}
  {1969} (\bibinfo {year} {2010})},\ \Eprint {http://arxiv.org/abs/0910.0760}
  {arXiv:0910.0760 [astro-ph.CO]} \BibitemShut {NoStop}%
\bibitem [{\citenamefont {{Vegetti}}\ \emph {et~al.}(2012)\citenamefont
  {{Vegetti}}, \citenamefont {{Lagattuta}}, \citenamefont {{McKean}},
  \citenamefont {{Auger}}, \citenamefont {{Fassnacht}},\ and\ \citenamefont
  {{Koopmans}}}]{vegetti_nature}%
  \BibitemOpen
  \bibfield  {author} {\bibinfo {author} {\bibfnamefont {S.}~\bibnamefont
  {{Vegetti}}}, \bibinfo {author} {\bibfnamefont {D.~J.}\ \bibnamefont
  {{Lagattuta}}}, \bibinfo {author} {\bibfnamefont {J.~P.}\ \bibnamefont
  {{McKean}}}, \bibinfo {author} {\bibfnamefont {M.~W.}\ \bibnamefont
  {{Auger}}}, \bibinfo {author} {\bibfnamefont {C.~D.}\ \bibnamefont
  {{Fassnacht}}}, \ and\ \bibinfo {author} {\bibfnamefont {L.~V.~E.}\
  \bibnamefont {{Koopmans}}},\ }\href {\doibase 10.1038/nature10669} {\bibfield
   {journal} {\bibinfo  {journal} {\nat}\ }\textbf {\bibinfo {volume} {481}},\
  \bibinfo {pages} {341} (\bibinfo {year} {2012})},\ \Eprint
  {http://arxiv.org/abs/1201.3643} {arXiv:1201.3643 [astro-ph.CO]} \BibitemShut
  {NoStop}%
\bibitem [{\citenamefont {Ritondale}\ \emph {et~al.}(2019)\citenamefont
  {Ritondale}, \citenamefont {Vegetti}, \citenamefont {Despali}, \citenamefont
  {Auger}, \citenamefont {Koopmans},\ and\ \citenamefont
  {McKean}}]{bells_2018}%
  \BibitemOpen
  \bibfield  {author} {\bibinfo {author} {\bibfnamefont {E.}~\bibnamefont
  {Ritondale}}, \bibinfo {author} {\bibfnamefont {S.}~\bibnamefont {Vegetti}},
  \bibinfo {author} {\bibfnamefont {G.}~\bibnamefont {Despali}}, \bibinfo
  {author} {\bibfnamefont {M.~W.}\ \bibnamefont {Auger}}, \bibinfo {author}
  {\bibfnamefont {L.~V.~E.}\ \bibnamefont {Koopmans}}, \ and\ \bibinfo {author}
  {\bibfnamefont {J.~P.}\ \bibnamefont {McKean}},\ }\href {\doibase
  10.1093/mnras/stz464} {\bibfield  {journal} {\bibinfo  {journal} {Mon. Not.
  Roy. Astron. Soc.}\ }\textbf {\bibinfo {volume} {485}},\ \bibinfo {pages}
  {2179} (\bibinfo {year} {2019})},\ \Eprint {http://arxiv.org/abs/1811.03627}
  {arXiv:1811.03627 [astro-ph.CO]} \BibitemShut {NoStop}%
\bibitem [{\citenamefont {{Vegetti}}\ \emph {et~al.}(2014)\citenamefont
  {{Vegetti}}, \citenamefont {{Koopmans}}, \citenamefont {{Auger}},
  \citenamefont {{Treu}},\ and\ \citenamefont
  {{Bolton}}}]{2014MNRAS.442.2017V}%
  \BibitemOpen
  \bibfield  {author} {\bibinfo {author} {\bibfnamefont {S.}~\bibnamefont
  {{Vegetti}}}, \bibinfo {author} {\bibfnamefont {L.~V.~E.}\ \bibnamefont
  {{Koopmans}}}, \bibinfo {author} {\bibfnamefont {M.~W.}\ \bibnamefont
  {{Auger}}}, \bibinfo {author} {\bibfnamefont {T.}~\bibnamefont {{Treu}}}, \
  and\ \bibinfo {author} {\bibfnamefont {A.~S.}\ \bibnamefont {{Bolton}}},\
  }\href {\doibase 10.1093/mnras/stu943} {\bibfield  {journal} {\bibinfo
  {journal} {\mnras}\ }\textbf {\bibinfo {volume} {442}},\ \bibinfo {pages}
  {2017} (\bibinfo {year} {2014})},\ \Eprint {http://arxiv.org/abs/1405.3666}
  {arXiv:1405.3666 [astro-ph.GA]} \BibitemShut {NoStop}%
\bibitem [{\citenamefont {Vegetti}\ \emph {et~al.}(2018)\citenamefont
  {Vegetti}, \citenamefont {Despali}, \citenamefont {Lovell},\ and\
  \citenamefont {Enzi}}]{Vegetti:2018dly}%
  \BibitemOpen
  \bibfield  {author} {\bibinfo {author} {\bibfnamefont {S.}~\bibnamefont
  {Vegetti}}, \bibinfo {author} {\bibfnamefont {G.}~\bibnamefont {Despali}},
  \bibinfo {author} {\bibfnamefont {M.~R.}\ \bibnamefont {Lovell}}, \ and\
  \bibinfo {author} {\bibfnamefont {W.}~\bibnamefont {Enzi}},\ }\href {\doibase
  10.1093/mnras/sty2393} {\bibfield  {journal} {\bibinfo  {journal} {Mon. Not.
  Roy. Astron. Soc.}\ }\textbf {\bibinfo {volume} {481}},\ \bibinfo {pages}
  {3661} (\bibinfo {year} {2018})},\ \Eprint {http://arxiv.org/abs/1801.01505}
  {arXiv:1801.01505 [astro-ph.CO]} \BibitemShut {NoStop}%
\bibitem [{\citenamefont {Li}\ \emph {et~al.}(2017)\citenamefont {Li},
  \citenamefont {Frenk}, \citenamefont {Cole}, \citenamefont {Wang},\ and\
  \citenamefont {Gao}}]{Li_los}%
  \BibitemOpen
  \bibfield  {author} {\bibinfo {author} {\bibfnamefont {R.}~\bibnamefont
  {Li}}, \bibinfo {author} {\bibfnamefont {C.~S.}\ \bibnamefont {Frenk}},
  \bibinfo {author} {\bibfnamefont {S.}~\bibnamefont {Cole}}, \bibinfo {author}
  {\bibfnamefont {Q.}~\bibnamefont {Wang}}, \ and\ \bibinfo {author}
  {\bibfnamefont {L.}~\bibnamefont {Gao}},\ }\href {\doibase
  10.1093/mnras/stx554} {\bibfield  {journal} {\bibinfo  {journal} {Mon. Not.
  Roy. Astron. Soc.}\ }\textbf {\bibinfo {volume} {468}},\ \bibinfo {pages}
  {1426} (\bibinfo {year} {2017})},\ \Eprint {http://arxiv.org/abs/1612.06227}
  {arXiv:1612.06227 [astro-ph.CO]} \BibitemShut {NoStop}%
\bibitem [{\citenamefont {{D'Aloisio}}\ and\ \citenamefont
  {{Natarajan}}(2011)}]{daloisio11}%
  \BibitemOpen
  \bibfield  {author} {\bibinfo {author} {\bibfnamefont {A.}~\bibnamefont
  {{D'Aloisio}}}\ and\ \bibinfo {author} {\bibfnamefont {P.}~\bibnamefont
  {{Natarajan}}},\ }\href {\doibase 10.1111/j.1365-2966.2010.17795.x}
  {\bibfield  {journal} {\bibinfo  {journal} {\mnras}\ }\textbf {\bibinfo
  {volume} {411}},\ \bibinfo {pages} {1628} (\bibinfo {year} {2011})},\ \Eprint
  {http://arxiv.org/abs/1010.0004} {arXiv:1010.0004 [astro-ph.CO]} \BibitemShut
  {NoStop}%
\bibitem [{\citenamefont {McCully}\ \emph {et~al.}(2017)\citenamefont
  {McCully}, \citenamefont {Keeton}, \citenamefont {Wong},\ and\ \citenamefont
  {Zabludoff}}]{McCully:2016yfe}%
  \BibitemOpen
  \bibfield  {author} {\bibinfo {author} {\bibfnamefont {C.}~\bibnamefont
  {McCully}}, \bibinfo {author} {\bibfnamefont {C.~R.}\ \bibnamefont {Keeton}},
  \bibinfo {author} {\bibfnamefont {K.~C.}\ \bibnamefont {Wong}}, \ and\
  \bibinfo {author} {\bibfnamefont {A.~I.}\ \bibnamefont {Zabludoff}},\ }\href
  {\doibase 10.3847/1538-4357/836/1/141} {\bibfield  {journal} {\bibinfo
  {journal} {Astrophys. J.}\ }\textbf {\bibinfo {volume} {836}},\ \bibinfo
  {pages} {141} (\bibinfo {year} {2017})},\ \Eprint
  {http://arxiv.org/abs/1601.05417} {arXiv:1601.05417 [astro-ph.CO]}
  \BibitemShut {NoStop}%
\bibitem [{\citenamefont {Despali}\ \emph {et~al.}(2018)\citenamefont
  {Despali}, \citenamefont {Vegetti}, \citenamefont {White}, \citenamefont
  {Giocoli},\ and\ \citenamefont {van~den Bosch}}]{Despali_los}%
  \BibitemOpen
  \bibfield  {author} {\bibinfo {author} {\bibfnamefont {G.}~\bibnamefont
  {Despali}}, \bibinfo {author} {\bibfnamefont {S.}~\bibnamefont {Vegetti}},
  \bibinfo {author} {\bibfnamefont {S.~D.~M.}\ \bibnamefont {White}}, \bibinfo
  {author} {\bibfnamefont {C.}~\bibnamefont {Giocoli}}, \ and\ \bibinfo
  {author} {\bibfnamefont {F.~C.}\ \bibnamefont {van~den Bosch}},\ }\href
  {\doibase 10.1093/mnras/sty159} {\bibfield  {journal} {\bibinfo  {journal}
  {Mon. Not. Roy. Astron. Soc.}\ }\textbf {\bibinfo {volume} {475}},\ \bibinfo
  {pages} {5424} (\bibinfo {year} {2018})},\ \Eprint
  {http://arxiv.org/abs/1710.05029} {arXiv:1710.05029 [astro-ph.CO]}
  \BibitemShut {NoStop}%
\bibitem [{\citenamefont {Gilman}\ \emph {et~al.}(2019)\citenamefont {Gilman},
  \citenamefont {Birrer}, \citenamefont {Treu}, \citenamefont {Nierenberg},\
  and\ \citenamefont {Benson}}]{Gilman:2019vca}%
  \BibitemOpen
  \bibfield  {author} {\bibinfo {author} {\bibfnamefont {D.}~\bibnamefont
  {Gilman}}, \bibinfo {author} {\bibfnamefont {S.}~\bibnamefont {Birrer}},
  \bibinfo {author} {\bibfnamefont {T.}~\bibnamefont {Treu}}, \bibinfo {author}
  {\bibfnamefont {A.}~\bibnamefont {Nierenberg}}, \ and\ \bibinfo {author}
  {\bibfnamefont {A.}~\bibnamefont {Benson}},\ }\href {\doibase
  10.1093/mnras/stz1593} {\bibfield  {journal} {\bibinfo  {journal} {Mon. Not.
  Roy. Astron. Soc.}\ }\textbf {\bibinfo {volume} {487}},\ \bibinfo {pages}
  {5721} (\bibinfo {year} {2019})},\ \Eprint {http://arxiv.org/abs/1901.11031}
  {arXiv:1901.11031 [astro-ph.CO]} \BibitemShut {NoStop}%
\bibitem [{\citenamefont {Gilman}\ \emph
  {et~al.}(2020{\natexlab{a}})\citenamefont {Gilman}, \citenamefont {Birrer},
  \citenamefont {Nierenberg}, \citenamefont {Treu}, \citenamefont {Du},\ and\
  \citenamefont {Benson}}]{Gilman:2019nap}%
  \BibitemOpen
  \bibfield  {author} {\bibinfo {author} {\bibfnamefont {D.}~\bibnamefont
  {Gilman}}, \bibinfo {author} {\bibfnamefont {S.}~\bibnamefont {Birrer}},
  \bibinfo {author} {\bibfnamefont {A.}~\bibnamefont {Nierenberg}}, \bibinfo
  {author} {\bibfnamefont {T.}~\bibnamefont {Treu}}, \bibinfo {author}
  {\bibfnamefont {X.}~\bibnamefont {Du}}, \ and\ \bibinfo {author}
  {\bibfnamefont {A.}~\bibnamefont {Benson}},\ }\href {\doibase
  10.1093/mnras/stz3480} {\bibfield  {journal} {\bibinfo  {journal} {Mon. Not.
  Roy. Astron. Soc.}\ }\textbf {\bibinfo {volume} {491}},\ \bibinfo {pages}
  {6077} (\bibinfo {year} {2020}{\natexlab{a}})},\ \Eprint
  {http://arxiv.org/abs/1908.06983} {arXiv:1908.06983 [astro-ph.CO]}
  \BibitemShut {NoStop}%
\bibitem [{\citenamefont {Gilman}\ \emph
  {et~al.}(2020{\natexlab{b}})\citenamefont {Gilman}, \citenamefont {Du},
  \citenamefont {Benson}, \citenamefont {Birrer}, \citenamefont {Nierenberg},\
  and\ \citenamefont {Treu}}]{Gilman:2019bdm}%
  \BibitemOpen
  \bibfield  {author} {\bibinfo {author} {\bibfnamefont {D.}~\bibnamefont
  {Gilman}}, \bibinfo {author} {\bibfnamefont {X.}~\bibnamefont {Du}}, \bibinfo
  {author} {\bibfnamefont {A.}~\bibnamefont {Benson}}, \bibinfo {author}
  {\bibfnamefont {S.}~\bibnamefont {Birrer}}, \bibinfo {author} {\bibfnamefont
  {A.}~\bibnamefont {Nierenberg}}, \ and\ \bibinfo {author} {\bibfnamefont
  {T.}~\bibnamefont {Treu}},\ }\href {\doibase 10.1093/mnrasl/slz173}
  {\bibfield  {journal} {\bibinfo  {journal} {Mon. Not. Roy. Astron. Soc.}\
  }\textbf {\bibinfo {volume} {492}},\ \bibinfo {pages} {L12} (\bibinfo {year}
  {2020}{\natexlab{b}})},\ \Eprint {http://arxiv.org/abs/1909.02573}
  {arXiv:1909.02573 [astro-ph.CO]} \BibitemShut {NoStop}%
\bibitem [{\citenamefont {Hsueh}\ \emph {et~al.}(2020)\citenamefont {Hsueh},
  \citenamefont {Enzi}, \citenamefont {Vegetti}, \citenamefont {Auger},
  \citenamefont {Fassnacht}, \citenamefont {Despali}, \citenamefont
  {Koopmans},\ and\ \citenamefont {McKean}}]{Hsueh:2019ynk}%
  \BibitemOpen
  \bibfield  {author} {\bibinfo {author} {\bibfnamefont {J.-W.}\ \bibnamefont
  {Hsueh}}, \bibinfo {author} {\bibfnamefont {W.}~\bibnamefont {Enzi}},
  \bibinfo {author} {\bibfnamefont {S.}~\bibnamefont {Vegetti}}, \bibinfo
  {author} {\bibfnamefont {M.}~\bibnamefont {Auger}}, \bibinfo {author}
  {\bibfnamefont {C.~D.}\ \bibnamefont {Fassnacht}}, \bibinfo {author}
  {\bibfnamefont {G.}~\bibnamefont {Despali}}, \bibinfo {author} {\bibfnamefont
  {L.~V.~E.}\ \bibnamefont {Koopmans}}, \ and\ \bibinfo {author} {\bibfnamefont
  {J.~P.}\ \bibnamefont {McKean}},\ }\href {\doibase 10.1093/mnras/stz3177}
  {\bibfield  {journal} {\bibinfo  {journal} {Mon. Not. Roy. Astron. Soc.}\
  }\textbf {\bibinfo {volume} {492}},\ \bibinfo {pages} {3047} (\bibinfo {year}
  {2020})},\ \Eprint {http://arxiv.org/abs/1905.04182} {arXiv:1905.04182
  [astro-ph.CO]} \BibitemShut {NoStop}%
\bibitem [{\citenamefont {Wong}\ \emph {et~al.}(2019)\citenamefont {Wong} \emph
  {et~al.}}]{Wong:2019kwg}%
  \BibitemOpen
  \bibfield  {author} {\bibinfo {author} {\bibfnamefont {K.~C.}\ \bibnamefont
  {Wong}} \emph {et~al.},\ }\href@noop {} {\  (\bibinfo {year} {2019})},\
  \Eprint {http://arxiv.org/abs/1907.04869} {arXiv:1907.04869 [astro-ph.CO]}
  \BibitemShut {NoStop}%
\bibitem [{\citenamefont {Rusu}\ \emph {et~al.}(2019)\citenamefont {Rusu} \emph
  {et~al.}}]{Rusu:2019xrq}%
  \BibitemOpen
  \bibfield  {author} {\bibinfo {author} {\bibfnamefont {C.~E.}\ \bibnamefont
  {Rusu}} \emph {et~al.},\ }\href@noop {} {\  (\bibinfo {year} {2019})},\
  \Eprint {http://arxiv.org/abs/1905.09338} {arXiv:1905.09338 [astro-ph.CO]}
  \BibitemShut {NoStop}%
\bibitem [{\citenamefont {Chen}\ \emph {et~al.}(2019)\citenamefont {Chen} \emph
  {et~al.}}]{Chen:2019ejq}%
  \BibitemOpen
  \bibfield  {author} {\bibinfo {author} {\bibfnamefont {G.~C.~F.}\
  \bibnamefont {Chen}} \emph {et~al.},\ }\href {\doibase 10.1093/mnras/stz2547}
  {\bibfield  {journal} {\bibinfo  {journal} {Mon. Not. Roy. Astron. Soc.}\
  }\textbf {\bibinfo {volume} {490}},\ \bibinfo {pages} {1743} (\bibinfo {year}
  {2019})},\ \Eprint {http://arxiv.org/abs/1907.02533} {arXiv:1907.02533
  [astro-ph.CO]} \BibitemShut {NoStop}%
\bibitem [{\citenamefont {Díaz~Rivero}\ \emph {et~al.}(2018)\citenamefont
  {Díaz~Rivero}, \citenamefont {Dvorkin}, \citenamefont {Cyr-Racine},
  \citenamefont {Zavala},\ and\ \citenamefont {Vogelsberger}}]{powerspec2}%
  \BibitemOpen
  \bibfield  {author} {\bibinfo {author} {\bibfnamefont {A.}~\bibnamefont
  {Díaz~Rivero}}, \bibinfo {author} {\bibfnamefont {C.}~\bibnamefont
  {Dvorkin}}, \bibinfo {author} {\bibfnamefont {F.-Y.}\ \bibnamefont
  {Cyr-Racine}}, \bibinfo {author} {\bibfnamefont {J.}~\bibnamefont {Zavala}},
  \ and\ \bibinfo {author} {\bibfnamefont {M.}~\bibnamefont {Vogelsberger}},\
  }\href {\doibase 10.1103/PhysRevD.98.103517} {\bibfield  {journal} {\bibinfo
  {journal} {Phys. Rev.}\ }\textbf {\bibinfo {volume} {D98}},\ \bibinfo {pages}
  {103517} (\bibinfo {year} {2018})},\ \Eprint
  {http://arxiv.org/abs/1809.00004} {arXiv:1809.00004 [astro-ph.CO]}
  \BibitemShut {NoStop}%
\bibitem [{\citenamefont {Brennan}\ \emph {et~al.}(2019)\citenamefont
  {Brennan}, \citenamefont {Benson}, \citenamefont {Cyr-Racine}, \citenamefont
  {Keeton}, \citenamefont {Moustakas},\ and\ \citenamefont {Pullen}}]{Brennan}%
  \BibitemOpen
  \bibfield  {author} {\bibinfo {author} {\bibfnamefont {S.}~\bibnamefont
  {Brennan}}, \bibinfo {author} {\bibfnamefont {A.~J.}\ \bibnamefont {Benson}},
  \bibinfo {author} {\bibfnamefont {F.-Y.}\ \bibnamefont {Cyr-Racine}},
  \bibinfo {author} {\bibfnamefont {C.~R.}\ \bibnamefont {Keeton}}, \bibinfo
  {author} {\bibfnamefont {L.~A.}\ \bibnamefont {Moustakas}}, \ and\ \bibinfo
  {author} {\bibfnamefont {A.~R.}\ \bibnamefont {Pullen}},\ }\href {\doibase
  10.1093/mnras/stz1607} {\bibfield  {journal} {\bibinfo  {journal} {Mon. Not.
  Roy. Astron. Soc.}\ }\textbf {\bibinfo {volume} {488}},\ \bibinfo {pages}
  {5085} (\bibinfo {year} {2019})},\ \Eprint {http://arxiv.org/abs/1808.03501}
  {arXiv:1808.03501 [astro-ph.GA]} \BibitemShut {NoStop}%
\bibitem [{\citenamefont {{Schneider}}\ \emph {et~al.}(1992)\citenamefont
  {{Schneider}}, \citenamefont {{Ehlers}},\ and\ \citenamefont
  {{Falco}}}]{lensing_book}%
  \BibitemOpen
  \bibfield  {author} {\bibinfo {author} {\bibfnamefont {P.}~\bibnamefont
  {{Schneider}}}, \bibinfo {author} {\bibfnamefont {J.}~\bibnamefont
  {{Ehlers}}}, \ and\ \bibinfo {author} {\bibfnamefont {E.~E.}\ \bibnamefont
  {{Falco}}},\ }\href {\doibase 10.1007/978-3-662-03758-4} {\emph {\bibinfo
  {title} {{Gravitational Lenses}}}}\ (\bibinfo {year} {1992})\BibitemShut
  {NoStop}%
\bibitem [{\citenamefont {Birrer}\ \emph
  {et~al.}(2017{\natexlab{b}})\citenamefont {Birrer}, \citenamefont {Welschen},
  \citenamefont {Amara},\ and\ \citenamefont {Refregier}}]{CSB}%
  \BibitemOpen
  \bibfield  {author} {\bibinfo {author} {\bibfnamefont {S.}~\bibnamefont
  {Birrer}}, \bibinfo {author} {\bibfnamefont {C.}~\bibnamefont {Welschen}},
  \bibinfo {author} {\bibfnamefont {A.}~\bibnamefont {Amara}}, \ and\ \bibinfo
  {author} {\bibfnamefont {A.}~\bibnamefont {Refregier}},\ }\href {\doibase
  10.1088/1475-7516/2017/04/049} {\bibfield  {journal} {\bibinfo  {journal}
  {JCAP}\ }\textbf {\bibinfo {volume} {1704}},\ \bibinfo {pages} {049}
  (\bibinfo {year} {2017}{\natexlab{b}})},\ \Eprint
  {http://arxiv.org/abs/1610.01599} {arXiv:1610.01599 [astro-ph.CO]}
  \BibitemShut {NoStop}%
\bibitem [{\citenamefont {Baltz}\ \emph {et~al.}(2009)\citenamefont {Baltz},
  \citenamefont {Marshall},\ and\ \citenamefont {Oguri}}]{tnfw}%
  \BibitemOpen
  \bibfield  {author} {\bibinfo {author} {\bibfnamefont {E.~A.}\ \bibnamefont
  {Baltz}}, \bibinfo {author} {\bibfnamefont {P.}~\bibnamefont {Marshall}}, \
  and\ \bibinfo {author} {\bibfnamefont {M.}~\bibnamefont {Oguri}},\ }\href
  {\doibase 10.1088/1475-7516/2009/01/015} {\bibfield  {journal} {\bibinfo
  {journal} {JCAP}\ }\textbf {\bibinfo {volume} {0901}},\ \bibinfo {pages}
  {015} (\bibinfo {year} {2009})},\ \Eprint {http://arxiv.org/abs/0705.0682}
  {arXiv:0705.0682 [astro-ph]} \BibitemShut {NoStop}%
\bibitem [{\citenamefont {{Limber}}(1953)}]{1953ApJ...117..134L}%
  \BibitemOpen
  \bibfield  {author} {\bibinfo {author} {\bibfnamefont {D.~N.}\ \bibnamefont
  {{Limber}}},\ }\href {\doibase 10.1086/145672} {\bibfield  {journal}
  {\bibinfo  {journal} {\apj}\ }\textbf {\bibinfo {volume} {117}},\ \bibinfo
  {pages} {134} (\bibinfo {year} {1953})}\BibitemShut {NoStop}%
\bibitem [{\citenamefont {Dodelson}(2003)}]{dodelson:2003}%
  \BibitemOpen
  \bibfield  {author} {\bibinfo {author} {\bibfnamefont {S.}~\bibnamefont
  {Dodelson}},\ }\href@noop {} {\emph {\bibinfo {title} {{Modern Cosmology}}}}\
  (\bibinfo  {publisher} {Academic Press, Elsevier Science},\ \bibinfo {year}
  {2003})\BibitemShut {NoStop}%
\bibitem [{\citenamefont {Sheth}\ \emph {et~al.}(1999)\citenamefont {Sheth},
  \citenamefont {Mo},\ and\ \citenamefont {Tormen}}]{sheth}%
  \BibitemOpen
  \bibfield  {author} {\bibinfo {author} {\bibfnamefont {R.~K.}\ \bibnamefont
  {Sheth}}, \bibinfo {author} {\bibfnamefont {H.~J.}\ \bibnamefont {Mo}}, \
  and\ \bibinfo {author} {\bibfnamefont {G.}~\bibnamefont {Tormen}},\ }\href
  {\doibase 10.1046/j.1365-8711.2001.04006.x} {\  (\bibinfo {year} {1999}),\
  10.1046/j.1365-8711.2001.04006.x},\ \Eprint
  {http://arxiv.org/abs/arXiv:astro-ph/9907024} {arXiv:astro-ph/9907024}
  \BibitemShut {NoStop}%
\bibitem [{\citenamefont {{Reed}}\ \emph {et~al.}(2007)\citenamefont {{Reed}},
  \citenamefont {{Bower}}, \citenamefont {{Frenk}}, \citenamefont {{Jenkins}},\
  and\ \citenamefont {{Theuns}}}]{st_parameters}%
  \BibitemOpen
  \bibfield  {author} {\bibinfo {author} {\bibfnamefont {D.~S.}\ \bibnamefont
  {{Reed}}}, \bibinfo {author} {\bibfnamefont {R.}~\bibnamefont {{Bower}}},
  \bibinfo {author} {\bibfnamefont {C.~S.}\ \bibnamefont {{Frenk}}}, \bibinfo
  {author} {\bibfnamefont {A.}~\bibnamefont {{Jenkins}}}, \ and\ \bibinfo
  {author} {\bibfnamefont {T.}~\bibnamefont {{Theuns}}},\ }\href {\doibase
  10.1111/j.1365-2966.2006.11204.x} {\bibfield  {journal} {\bibinfo  {journal}
  {\mnras}\ }\textbf {\bibinfo {volume} {374}},\ \bibinfo {pages} {2} (\bibinfo
  {year} {2007})},\ \Eprint {http://arxiv.org/abs/astro-ph/0607150}
  {arXiv:astro-ph/0607150 [astro-ph]} \BibitemShut {NoStop}%
\bibitem [{\citenamefont {{Vegetti}}\ and\ \citenamefont
  {{Vogelsberger}}(2014)}]{subs_den_prof}%
  \BibitemOpen
  \bibfield  {author} {\bibinfo {author} {\bibfnamefont {S.}~\bibnamefont
  {{Vegetti}}}\ and\ \bibinfo {author} {\bibfnamefont {M.}~\bibnamefont
  {{Vogelsberger}}},\ }\href {\doibase 10.1093/mnras/stu1284} {\bibfield
  {journal} {\bibinfo  {journal} {\mnras}\ }\textbf {\bibinfo {volume} {442}},\
  \bibinfo {pages} {3598} (\bibinfo {year} {2014})},\ \Eprint
  {http://arxiv.org/abs/1406.1170} {arXiv:1406.1170 [astro-ph.CO]} \BibitemShut
  {NoStop}%
\bibitem [{\citenamefont {{Duffy}}\ \emph {et~al.}(2008)\citenamefont
  {{Duffy}}, \citenamefont {{Schaye}}, \citenamefont {{Kay}},\ and\
  \citenamefont {{Dalla Vecchia}}}]{halo_wmap}%
  \BibitemOpen
  \bibfield  {author} {\bibinfo {author} {\bibfnamefont {A.~R.}\ \bibnamefont
  {{Duffy}}}, \bibinfo {author} {\bibfnamefont {J.}~\bibnamefont {{Schaye}}},
  \bibinfo {author} {\bibfnamefont {S.~T.}\ \bibnamefont {{Kay}}}, \ and\
  \bibinfo {author} {\bibfnamefont {C.}~\bibnamefont {{Dalla Vecchia}}},\
  }\href {\doibase 10.1111/j.1745-3933.2008.00537.x} {\bibfield  {journal}
  {\bibinfo  {journal} {\mnras}\ }\textbf {\bibinfo {volume} {390}},\ \bibinfo
  {pages} {L64} (\bibinfo {year} {2008})},\ \Eprint
  {http://arxiv.org/abs/0804.2486} {arXiv:0804.2486 [astro-ph]} \BibitemShut
  {NoStop}%
\bibitem [{\citenamefont {{Giocoli}}\ \emph {et~al.}(2010)\citenamefont
  {{Giocoli}}, \citenamefont {{Tormen}}, \citenamefont {{Sheth}},\ and\
  \citenamefont {{van den Bosch}}}]{fsubredshift}%
  \BibitemOpen
  \bibfield  {author} {\bibinfo {author} {\bibfnamefont {C.}~\bibnamefont
  {{Giocoli}}}, \bibinfo {author} {\bibfnamefont {G.}~\bibnamefont {{Tormen}}},
  \bibinfo {author} {\bibfnamefont {R.~K.}\ \bibnamefont {{Sheth}}}, \ and\
  \bibinfo {author} {\bibfnamefont {F.~C.}\ \bibnamefont {{van den Bosch}}},\
  }\href {\doibase 10.1111/j.1365-2966.2010.16311.x} {\bibfield  {journal}
  {\bibinfo  {journal} {\mnras}\ }\textbf {\bibinfo {volume} {404}},\ \bibinfo
  {pages} {502} (\bibinfo {year} {2010})},\ \Eprint
  {http://arxiv.org/abs/0911.0436} {arXiv:0911.0436 [astro-ph.CO]} \BibitemShut
  {NoStop}%
\bibitem [{\citenamefont {{Okabe}}\ \emph {et~al.}(2014)\citenamefont
  {{Okabe}}, \citenamefont {{Futamase}}, \citenamefont {{Kajisawa}},\ and\
  \citenamefont {{Kuroshima}}}]{okabe14}%
  \BibitemOpen
  \bibfield  {author} {\bibinfo {author} {\bibfnamefont {N.}~\bibnamefont
  {{Okabe}}}, \bibinfo {author} {\bibfnamefont {T.}~\bibnamefont {{Futamase}}},
  \bibinfo {author} {\bibfnamefont {M.}~\bibnamefont {{Kajisawa}}}, \ and\
  \bibinfo {author} {\bibfnamefont {R.}~\bibnamefont {{Kuroshima}}},\ }\href
  {\doibase 10.1088/0004-637X/784/2/90} {\bibfield  {journal} {\bibinfo
  {journal} {\apj}\ }\textbf {\bibinfo {volume} {784}},\ \bibinfo {eid} {90}
  (\bibinfo {year} {2014})},\ \Eprint {http://arxiv.org/abs/1304.2399}
  {arXiv:1304.2399 [astro-ph.CO]} \BibitemShut {NoStop}%
\bibitem [{\citenamefont {{De Lucia}}\ \emph {et~al.}(2004)\citenamefont {{De
  Lucia}}, \citenamefont {{Kauffmann}}, \citenamefont {{Springel}},
  \citenamefont {{White}}, \citenamefont {{Lanzoni}}, \citenamefont {{Stoehr}},
  \citenamefont {{Tormen}},\ and\ \citenamefont {{Yoshida}}}]{delucia04}%
  \BibitemOpen
  \bibfield  {author} {\bibinfo {author} {\bibfnamefont {G.}~\bibnamefont {{De
  Lucia}}}, \bibinfo {author} {\bibfnamefont {G.}~\bibnamefont {{Kauffmann}}},
  \bibinfo {author} {\bibfnamefont {V.}~\bibnamefont {{Springel}}}, \bibinfo
  {author} {\bibfnamefont {S.~D.~M.}\ \bibnamefont {{White}}}, \bibinfo
  {author} {\bibfnamefont {B.}~\bibnamefont {{Lanzoni}}}, \bibinfo {author}
  {\bibfnamefont {F.}~\bibnamefont {{Stoehr}}}, \bibinfo {author}
  {\bibfnamefont {G.}~\bibnamefont {{Tormen}}}, \ and\ \bibinfo {author}
  {\bibfnamefont {N.}~\bibnamefont {{Yoshida}}},\ }\href {\doibase
  10.1111/j.1365-2966.2004.07372.x} {\bibfield  {journal} {\bibinfo  {journal}
  {\mnras}\ }\textbf {\bibinfo {volume} {348}},\ \bibinfo {pages} {333}
  (\bibinfo {year} {2004})},\ \Eprint {http://arxiv.org/abs/astro-ph/0306205}
  {arXiv:astro-ph/0306205 [astro-ph]} \BibitemShut {NoStop}%
\bibitem [{\citenamefont {{Madau}}\ \emph {et~al.}(2008)\citenamefont
  {{Madau}}, \citenamefont {{Diemand}},\ and\ \citenamefont
  {{Kuhlen}}}]{madau08}%
  \BibitemOpen
  \bibfield  {author} {\bibinfo {author} {\bibfnamefont {P.}~\bibnamefont
  {{Madau}}}, \bibinfo {author} {\bibfnamefont {J.}~\bibnamefont {{Diemand}}},
  \ and\ \bibinfo {author} {\bibfnamefont {M.}~\bibnamefont {{Kuhlen}}},\
  }\href {\doibase 10.1086/587545} {\bibfield  {journal} {\bibinfo  {journal}
  {\apj}\ }\textbf {\bibinfo {volume} {679}},\ \bibinfo {pages} {1260}
  (\bibinfo {year} {2008})},\ \Eprint {http://arxiv.org/abs/0802.2265}
  {arXiv:0802.2265 [astro-ph]} \BibitemShut {NoStop}%
\bibitem [{\citenamefont {{Boylan-Kolchin}}\ \emph {et~al.}(2010)\citenamefont
  {{Boylan-Kolchin}}, \citenamefont {{Springel}}, \citenamefont {{White}},\
  and\ \citenamefont {{Jenkins}}}]{bolyan10}%
  \BibitemOpen
  \bibfield  {author} {\bibinfo {author} {\bibfnamefont {M.}~\bibnamefont
  {{Boylan-Kolchin}}}, \bibinfo {author} {\bibfnamefont {V.}~\bibnamefont
  {{Springel}}}, \bibinfo {author} {\bibfnamefont {S.~D.~M.}\ \bibnamefont
  {{White}}}, \ and\ \bibinfo {author} {\bibfnamefont {A.}~\bibnamefont
  {{Jenkins}}},\ }\href {\doibase 10.1111/j.1365-2966.2010.16774.x} {\bibfield
  {journal} {\bibinfo  {journal} {\mnras}\ }\textbf {\bibinfo {volume} {406}},\
  \bibinfo {pages} {896} (\bibinfo {year} {2010})},\ \Eprint
  {http://arxiv.org/abs/0911.4484} {arXiv:0911.4484 [astro-ph.CO]} \BibitemShut
  {NoStop}%
\bibitem [{\citenamefont {{Gao}}\ \emph {et~al.}(2012)\citenamefont {{Gao}},
  \citenamefont {{Navarro}}, \citenamefont {{Frenk}}, \citenamefont
  {{Jenkins}}, \citenamefont {{Springel}},\ and\ \citenamefont
  {{White}}}]{gao12}%
  \BibitemOpen
  \bibfield  {author} {\bibinfo {author} {\bibfnamefont {L.}~\bibnamefont
  {{Gao}}}, \bibinfo {author} {\bibfnamefont {J.~F.}\ \bibnamefont
  {{Navarro}}}, \bibinfo {author} {\bibfnamefont {C.~S.}\ \bibnamefont
  {{Frenk}}}, \bibinfo {author} {\bibfnamefont {A.}~\bibnamefont {{Jenkins}}},
  \bibinfo {author} {\bibfnamefont {V.}~\bibnamefont {{Springel}}}, \ and\
  \bibinfo {author} {\bibfnamefont {S.~D.~M.}\ \bibnamefont {{White}}},\ }\href
  {\doibase 10.1111/j.1365-2966.2012.21564.x} {\bibfield  {journal} {\bibinfo
  {journal} {\mnras}\ }\textbf {\bibinfo {volume} {425}},\ \bibinfo {pages}
  {2169} (\bibinfo {year} {2012})},\ \Eprint {http://arxiv.org/abs/1201.1940}
  {arXiv:1201.1940 [astro-ph.CO]} \BibitemShut {NoStop}%
\bibitem [{\citenamefont {{Wu}}\ \emph {et~al.}(2013)\citenamefont {{Wu}},
  \citenamefont {{Hahn}}, \citenamefont {{Wechsler}}, \citenamefont
  {{Behroozi}},\ and\ \citenamefont {{Mao}}}]{wu13}%
  \BibitemOpen
  \bibfield  {author} {\bibinfo {author} {\bibfnamefont {H.-Y.}\ \bibnamefont
  {{Wu}}}, \bibinfo {author} {\bibfnamefont {O.}~\bibnamefont {{Hahn}}},
  \bibinfo {author} {\bibfnamefont {R.~H.}\ \bibnamefont {{Wechsler}}},
  \bibinfo {author} {\bibfnamefont {P.~S.}\ \bibnamefont {{Behroozi}}}, \ and\
  \bibinfo {author} {\bibfnamefont {Y.-Y.}\ \bibnamefont {{Mao}}},\ }\href
  {\doibase 10.1088/0004-637X/767/1/23} {\bibfield  {journal} {\bibinfo
  {journal} {\apj}\ }\textbf {\bibinfo {volume} {767}},\ \bibinfo {eid} {23}
  (\bibinfo {year} {2013})},\ \Eprint {http://arxiv.org/abs/1210.6358}
  {arXiv:1210.6358 [astro-ph.CO]} \BibitemShut {NoStop}%
\bibitem [{\citenamefont {{Springel}}\ \emph {et~al.}(2008)\citenamefont
  {{Springel}}, \citenamefont {{Wang}}, \citenamefont {{Vogelsberger}},
  \citenamefont {{Ludlow}}, \citenamefont {{Jenkins}}, \citenamefont {{Helmi}},
  \citenamefont {{Navarro}}, \citenamefont {{Frenk}},\ and\ \citenamefont
  {{White}}}]{aquarius08}%
  \BibitemOpen
  \bibfield  {author} {\bibinfo {author} {\bibfnamefont {V.}~\bibnamefont
  {{Springel}}}, \bibinfo {author} {\bibfnamefont {J.}~\bibnamefont {{Wang}}},
  \bibinfo {author} {\bibfnamefont {M.}~\bibnamefont {{Vogelsberger}}},
  \bibinfo {author} {\bibfnamefont {A.}~\bibnamefont {{Ludlow}}}, \bibinfo
  {author} {\bibfnamefont {A.}~\bibnamefont {{Jenkins}}}, \bibinfo {author}
  {\bibfnamefont {A.}~\bibnamefont {{Helmi}}}, \bibinfo {author} {\bibfnamefont
  {J.~F.}\ \bibnamefont {{Navarro}}}, \bibinfo {author} {\bibfnamefont {C.~S.}\
  \bibnamefont {{Frenk}}}, \ and\ \bibinfo {author} {\bibfnamefont {S.~D.~M.}\
  \bibnamefont {{White}}},\ }\href {\doibase 10.1111/j.1365-2966.2008.14066.x}
  {\bibfield  {journal} {\bibinfo  {journal} {\mnras}\ }\textbf {\bibinfo
  {volume} {391}},\ \bibinfo {pages} {1685} (\bibinfo {year} {2008})},\ \Eprint
  {http://arxiv.org/abs/0809.0898} {arXiv:0809.0898 [astro-ph]} \BibitemShut
  {NoStop}%
\bibitem [{\citenamefont {{Vogelsberger}}\ \emph {et~al.}(2016)\citenamefont
  {{Vogelsberger}}, \citenamefont {{Zavala}}, \citenamefont {{Cyr-Racine}},
  \citenamefont {{Pfrommer}}, \citenamefont {{Bringmann}},\ and\ \citenamefont
  {{Sigurdson}}}]{ethos16}%
  \BibitemOpen
  \bibfield  {author} {\bibinfo {author} {\bibfnamefont {M.}~\bibnamefont
  {{Vogelsberger}}}, \bibinfo {author} {\bibfnamefont {J.}~\bibnamefont
  {{Zavala}}}, \bibinfo {author} {\bibfnamefont {F.-Y.}\ \bibnamefont
  {{Cyr-Racine}}}, \bibinfo {author} {\bibfnamefont {C.}~\bibnamefont
  {{Pfrommer}}}, \bibinfo {author} {\bibfnamefont {T.}~\bibnamefont
  {{Bringmann}}}, \ and\ \bibinfo {author} {\bibfnamefont {K.}~\bibnamefont
  {{Sigurdson}}},\ }\href {\doibase 10.1093/mnras/stw1076} {\bibfield
  {journal} {\bibinfo  {journal} {\mnras}\ }\textbf {\bibinfo {volume} {460}},\
  \bibinfo {pages} {1399} (\bibinfo {year} {2016})},\ \Eprint
  {http://arxiv.org/abs/1512.05349} {arXiv:1512.05349 [astro-ph.CO]}
  \BibitemShut {NoStop}%
\bibitem [{\citenamefont {{Diemand}}\ \emph {et~al.}(2007)\citenamefont
  {{Diemand}}, \citenamefont {{Kuhlen}},\ and\ \citenamefont
  {{Madau}}}]{diemand07}%
  \BibitemOpen
  \bibfield  {author} {\bibinfo {author} {\bibfnamefont {J.}~\bibnamefont
  {{Diemand}}}, \bibinfo {author} {\bibfnamefont {M.}~\bibnamefont {{Kuhlen}}},
  \ and\ \bibinfo {author} {\bibfnamefont {P.}~\bibnamefont {{Madau}}},\ }\href
  {\doibase 10.1086/510736} {\bibfield  {journal} {\bibinfo  {journal} {\apj}\
  }\textbf {\bibinfo {volume} {657}},\ \bibinfo {pages} {262} (\bibinfo {year}
  {2007})},\ \Eprint {http://arxiv.org/abs/astro-ph/0611370}
  {arXiv:astro-ph/0611370 [astro-ph]} \BibitemShut {NoStop}%
\bibitem [{\citenamefont {{Auger}}\ \emph {et~al.}(2009)\citenamefont
  {{Auger}}, \citenamefont {{Treu}}, \citenamefont {{Bolton}}, \citenamefont
  {{Gavazzi}}, \citenamefont {{Koopmans}}, \citenamefont {{Marshall}},
  \citenamefont {{Bundy}},\ and\ \citenamefont {{Moustakas}}}]{auger09}%
  \BibitemOpen
  \bibfield  {author} {\bibinfo {author} {\bibfnamefont {M.~W.}\ \bibnamefont
  {{Auger}}}, \bibinfo {author} {\bibfnamefont {T.}~\bibnamefont {{Treu}}},
  \bibinfo {author} {\bibfnamefont {A.~S.}\ \bibnamefont {{Bolton}}}, \bibinfo
  {author} {\bibfnamefont {R.}~\bibnamefont {{Gavazzi}}}, \bibinfo {author}
  {\bibfnamefont {L.~V.~E.}\ \bibnamefont {{Koopmans}}}, \bibinfo {author}
  {\bibfnamefont {P.~J.}\ \bibnamefont {{Marshall}}}, \bibinfo {author}
  {\bibfnamefont {K.}~\bibnamefont {{Bundy}}}, \ and\ \bibinfo {author}
  {\bibfnamefont {L.~A.}\ \bibnamefont {{Moustakas}}},\ }\href {\doibase
  10.1088/0004-637X/705/2/1099} {\bibfield  {journal} {\bibinfo  {journal}
  {\apj}\ }\textbf {\bibinfo {volume} {705}},\ \bibinfo {pages} {1099}
  (\bibinfo {year} {2009})},\ \Eprint {http://arxiv.org/abs/0911.2471}
  {arXiv:0911.2471 [astro-ph.CO]} \BibitemShut {NoStop}%
\bibitem [{\citenamefont {{Press}}\ and\ \citenamefont
  {{Schechter}}(1974)}]{press}%
  \BibitemOpen
  \bibfield  {author} {\bibinfo {author} {\bibfnamefont {W.~H.}\ \bibnamefont
  {{Press}}}\ and\ \bibinfo {author} {\bibfnamefont {P.}~\bibnamefont
  {{Schechter}}},\ }\href {\doibase 10.1086/152650} {\bibfield  {journal}
  {\bibinfo  {journal} {\apj}\ }\textbf {\bibinfo {volume} {187}},\ \bibinfo
  {pages} {425} (\bibinfo {year} {1974})}\BibitemShut {NoStop}%
\bibitem [{\citenamefont {{Warren}}\ \emph {et~al.}(2006)\citenamefont
  {{Warren}}, \citenamefont {{Abazajian}}, \citenamefont {{Holz}},\ and\
  \citenamefont {{Teodoro}}}]{warren06}%
  \BibitemOpen
  \bibfield  {author} {\bibinfo {author} {\bibfnamefont {M.~S.}\ \bibnamefont
  {{Warren}}}, \bibinfo {author} {\bibfnamefont {K.}~\bibnamefont
  {{Abazajian}}}, \bibinfo {author} {\bibfnamefont {D.~E.}\ \bibnamefont
  {{Holz}}}, \ and\ \bibinfo {author} {\bibfnamefont {L.}~\bibnamefont
  {{Teodoro}}},\ }\href {\doibase 10.1086/504962} {\bibfield  {journal}
  {\bibinfo  {journal} {\apj}\ }\textbf {\bibinfo {volume} {646}},\ \bibinfo
  {pages} {881} (\bibinfo {year} {2006})},\ \Eprint
  {http://arxiv.org/abs/astro-ph/0506395} {arXiv:astro-ph/0506395 [astro-ph]}
  \BibitemShut {NoStop}%
\bibitem [{\citenamefont {{Tinker}}\ \emph {et~al.}(2008)\citenamefont
  {{Tinker}}, \citenamefont {{Kravtsov}}, \citenamefont {{Klypin}},
  \citenamefont {{Abazajian}}, \citenamefont {{Warren}}, \citenamefont
  {{Yepes}}, \citenamefont {{Gottl{\"o}ber}},\ and\ \citenamefont
  {{Holz}}}]{tinker08}%
  \BibitemOpen
  \bibfield  {author} {\bibinfo {author} {\bibfnamefont {J.}~\bibnamefont
  {{Tinker}}}, \bibinfo {author} {\bibfnamefont {A.~V.}\ \bibnamefont
  {{Kravtsov}}}, \bibinfo {author} {\bibfnamefont {A.}~\bibnamefont
  {{Klypin}}}, \bibinfo {author} {\bibfnamefont {K.}~\bibnamefont
  {{Abazajian}}}, \bibinfo {author} {\bibfnamefont {M.}~\bibnamefont
  {{Warren}}}, \bibinfo {author} {\bibfnamefont {G.}~\bibnamefont {{Yepes}}},
  \bibinfo {author} {\bibfnamefont {S.}~\bibnamefont {{Gottl{\"o}ber}}}, \ and\
  \bibinfo {author} {\bibfnamefont {D.~E.}\ \bibnamefont {{Holz}}},\ }\href
  {\doibase 10.1086/591439} {\bibfield  {journal} {\bibinfo  {journal} {\apj}\
  }\textbf {\bibinfo {volume} {688}},\ \bibinfo {pages} {709} (\bibinfo {year}
  {2008})},\ \Eprint {http://arxiv.org/abs/0803.2706} {arXiv:0803.2706
  [astro-ph]} \BibitemShut {NoStop}%
\bibitem [{\citenamefont {{Castro}}\ \emph {et~al.}(2016)\citenamefont
  {{Castro}}, \citenamefont {{Marra}},\ and\ \citenamefont
  {{Quartin}}}]{castro16}%
  \BibitemOpen
  \bibfield  {author} {\bibinfo {author} {\bibfnamefont {T.}~\bibnamefont
  {{Castro}}}, \bibinfo {author} {\bibfnamefont {V.}~\bibnamefont {{Marra}}}, \
  and\ \bibinfo {author} {\bibfnamefont {M.}~\bibnamefont {{Quartin}}},\ }\href
  {\doibase 10.1093/mnras/stw2072} {\bibfield  {journal} {\bibinfo  {journal}
  {\mnras}\ }\textbf {\bibinfo {volume} {463}},\ \bibinfo {pages} {1666}
  (\bibinfo {year} {2016})},\ \Eprint {http://arxiv.org/abs/1605.07548}
  {arXiv:1605.07548 [astro-ph.CO]} \BibitemShut {NoStop}%
\bibitem [{\citenamefont {Hezaveh}\ \emph
  {et~al.}(2016{\natexlab{b}})\citenamefont {Hezaveh} \emph
  {et~al.}}]{Hezaveh:2016ltk}%
  \BibitemOpen
  \bibfield  {author} {\bibinfo {author} {\bibfnamefont {Y.~D.}\ \bibnamefont
  {Hezaveh}} \emph {et~al.},\ }\href {\doibase 10.3847/0004-637X/823/1/37}
  {\bibfield  {journal} {\bibinfo  {journal} {Astrophys. J.}\ }\textbf
  {\bibinfo {volume} {823}},\ \bibinfo {pages} {37} (\bibinfo {year}
  {2016}{\natexlab{b}})},\ \Eprint {http://arxiv.org/abs/1601.01388}
  {arXiv:1601.01388 [astro-ph.CO]} \BibitemShut {NoStop}%
\bibitem [{\citenamefont {{Brownstein}}\ \emph {et~al.}(2012)\citenamefont
  {{Brownstein}}, \citenamefont {{Bolton}}, \citenamefont {{Schlegel}},
  \citenamefont {{Eisenstein}}, \citenamefont {{Kochanek}}, \citenamefont
  {{Connolly}}, \citenamefont {{Maraston}}, \citenamefont {{Pandey}},
  \citenamefont {{Seitz}}, \citenamefont {{Wake}}, \citenamefont
  {{Wood-Vasey}}, \citenamefont {{Brinkmann}}, \citenamefont {{Schneider}},\
  and\ \citenamefont {{Weaver}}}]{bells_i}%
  \BibitemOpen
  \bibfield  {author} {\bibinfo {author} {\bibfnamefont {J.~R.}\ \bibnamefont
  {{Brownstein}}}, \bibinfo {author} {\bibfnamefont {A.~S.}\ \bibnamefont
  {{Bolton}}}, \bibinfo {author} {\bibfnamefont {D.~J.}\ \bibnamefont
  {{Schlegel}}}, \bibinfo {author} {\bibfnamefont {D.~J.}\ \bibnamefont
  {{Eisenstein}}}, \bibinfo {author} {\bibfnamefont {C.~S.}\ \bibnamefont
  {{Kochanek}}}, \bibinfo {author} {\bibfnamefont {N.}~\bibnamefont
  {{Connolly}}}, \bibinfo {author} {\bibfnamefont {C.}~\bibnamefont
  {{Maraston}}}, \bibinfo {author} {\bibfnamefont {P.}~\bibnamefont
  {{Pandey}}}, \bibinfo {author} {\bibfnamefont {S.}~\bibnamefont {{Seitz}}},
  \bibinfo {author} {\bibfnamefont {D.~A.}\ \bibnamefont {{Wake}}}, \bibinfo
  {author} {\bibfnamefont {W.~M.}\ \bibnamefont {{Wood-Vasey}}}, \bibinfo
  {author} {\bibfnamefont {J.}~\bibnamefont {{Brinkmann}}}, \bibinfo {author}
  {\bibfnamefont {D.~P.}\ \bibnamefont {{Schneider}}}, \ and\ \bibinfo {author}
  {\bibfnamefont {B.~A.}\ \bibnamefont {{Weaver}}},\ }\href {\doibase
  10.1088/0004-637X/744/1/41} {\bibfield  {journal} {\bibinfo  {journal}
  {\apj}\ }\textbf {\bibinfo {volume} {744}},\ \bibinfo {eid} {41} (\bibinfo
  {year} {2012})},\ \Eprint {http://arxiv.org/abs/1112.3683} {arXiv:1112.3683
  [astro-ph.CO]} \BibitemShut {NoStop}%
\bibitem [{\citenamefont {{Auger}}\ \emph {et~al.}(2010)\citenamefont
  {{Auger}}, \citenamefont {{Treu}}, \citenamefont {{Bolton}}, \citenamefont
  {{Gavazzi}}, \citenamefont {{Koopmans}}, \citenamefont {{Marshall}},
  \citenamefont {{Moustakas}},\ and\ \citenamefont {{Burles}}}]{slacs2}%
  \BibitemOpen
  \bibfield  {author} {\bibinfo {author} {\bibfnamefont {M.~W.}\ \bibnamefont
  {{Auger}}}, \bibinfo {author} {\bibfnamefont {T.}~\bibnamefont {{Treu}}},
  \bibinfo {author} {\bibfnamefont {A.~S.}\ \bibnamefont {{Bolton}}}, \bibinfo
  {author} {\bibfnamefont {R.}~\bibnamefont {{Gavazzi}}}, \bibinfo {author}
  {\bibfnamefont {L.~V.~E.}\ \bibnamefont {{Koopmans}}}, \bibinfo {author}
  {\bibfnamefont {P.~J.}\ \bibnamefont {{Marshall}}}, \bibinfo {author}
  {\bibfnamefont {L.~A.}\ \bibnamefont {{Moustakas}}}, \ and\ \bibinfo {author}
  {\bibfnamefont {S.}~\bibnamefont {{Burles}}},\ }\href {\doibase
  10.1088/0004-637X/724/1/511} {\bibfield  {journal} {\bibinfo  {journal}
  {\apj}\ }\textbf {\bibinfo {volume} {724}},\ \bibinfo {pages} {511} (\bibinfo
  {year} {2010})},\ \Eprint {http://arxiv.org/abs/1007.2880} {arXiv:1007.2880
  [astro-ph.CO]} \BibitemShut {NoStop}%
\bibitem [{\citenamefont {{Robertson}}\ \emph {et~al.}(2020)\citenamefont
  {{Robertson}}, \citenamefont {{Smith}}, \citenamefont {{Massey}},
  \citenamefont {{Eke}}, \citenamefont {{Jauzac}}, \citenamefont {{Bianconi}},\
  and\ \citenamefont {{Ryczanowski}}}]{robertson20}%
  \BibitemOpen
  \bibfield  {author} {\bibinfo {author} {\bibfnamefont {A.}~\bibnamefont
  {{Robertson}}}, \bibinfo {author} {\bibfnamefont {G.~P.}\ \bibnamefont
  {{Smith}}}, \bibinfo {author} {\bibfnamefont {R.}~\bibnamefont {{Massey}}},
  \bibinfo {author} {\bibfnamefont {V.}~\bibnamefont {{Eke}}}, \bibinfo
  {author} {\bibfnamefont {M.}~\bibnamefont {{Jauzac}}}, \bibinfo {author}
  {\bibfnamefont {M.}~\bibnamefont {{Bianconi}}}, \ and\ \bibinfo {author}
  {\bibfnamefont {D.}~\bibnamefont {{Ryczanowski}}},\ }\href@noop {} {\bibfield
   {journal} {\bibinfo  {journal} {arXiv e-prints}\ ,\ \bibinfo {eid}
  {arXiv:2002.01479}} (\bibinfo {year} {2020})},\ \Eprint
  {http://arxiv.org/abs/2002.01479} {arXiv:2002.01479 [astro-ph.CO]}
  \BibitemShut {NoStop}%
\bibitem [{\citenamefont {Birrer}\ and\ \citenamefont
  {Amara}(2018)}]{lenstronomy}%
  \BibitemOpen
  \bibfield  {author} {\bibinfo {author} {\bibfnamefont {S.}~\bibnamefont
  {Birrer}}\ and\ \bibinfo {author} {\bibfnamefont {A.}~\bibnamefont {Amara}},\
  }\href {\doibase 10.1016/j.dark.2018.11.002} {\  (\bibinfo {year} {2018}),\
  10.1016/j.dark.2018.11.002},\ \Eprint {http://arxiv.org/abs/arXiv:1803.09746}
  {arXiv:1803.09746} \BibitemShut {NoStop}%
\bibitem [{\citenamefont {{Cyr-Racine}}\ \emph {et~al.}(2019)\citenamefont
  {{Cyr-Racine}}, \citenamefont {{Keeton}},\ and\ \citenamefont
  {{Moustakas}}}]{beyond_subhalos}%
  \BibitemOpen
  \bibfield  {author} {\bibinfo {author} {\bibfnamefont {F.-Y.}\ \bibnamefont
  {{Cyr-Racine}}}, \bibinfo {author} {\bibfnamefont {C.~R.}\ \bibnamefont
  {{Keeton}}}, \ and\ \bibinfo {author} {\bibfnamefont {L.~A.}\ \bibnamefont
  {{Moustakas}}},\ }\href {\doibase 10.1103/PhysRevD.100.023013} {\bibfield
  {journal} {\bibinfo  {journal} {\prd}\ }\textbf {\bibinfo {volume} {100}},\
  \bibinfo {eid} {023013} (\bibinfo {year} {2019})},\ \Eprint
  {http://arxiv.org/abs/1806.07897} {arXiv:1806.07897 [astro-ph.CO]}
  \BibitemShut {NoStop}%
\bibitem [{\citenamefont {{Bayer}}\ \emph {et~al.}(2018)\citenamefont
  {{Bayer}}, \citenamefont {{Chatterjee}}, \citenamefont {{Koopmans}},
  \citenamefont {{Vegetti}}, \citenamefont {{McKean}}, \citenamefont {{Treu}},\
  and\ \citenamefont {{Fassnacht}}}]{bayer_power}%
  \BibitemOpen
  \bibfield  {author} {\bibinfo {author} {\bibfnamefont {D.}~\bibnamefont
  {{Bayer}}}, \bibinfo {author} {\bibfnamefont {S.}~\bibnamefont
  {{Chatterjee}}}, \bibinfo {author} {\bibfnamefont {L.~V.~E.}\ \bibnamefont
  {{Koopmans}}}, \bibinfo {author} {\bibfnamefont {S.}~\bibnamefont
  {{Vegetti}}}, \bibinfo {author} {\bibfnamefont {J.~P.}\ \bibnamefont
  {{McKean}}}, \bibinfo {author} {\bibfnamefont {T.}~\bibnamefont {{Treu}}}, \
  and\ \bibinfo {author} {\bibfnamefont {C.~D.}\ \bibnamefont {{Fassnacht}}},\
  }\href@noop {} {\bibfield  {journal} {\bibinfo  {journal} {arXiv e-prints}\
  ,\ \bibinfo {eid} {arXiv:1803.05952}} (\bibinfo {year} {2018})},\ \Eprint
  {http://arxiv.org/abs/1803.05952} {arXiv:1803.05952 [astro-ph.GA]}
  \BibitemShut {NoStop}%
\bibitem [{\citenamefont {Alexander}\ \emph {et~al.}(2020)\citenamefont
  {Alexander}, \citenamefont {Gleyzer}, \citenamefont {McDonough},
  \citenamefont {Toomey},\ and\ \citenamefont {Usai}}]{Alexander:2019puy}%
  \BibitemOpen
  \bibfield  {author} {\bibinfo {author} {\bibfnamefont {S.}~\bibnamefont
  {Alexander}}, \bibinfo {author} {\bibfnamefont {S.}~\bibnamefont {Gleyzer}},
  \bibinfo {author} {\bibfnamefont {E.}~\bibnamefont {McDonough}}, \bibinfo
  {author} {\bibfnamefont {M.~W.}\ \bibnamefont {Toomey}}, \ and\ \bibinfo
  {author} {\bibfnamefont {E.}~\bibnamefont {Usai}},\ }\href {\doibase
  10.3847/1538-4357/ab7925} {\bibfield  {journal} {\bibinfo  {journal}
  {Astrophys. J.}\ }\textbf {\bibinfo {volume} {893}},\ \bibinfo {pages} {15}
  (\bibinfo {year} {2020})},\ \Eprint {http://arxiv.org/abs/1909.07346}
  {arXiv:1909.07346 [astro-ph.CO]} \BibitemShut {NoStop}%
\bibitem [{\citenamefont {Varma}\ \emph {et~al.}(2020)\citenamefont {Varma},
  \citenamefont {Fairbairn},\ and\ \citenamefont {Figueroa}}]{Varma:2020kbq}%
  \BibitemOpen
  \bibfield  {author} {\bibinfo {author} {\bibfnamefont {S.}~\bibnamefont
  {Varma}}, \bibinfo {author} {\bibfnamefont {M.}~\bibnamefont {Fairbairn}}, \
  and\ \bibinfo {author} {\bibfnamefont {J.}~\bibnamefont {Figueroa}},\
  }\href@noop {} {\  (\bibinfo {year} {2020})},\ \Eprint
  {http://arxiv.org/abs/2005.05353} {arXiv:2005.05353 [astro-ph.CO]}
  \BibitemShut {NoStop}%
\bibitem [{\citenamefont {{Oguri}}\ and\ \citenamefont
  {{Marshall}}(2010)}]{LSST_lenses}%
  \BibitemOpen
  \bibfield  {author} {\bibinfo {author} {\bibfnamefont {M.}~\bibnamefont
  {{Oguri}}}\ and\ \bibinfo {author} {\bibfnamefont {P.~J.}\ \bibnamefont
  {{Marshall}}},\ }\href {\doibase 10.1111/j.1365-2966.2010.16639.x} {\bibfield
   {journal} {\bibinfo  {journal} {\mnras}\ }\textbf {\bibinfo {volume}
  {405}},\ \bibinfo {pages} {2579} (\bibinfo {year} {2010})},\ \Eprint
  {http://arxiv.org/abs/1001.2037} {arXiv:1001.2037 [astro-ph.CO]} \BibitemShut
  {NoStop}%
\bibitem [{\citenamefont {{Pawase}}\ \emph {et~al.}(2014)\citenamefont
  {{Pawase}}, \citenamefont {{Courbin}}, \citenamefont {{Faure}}, \citenamefont
  {{Kokotanekova}},\ and\ \citenamefont {{Meylan}}}]{2014MNRAS.439.3392P}%
  \BibitemOpen
  \bibfield  {author} {\bibinfo {author} {\bibfnamefont {R.~S.}\ \bibnamefont
  {{Pawase}}}, \bibinfo {author} {\bibfnamefont {F.}~\bibnamefont {{Courbin}}},
  \bibinfo {author} {\bibfnamefont {C.}~\bibnamefont {{Faure}}}, \bibinfo
  {author} {\bibfnamefont {R.}~\bibnamefont {{Kokotanekova}}}, \ and\ \bibinfo
  {author} {\bibfnamefont {G.}~\bibnamefont {{Meylan}}},\ }\href {\doibase
  10.1093/mnras/stu179} {\bibfield  {journal} {\bibinfo  {journal} {\mnras}\
  }\textbf {\bibinfo {volume} {439}},\ \bibinfo {pages} {3392} (\bibinfo {year}
  {2014})},\ \Eprint {http://arxiv.org/abs/1206.3412} {arXiv:1206.3412
  [astro-ph.CO]} \BibitemShut {NoStop}%
\bibitem [{\citenamefont {Collett}(2015)}]{Collett_2015}%
  \BibitemOpen
  \bibfield  {author} {\bibinfo {author} {\bibfnamefont {T.~E.}\ \bibnamefont
  {Collett}},\ }\href {\doibase 10.1088/0004-637x/811/1/20} {\bibfield
  {journal} {\bibinfo  {journal} {The Astrophysical Journal}\ }\textbf
  {\bibinfo {volume} {811}},\ \bibinfo {pages} {20} (\bibinfo {year}
  {2015})}\BibitemShut {NoStop}%
\bibitem [{\citenamefont {Sauer}(2013)}]{sauer2013numerical}%
  \BibitemOpen
  \bibfield  {author} {\bibinfo {author} {\bibfnamefont {T.}~\bibnamefont
  {Sauer}},\ }\href {https://books.google.com/books?id=NTxungEACAAJ} {\emph
  {\bibinfo {title} {Numerical Analysis}}},\ Pearson custom library\ (\bibinfo
  {publisher} {Pearson},\ \bibinfo {year} {2013})\BibitemShut {NoStop}%
\end{thebibliography}%
